\documentclass[reprint, twocolumn, amsfonts, amsmath, amssymb, superscriptaddress, prx]{revtex4-2}

\usepackage{graphicx}
\usepackage{dcolumn}
\usepackage{bm}
\usepackage{physics}
\usepackage{layouts}
\usepackage[normalem]{ulem}

\usepackage[dvipsnames]{xcolor}
\usepackage{xfrac}
\usepackage[linkcolor=Blue,citecolor=Blue,urlcolor=Blue,colorlinks=true]{hyperref}
\usepackage{notes2bib}
\bibnotesetup{ note-name = , use-sort-key = false}

\usepackage{mathtools}
\usepackage{xpatch}
\usepackage{siunitx}

\usepackage{cancel}

\usepackage[inline]{enumitem}
\usepackage{xcolor}
\usepackage{lipsum}
\usepackage{ulem}
\definecolor{myGreen}{RGB}{89,132,63}
\definecolor{cadmiumorange}{rgb}{0.93, 0.53, 0.18}
\definecolor{olivegreen}{rgb}{0.43, 0.72, 0.48}
\definecolor{wildstrawberry}{rgb}{1., 0.26, 0.64}
\definecolor{mypurple}{rgb}{50.2, 0., 50.2}
\definecolor{darkpastelgreen}{rgb}{0.1, 0.75, 0.24}

\usepackage[normalem]{ulem}
\renewcommand{\sout}[1]{\unskip}   

\usepackage{soul}
\newcommand{\soutr}[1]{\st{#1}}
\renewcommand{\soutr}[1]{\unskip}   

\usepackage{graphicx}
\usepackage{dcolumn}
\usepackage{bm}
\usepackage{mathtools}
\usepackage[dvipsnames]{xcolor}
\usepackage[linkcolor=Blue,citecolor=Blue,urlcolor=Blue,colorlinks=true]{hyperref}

\begin{document}

\title{Design principles for fast and efficient self-assembly processes}

\author{Florian M. Gartner}
\affiliation{%
Arnold Sommerfeld Center for Theoretical Physics (ASC) and Center for NanoScience (CeNS), Department of Physics,
Ludwig-Maximilians-Universit\"at M\"unchen, Theresienstra\ss e 37, D-80333 Munich, Germany
}%
\author{Erwin Frey}%
\email{frey@lmu.de}
\affiliation{%
Arnold Sommerfeld Center for Theoretical Physics (ASC) and Center for NanoScience (CeNS), Department of Physics,
Ludwig-Maximilians-Universit\"at M\"unchen, Theresienstra\ss e 37, D-80333 Munich, Germany
}%
\affiliation{%
Max Planck School Matter to Life, Hofgartenstra\ss e 8, D-80539 Munich, Germany
}%

\date{\today}

\begin{abstract}
Self-assembly is a fundamental concept in biology and of significant interest to nanotechnology. 
Significant progress has been made in characterizing and controlling the properties of the resulting structures, both experimentally and theoretically. 
However, much less is known about kinetic constraints and determinants of dynamical properties like time efficiency, although these constraints can become severe limiting factors of self-assembly processes.
Here we investigate how the time efficiency and other dynamical properties of reversible self-assembly depend on the morphology (shape) of the building blocks for systems in which the binding energy between the constituents is large. 
As paradigmatic examples, we stochastically simulate the self-assembly of constituents with triangular, square, and hexagonal morphology into two-dimensional structures of a specified size.
We find that the constituents' morphology critically determines the assembly time and how it scales with the size of the target structure. 
Our analysis reveals three key structural parameters defined by the morphology:
The nucleation size and attachment order, which describe the effective order of the chemical reactions by which clusters nucleate and grow, respectively, and the growth exponent, which determines how the growth rate of an emerging structure scales with its size. 
Using this characterization, we formulate an effective theory of the self-assembly kinetics, which we show exhibits an inherent scale invariance. This allows us to identify general scaling laws that describe the minimal assembly time as a function of the size of the target structure. 
We show how these insights on the kinetics of self-assembly processes can be used to design assembly schemes that could significantly increase the time efficiency and robustness of artificial self-assembly processes.
\end{abstract}

\maketitle

\section{Introduction}
\label{sec:introduction} 

\begin{figure*} 
\centering
\includegraphics[width=\linewidth]{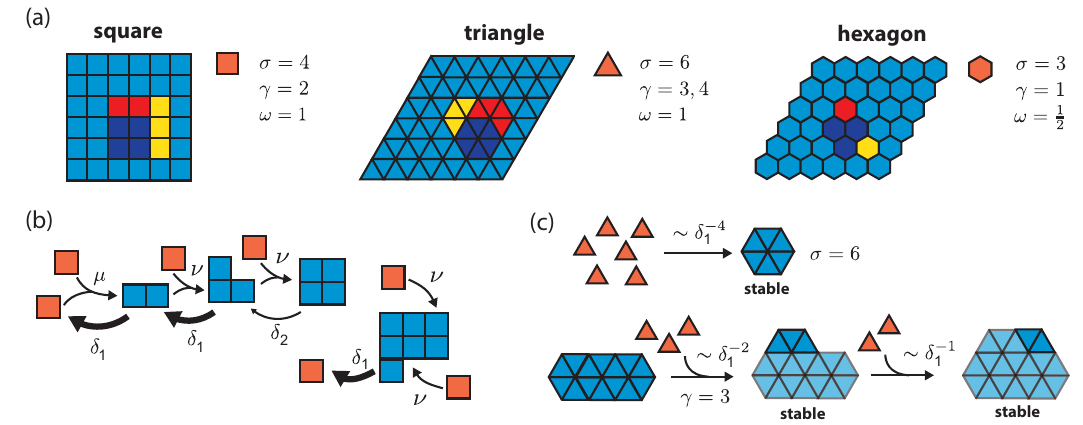}
\caption{
\textbf{Illustration of the model}. 
(a) We consider the self-assembly of two-dimensional structures of volume (total number of constituents) $S$ from square-, triangle- or hexagon-shaped building blocks and ask for the time the assembly processes take to reach substantial yields. The configurations shaded in dark blue, red, and yellow illustrate the first intermediate assembly states with enhanced stability (wherein each particle has at least two bonds). Typical assembly pathways consecutively pass through these (or equivalent) intermediate states. 
The dynamics can thus be effectively described by the three parameters $\sigma$, $\gamma$, and $\omega$, which are listed for the three cases and denote the \textit{nucleation size}, \textit{attachment order},  and \textit{growth exponent}, respectively (see panel (c) and main text Secs.~\ref{sec:predominant_assembly_pathways} and \ref{sec:effective_model}). 
(b) Self-assembly dynamics is modeled as follows (illustrated for the system with square-shaped monomers): 
Two monomers dimerize at rate $\mu$ along any of their edges. 
Subsequently, monomers attach at rate $\nu$ to any free neighboring site until a square consisting of $S$ monomers with edge length $L=\sqrt{S}$ ($L=\sqrt{S/2}$ in the case of triangular monomers) is completed as illustrated in (a) assuming periodic boundary conditions between top and bottom and left and right. 
At the same time, particles can also detach from a given structure at rate ${\delta_n \sim \exp[-n E_B / k_\text{B} T]}$, with $n$ denoting the number of bonds between the particle and the structure and $E_B$ the binding energy per bond. 
(c) Coarse-grained effective dynamics (illustrated for the system with triangle-shaped monomers): 
Assembly starts with forming a stable nucleus (dark blue shaded states in A), which requires a minimum number of $\sigma$ particles (nucleation size). 
The effective nucleation rate scales as $1/\delta_1^{\sigma-2}$ since $\sigma\,{-}\,2$ unstable states must be traversed to form the nucleus. 
Analogously, attachment processes are initiated by an effective reaction of order $\gamma$, (attachment order), after which a series of reactions of smaller (subleading) order may follow. As a particularity of triangular monomers, the first attachment reaction right after nucleation has attachment order $\gamma=4$, while all subsequent attachment reactions have a smaller attachment order $\gamma=3$. }
\label{fig:model}
\end{figure*}

Self-assembly is an essential concept in biology that explains the formation of large functional structures from smaller subunits and thereby `bridges the gap’ between different length scales \cite{kushner1969, Whitesides2002a}. 
Artificial self-assembly has recently attracted considerable interest, promising numerous applications in nanotechnology, medicine, and biology~\cite{Whitesides1991, Whitesides2002a, Zhang2003a, Mendes2013, Sigl2021}.
Typically, the formation of artificial structures \cite{Zhang2003a}, as well as many biological structures such as virus capsids \cite{mateu2013assembly, johnston2010modelling} and micelles \cite{menger1979structure}, is promoted by weak and reversible interactions between subunits \cite{Whitesides2002b, Freeman2018, Ceres2002, Grant2011, Zlotnick2003, Rapaport2008, Hagan2014, Perlmutter2015, gelbart2012micelles}. 
Consequently, single bonds between subunits are usually unstable, and a large number of mutual interactions is required before an emerging larger structure becomes stabilized as a whole. 
In this way, kinetic traps are eluded that would arise if too many stable intermediates are formed that cannot be completed with the available amount of resources~\cite{Endres2002, Hagan2011, Deeds2012, Perlmutter2015, Whitelam2015, Jacobs2016}. 
While there are various strategies to evade kinetic traps even in irreversible self-assembly processes~\cite{Baek2015, Murugan2015, Gartner2021}, here we focus exclusively on reversible reaction kinetics, which plays a pivotal role in nanotechnology and biology~\cite{Wagenbauer2017, Grant2011, Rapaport2008}. 

Reversible self-assembly (in thermal equilibrium systems) can be viewed as a process in which the system minimizes its free energy and thereby reaches a state in which a large fraction of the monomers are tied to complete structures \cite{Katen2009, Grzybowski2009, Hagan2014, Whitelam2015}. 
This notion motivates using ensemble theory from statistical physics to characterize the final steady state \cite{gelbart2012micelles, Zlotnick1994, Hedges2014, Hagan2011, Hagan2021}. 
Indeed, equilibrium methods have been applied successfully to a broad class of self-assembling systems \cite{Israelachvili1976,hiemenz2007polymer, Katen2009, Evans2017, lenz2017geometrical} and---when applicable---represent powerful and convenient tools to characterize the resulting final structure and the yield.  

However, equilibrium methods have a clear limitation in that they do not provide insights into the dynamic aspects of self-assembly processes.
In particular, they do not inform about the time the system requires to achieve a specified yield. 
However, understanding the kinetics of self-assembly processes is important, as experiments clearly show that the time required to complete self-assembly processes can become a severe obstacle, especially when the target structures are large \cite{Sigl2021}.  
Hence, it is indispensable for the advancement of the field to gain a deeper understanding of those factors that control kinetic properties, like the time efficiency of self-assembly. 

An essential feature of self-assembling systems is the morphology (shape) of the building blocks. For example, several icosahedral virus capsids are assembled from triangular or trapezoidal capsid proteins \cite{masson2012viralzone}. 
Carboxysomes---another example of biological structures with icosahedral symmetry---emerge from pentagon- and hexagon-shaped building blocks \cite{Kerfeld2016}. 
In DNA nanotechnology, simple regular shapes are typically used like squares in algorithmic self-assembly \cite{Evans2017} or triangles to build artificial capsids \cite{Sigl2021}. 
The shape of the constituents might play an important role as it determines how the subunits connect with each other when forming higher-order assemblies. 
Does this ``connectivity'' and thus the shape of the subunits crucially influence the speed of self-assembly? Do the typical morphologies of the above examples support efficient self-assembly, or could constituents with another morphology assemble even faster? 

We use a generic conceptual model for reversible self-assembly to address these basic questions.
As paradigmatic examples, we perform stochastic simulations of systems composed of monomers with triangular, square, and hexagonal shapes that self-assemble into two-dimensional structures of a specified size [Fig.~\ref{fig:model}(a)]. 
As an essential and informative characteristic, we thereby analyze the 'time complexity,' describing how the minimally required assembly time scales with the size of the target structure. 
A similar concept was applied previously by us to quantify the time efficiency of different (reversible and irreversible) self-assembly scenarios disregarding effects arising due to the morphology of the constituents~\cite{Gartner2021}.
Here we show that the time efficiency and its scaling as a function of structure size (\textit{time complexity}) for reversible self-assembly processes depend strongly on the morphology of the constituents. 
Using an effective kinetic model, we determine analytic estimates for these scaling laws describing how the assembly time scales with the size of the target structure and the detachment rate. 
These findings suggest that the morphology of the building blocks is a major determinant of the time efficiency of self-assembly processes and implies various possibilities to enhance the time efficiency of artificial self-assembly: 
For example, by enforcing a hierarchical assembly step, monomers can be induced to first form higher-order assemblies with a more suitable morphology, which subsequently assembles significantly faster than the original monomers. 
We simulate such assembly schemes explicitly and show that, in this way, the assembly time can be decreased by up to several orders of magnitude. 
This is an important result because it shows that a putatively unimportant detail, like the precise shape of the building blocks, can critically influence the time efficiency and even its scaling behavior in non-equilibrium processes.  

This paper is structured as follows: 
In Sec.~II, we introduce the mathematical model and the central concepts. 
In Sec.~III, we discuss the phenomenology of the self-assembly dynamics for the case in which the binding energy between the subunits is large, and thus only leading order detachment reactions need to be considered (strong binding limit of self-assembly). 
As a major result, we show that the required assembly time exhibits an interesting scaling behavior as a function of the target structure size, depending on the morphology of the building blocks. 
In Sec.~IV, we formulate an effective mathematical model for self-assembly in the strong binding limit, and in Sec.~V, we show that this theory reveals an underlying scale invariance, thereby explaining the observed scaling behavior. 
Finally, in Sec.~VI, we summarize the results and discuss their implications and possible applications in biology and nanotechnology.

\section{Model description} 
\label{sec:model_description}

We consider the self-assembly of $N$ identical particles (building blocks, monomers) of triangular, square, and hexagonal shapes into two-dimensional square-shaped structures of size 
(number of constituents) $S$ assuming chemical reaction kinetics in a well-mixed fluid environment [Fig.~\ref{fig:model}(a)]. 
We assume that during the self-assembly process monomers from a well-mixed, dilute solution bind to sparse clusters. These conditions typically apply to a broad class of biological and experimental self-assembly systems like, e.g., virus capsid assembly for various virus species \cite{hagan2014modeling}, self-assembly of artificial DNA nano bricks \cite{ke2012three}, and self-assembly of various cellular organelles \cite{alberts2017molecular}. 
Our model does not cover systems in which spatial degrees of freedom or scaffolding play a significant role (e.g., as studied in Refs.~\cite{ianiro2019liquid,Perlmutter2015,Li2018}) or where monomers organize into spatial aggregates before forming the final products \cite{ye2013competition,fang2020two,auer2001prediction}.
Furthermore, we assume that all binding reactions are specific and result in correctly assembled intermediates.
This excludes systems with a significant rate for incorrect bonding \cite{hayakawa2022geometrically, Hagan2006a} or where particles have multiple different binding partners (e.g., see Ref.~\cite{evans2024pattern}); albeit we expect that the model can be generalized in future work to include these additional effects.
To simplify the analysis, we assume that the final structures have periodic boundary conditions, i.e., they form closed tori.
The boundary conditions, representing the geometry of the target structures, induce a mechanism for self-limiting growth of the clusters \cite{Hagan2021}. 
Below, we show that our results apply identically to other kinds of boundary conditions and even to systems with unlimited cluster growth. 
In the case of periodic boundary conditions, it can be shown mathematically [cf.~App.~Sec.~\ref{app:sec:heterogen_identical_rates}] that in the limit of large particle numbers, it is irrelevant for the assembly dynamics whether the particles are heterogeneous and bind only with specific neighbors (`information-rich' systems \cite{ke2012three,wei2012complex,Jacobs2016,jacobs2015rational,reinhardt2016effects}) or whether all particles are identical \cite{Wagenbauer2017} and bind indistinctly with each other \cite{gartner2020stochastic, Gartner2021}. 
For convenience, throughout the main text, we therefore assume homogeneous systems in which the particles are indistinguishable. In App.~Sec.~\ref{app:sec:heterogeneous_systems}, we show that our results apply equally well to heterogeneous (`information-rich') systems if the reaction rate constants and concentrations of all species are the same, and we discuss several special cases (e.g., where the concentrations or reaction rates of the species differ strongly). 
In both the homogeneous and the heterogeneous case, we do not consider the possibility of incorrect binding of the monomers, e.g., binding in a wrong orientation or with the wrong binding partners.
The initial concentration of monomers is denoted by ${C=N/V}$, with $V$ being the reaction volume.

The self-assembly dynamics of the model are defined by the following reaction kinetics [Fig.~\ref{fig:model}(b)]: 
Any two building blocks (monomers) can bind with rate $\mu$ along any edge to form a dimer, which can subsequently grow at rate $\nu$ by further attachment of monomers at any free neighboring site. 
We formally distinguish the dimerization rate from the attachment rate because cooperative binding effects might disfavor the dimerization of monomers over the attachment of monomers to larger structures in biological or experimental systems. 
However, we generally set ${\mu = \nu}$ in the simulations. 
Following the assumptions of classical or ideal aggregation theory \cite{Hagan2021}, we assume that structures grow only by the attachment of single monomers, while interactions among larger oligomers are neglected~ \cite{Chen2008, Ben-Shaul1994}. 
This assumption is generally justified since the concentration of monomers is usually much larger than those of oligomers. 
However, we will assess below to which extent ideal aggregation theory is a reasonable assumption specifically for the different particle morphologies considered here. 

Furthermore, to account for the reversibility of the binding reactions, it is assumed that single monomers detach from existing structures at a rate $\delta_n$ that decreases exponentially with the number $n$ of bonds that need to be broken (Arrhenius’ law for diffusion-limited processes \cite{arrhenius1889reaktionsgeschwindigkeit, laidler1984development}): 
\begin{align}  
\label{definition_delta_n}
    \delta_n 
    = 
    A \, \exp \big[ -n \, E_B / k_\text{B} T \big]
    \, .
\end{align}
Here, $E_B$ is the binding energy per bond formed between the building blocks, and $A$ is a constant that sets the unbinding rate's overall scale. 
The building blocks of triangular, square, and hexagonal shapes differ in the maximum number $z_\text{bond}$ of bonds they can form with other building blocks: ${z_\text{bond}=3}$, $4$, and $6$, respectively.
In particular, if complete structures are to be stable, the binding energy must be large enough so that the detachment rate $\delta_{z_\text{bond}}$ is sufficiently small. 

The initial number $N$ of monomer determines the maximum number $N/S$ of complete structures that can be formed. The yield at time $t$ is defined as the actual number of complete structures divided by their maximum number. In other words, it represents the fraction of resources bound into complete structures.  
Furthermore, to quantify the time efficiency of the self-assembly processes, we define the assembly time $T_{90}$, which measures the time at which the yield first surpasses a value of 90\%.
In particular, we will investigate how the minimal assembly time $T_{90}^\text{min}$ that can be achieved by tuning the experimental control parameters depends on the structure size $S$, and we quantify its asymptotic scaling behavior.     

The dynamics of the assembly process can be controlled by changing the ratio between the frequencies of detachment and attachment events, i.e., by changing the detachment rates relative to the overall scale of the reaction kinetics: ${\tilde \delta_n:= \delta_n / (C \nu)}$. 
Assuming detailed balance, the control parameters can also be written as ${\tilde \delta_n \sim \exp \big[\Delta G_n / k_\text{B} T \big]}$, where $\Delta G_n$ is the free energy difference between an $n$-fold bound subunit and a free subunit.   
Experimentally, this implies various possibilities to control the assembly process. 
For example, it can be controlled thermodynamically by changing parameters that affect the free energy difference $\Delta G_n$, like temperature, binding energy, pH value, or salt concentration, in addition to changing the monomer concentration $C$. 
Additional possibilities to control the assembly process kinetically arise, assuming that the monomers have different internal states, among which they switch under energy consumption, whereby detailed balance may be broken \cite{ben2021nonequilibrium, Gartner2021, faran2023nonequilibrium}.    
In this kinetic approach \cite{krapivsky2010kinetic}, therefore, we describe the system's behavior in its most general form in terms of the dimensionless control parameters ${\tilde \delta_n =\delta_n/(C\nu)}$. 
Similarly, the resulting assembly time is measured relative to the reactive time scale in units $(C\nu)^{-1}$. 
We simulate the stochastic dynamics of the system using Gillespie's algorithm \cite{Gillespie2007}; 
see App.~\ref{app_stochastic_simulation} for details on how the simulations were implemented.

\section{Scaling laws}
\label{sec:irreversible_limit}

\begin{figure}[thb]
\centering
\includegraphics[width=1.0\columnwidth]{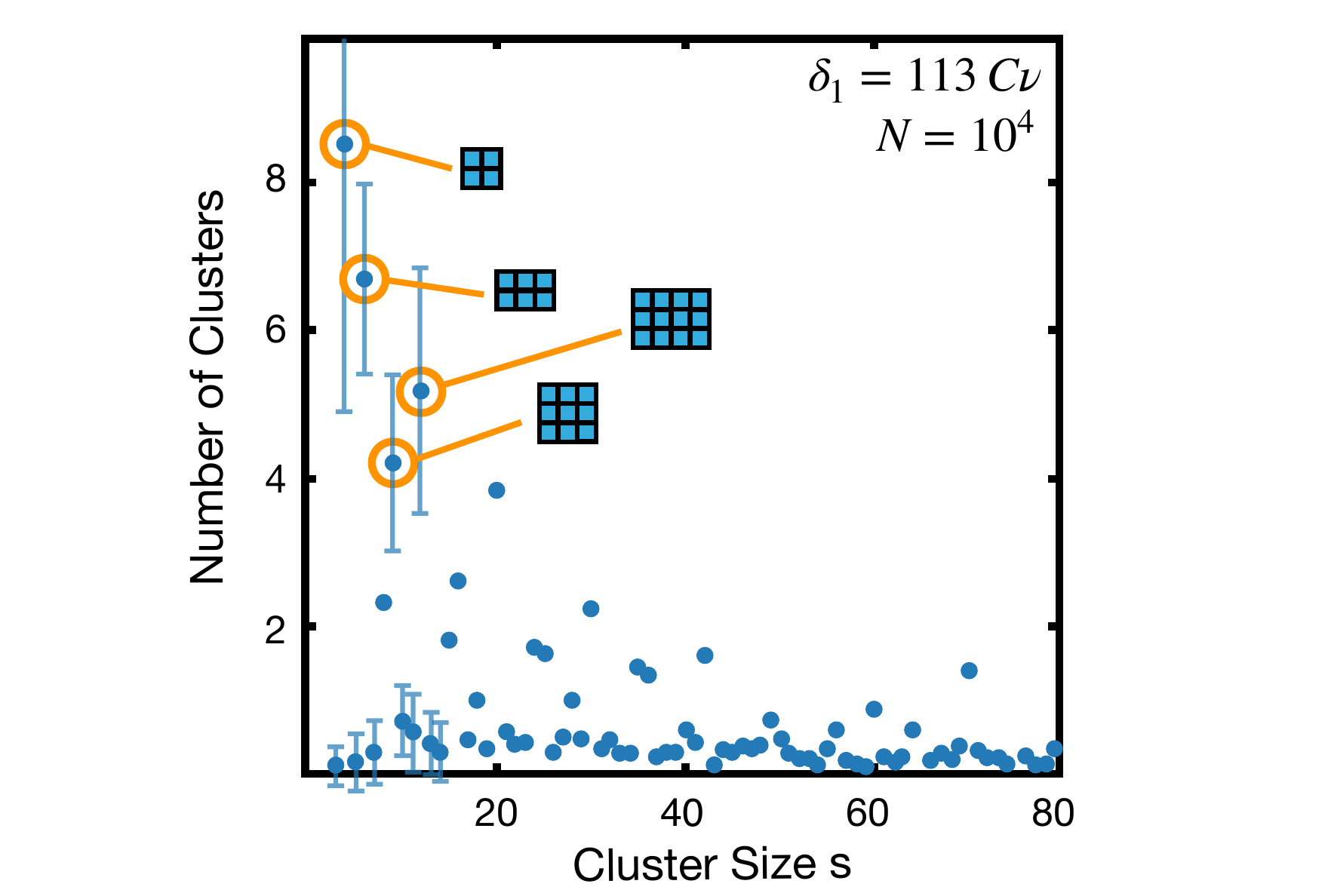}
\caption{
\textbf{Typical cluster size distribution} of a system with square-shaped monomers at a fixed, early time point (${T = 0.1 \cdot T_{90}}$) averaged over $100$ independent simulation runs with target structure size ${S = 81}$. 
Parameters: ${E_B = 10 \ k_\text{B} T}$ and ${A = 2.5 \cdot 10^6 \, C \nu}$. Error bars show the sample standard deviation for a few selected cluster sizes.
Structure sizes that allow for stable configurations are significantly more frequent on average than those which do not allow for a stable configuration. 
A few of these stable configurations are illustrated explicitly. 
In an effective theory, one can exploit this feature and use a reduced state space that considers only the most likely connection pathways between these stable configurations; see Section~\ref{sec:effective_model}.
}
\label{fig:cluster_size_distribution}
\end{figure}

In systems with sufficiently strong bonds between the building blocks, ${E_B \gg k_\text{B} T}$, the detachment rate $\delta_1$ is significantly larger than the higher order detachment rates $\delta_n$ with ${n \geq 2}$. 
The resulting separation of time scales implies that cluster configurations where each particle forms at least two bonds with neighboring particles are significantly more stable than configurations in which particles are connected with only one bond. 
Consequently, the self-assembly process typically proceeds through a path that leads through these intermediate states of enhanced stability. 
Hallmarks of these dynamics can be inferred from the typical distribution of cluster sizes in the assembly process. 
For example, Fig.~\ref{fig:cluster_size_distribution} shows the cluster size distribution for square-shaped monomers at a fixed time point, averaged over $100$ independent runs of the stochastic simulation.
We observe that cluster sizes that allow for configurations of enhanced stability are significantly more frequent on average than those that do not [similar results based on molecular dynamics simulations are reported in Ref.~\cite{Rapaport2008}; Refs.~\cite{jacobs2015rational,reinhardt2016effects} furthermore demonstrate that the prevalence of particular assembly pathways can be shown with calculations of the free energy landscape]. 
This suggests that the detachment rate $\delta_1$ plays a key role in the assembly dynamics as it determines the prevalence of intermediate configurations and controls the assembly pathway. 

Throughout the rest of this paper, we consider the limit where the higher order detachment rates $\delta_n$ with ${n \geq 2}$ are so small that they effectively do not affect the assembly dynamics on the relevant time scales and can thus be neglected. 
This limit can also be understood as the \textit{leading order} in an expansion in which one truncates the sequence of detachment rates ${\delta_1 \ll \delta_2 \ll \delta_3 \ldots}$ at successively higher order. 
The idea is that---also for intermediate or small binding energies---the leading order effect already describes essential features of the kinetics of self-assembly processes. 
In a follow-up paper, we will investigate the diametric case of low binding energies and analyze the resulting effect of higher-order detachment rates on the self-assembly dynamics.
Note that if ${\delta_n = 0}$ for ${n \geq 2}$, configurations in which each particle has two bonds cannot disassemble anymore; hence the assembly process contains irreversible steps. 
We will refer to this case, which is the subject of this paper, also as the strong binding limit of self-assembly dynamics. 

For a given binding energy $E_B$, a characteristic time scale ${\tau_\text{irr} \approx \delta_2^{-1} = \delta_1^{-1} \, e^{\beta E_B}}$ with ${\beta = 1 / k_\text{B} T}$ can be defined, below which higher-order detachment processes are negligible, and the dynamics will be accurately described by the strong binding limit. 
Conversely, for any given time scale $\tau$,  one can specify a minimal binding energy ${\beta E_{\text{irr}} \approx \log{(\tau \, \delta_1)}}$, so that on time scales smaller than $\tau$ the systems behaves effectively as described by the strong binding limit.

\subsection{Predominant assembly pathways}
\label{sec:predominant_assembly_pathways}

In the strong binding limit, one can characterize the self-assembly pathways as follows:
The assembly process begins with the formation of a nucleus, which requires a minimum number of $\sigma$ building blocks to achieve a stable configuration where each building block has two binding partners.
The \textit{nucleation size} $\sigma$ depends on the morphology of the building blocks [Fig.~\ref{fig:model}(a,c)]. 
Subsequent attachment processes proceed in two steps, as illustrated in Fig.~\ref{fig:model}(c): 
First, a leading order reaction involving $\gamma$ building blocks must occur, where we refer to $\gamma$ as the \textit{attachment order}, to reach a new stable configuration. 
Again, the morphology of the building blocks determines the value of the attachment order \footnote{Note that in the system with triangle-shaped monomers, the attachment order for the first assembly step following nucleation is larger, ${\gamma=4}$, compared to all subsequent assembly steps, where ${\gamma=3}$.}.
Second, for larger clusters, this leading-order attachment event may be followed by a series of lower-order reactions where the cluster traverses further stable configurations until another leading-order $\gamma$ reaction is required. 
The probability of all the above processes is proportional to $\tilde \delta_1^{-u}$, where $u$ denotes the number of unstable states passed through before the next stable configuration is reached. 
This will be the basis for the effective kinetic theory that we derive in Section~\ref{sec:effective_model}. 
Note that, during the nucleation process, the system passes through ${u_\text{nuc} = \sigma-2}$ unstable states, while in a leading attachment reaction, ${u_\text{att} = \gamma-1}$ unstable states are passed through. 
Therefore, if ${\sigma > \gamma+1}$, the effective nucleation rate can be slowed down relative to the growth rate of existing structures by tuning the control parameter $\tilde \delta_1$. 
Since a reduced nucleation rate allows for more structures to get completed, the yield increases as a function of $\tilde \delta_1$ as we discuss in the next section. 

We would like to remark that the concept of morphology is more general than merely the shape of the monomers. For example, we discuss in Sec.~\ref{sec:summary_and_applications} how square-shaped monomers with six instead of four binding sites can have the same morphological parameters $\sigma$ and $\gamma$ as hexagonal monomers (cf.~Fig.~\ref{fig:optimal_morphologies_examples}(a)). Furthermore, icosahedral virus capsids often self-assemble from triangle-shaped monomers but the assembly process starts with the formation of a pentamer ($\sigma=5$) instead of a hexamer ($\sigma=6$) as in our case \cite{kushner1969,masson2012viralzone}. Hence, the parameters $\sigma$ and $\gamma$ do not uniquely relate neither with the shape of the monomers nor their number of binding sites. Nevertheless, in the context of our specific system, we will often use the terms `shape' and `morphology' interchangeably.

\subsection{Scaling of the yield curves}

\begin{figure}[!t]
\centering
\includegraphics[width=\linewidth]{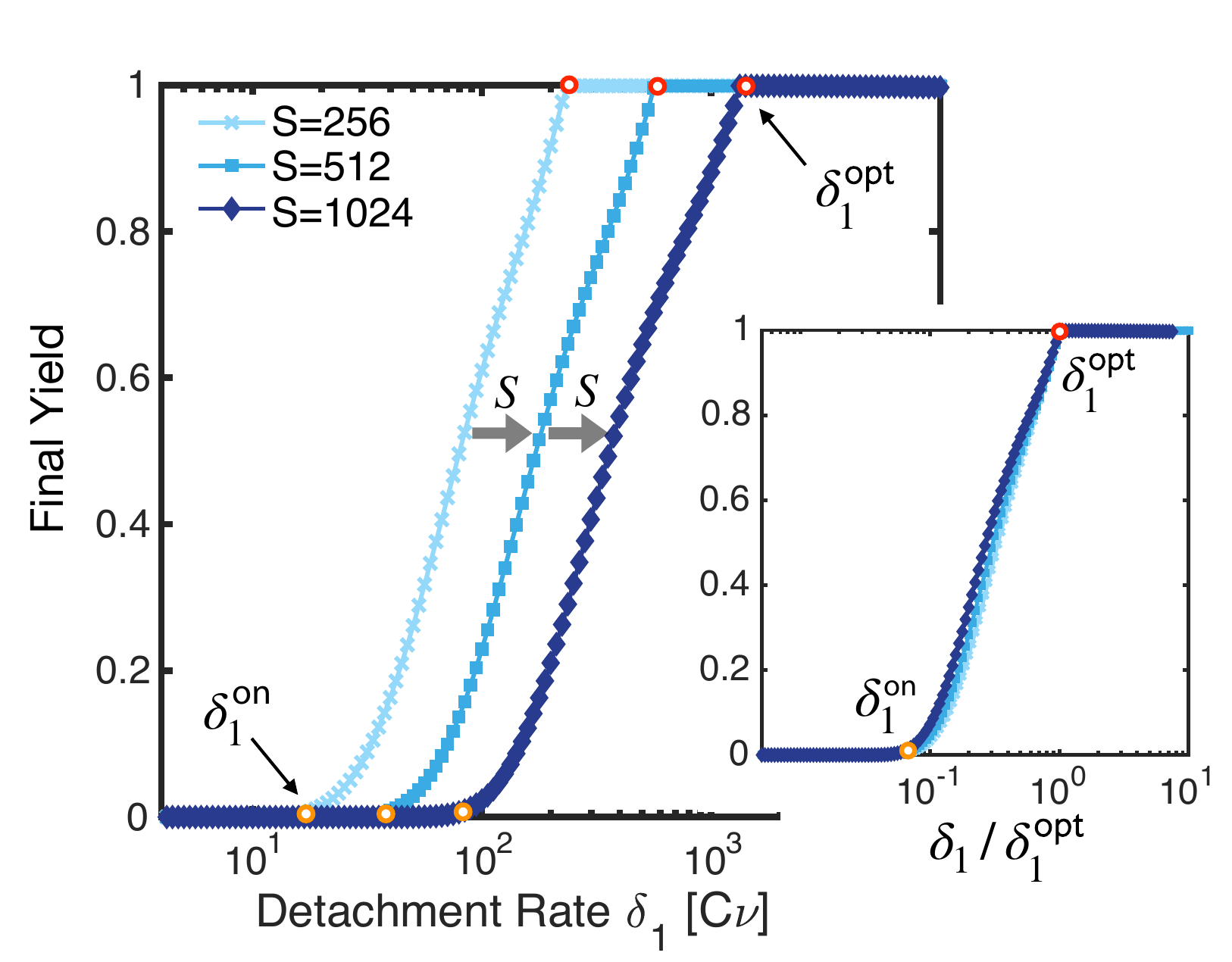}
\caption{
\textbf{Final yield in the strong binding limit} as a function of the detachment rate $\tilde \delta_1$ ($\delta_1$ in units of $C\,\nu$) for different target structure sizes $S$ as indicated in the graph. Data points represent averages over ten independent runs of the stochastic simulation performed with square-shaped monomers and particle numbers $N=200\,S$.
By increasing $\tilde \delta_1$, the final yield increases from 0 to a perfect value of 1. Larger target structure sizes require a larger value of $\tilde \delta_1$ to achieve a specified yield. 
\textit{Inset:} 
Rescaling of the detachment rate $\tilde \delta_1$ by the optimal detachment rate $\tilde \delta_{1}^\text{opt} (S)$ leads to a data collapse of the yield curves. 
}
\label{fig:yield_scaling}
\end{figure}

For reaction kinetics where the detachment rates are ${\delta_n = 0}$ for ${n \geq 2}$, stable intermediate configurations no longer decay, and the assembly process thus contains irreversible steps (strong binding limit).
Consequently, the system will eventually reach an absorbing state where all monomers are bound into complete structures or stable intermediate states. 
The self-assembly process can therefore be uniquely characterized by its final yield, which is found to be a monotonic function of $\tilde \delta_1$. Our stochastic simulations reveal the following characteristic features [Fig.~\ref{fig:yield_scaling}]:
If the ratio $\tilde \delta_1$ is small, structures nucleate very quickly, and there are kinetic traps, resulting in a poor final yield.
Increasing $\tilde \delta_1$ slows down nucleation relative to cluster growth (provided ${\sigma > \gamma+1}$), thereby increasing the number of structures that get completed but also the time required to reach the final state. 

The optimal values of the detachment rate, $\tilde \delta_1^\text{opt} (S)$, with which one can achieve a 90\% yield in the smallest amount of time, are shown as open red circles on the yield curves in Fig.~\ref{fig:yield_scaling} for different target sizes $S$.
The optimal values happen to approximately coincide with the values of $\tilde \delta_1$ for which 100\% yield is reached for the first time.
Furthermore, the optimal values are an increasing function of the target structure size $S$. 
Remarkably, rescaling the detachment rate $\tilde \delta_1$ by the optimal detachment rate $\tilde \delta_{1}^\text{opt}$ leads to a data collapse of the yield curves [inset of Fig.~\ref{fig:yield_scaling}].
In particular, this implies that $\tilde \delta_1^\text{opt} (S)$ scales identically with the structure size $S$ as the onset rate for finite yield $\tilde \delta_1^\text{on} (S)$, i.e., the value of the control parameter at which the final yield is larger than zero for the first time (shown as orange circles in Fig.~\ref{fig:yield_scaling}). 
Using this scaling equivalence will allow us to simplify the analysis in Section~\ref{sec:onset_criterion} since it permits us to work in a regime in which the yield is negligible. 
Further below, in Section~\ref{sec:general_scaling_theory}, we will prove the validity of the phenomenological scaling law for the yield curves based on a scale invariance of the underlying reaction kinetics.

\begin{figure*}[!t]
\centering
\includegraphics[width=1.0\linewidth]{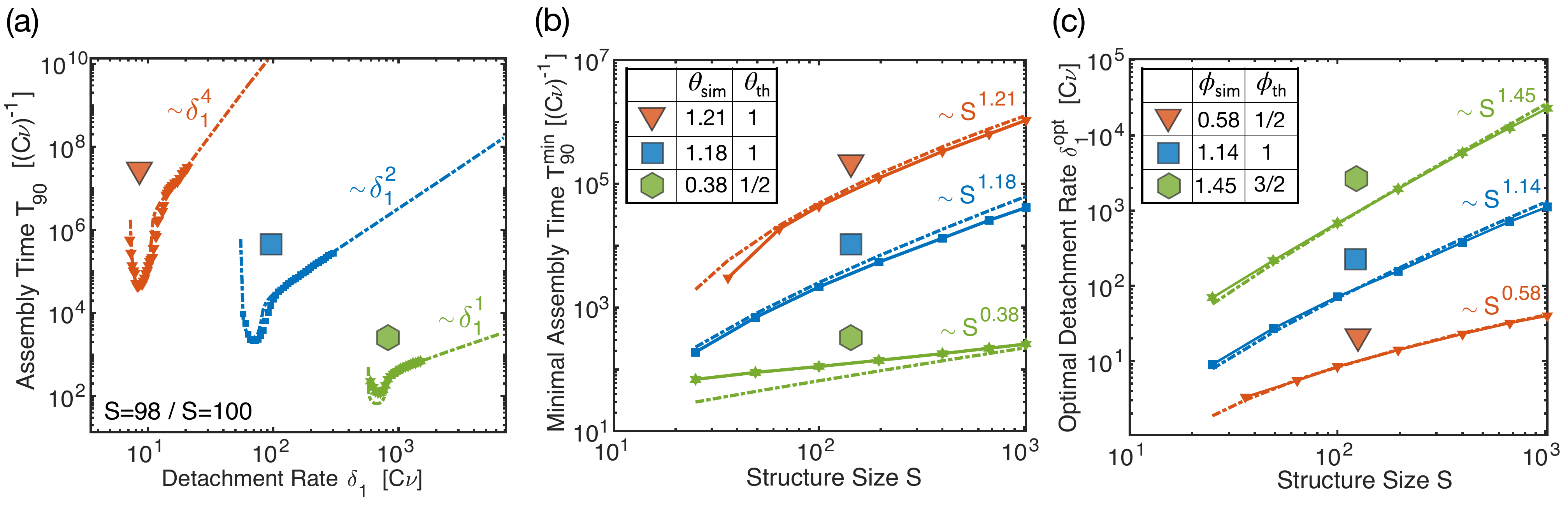}
\caption{
\textbf{Assembly time and optimal detachment parameter in the strong binding limit}.  
(a) The time $T_{90}$ required to achieve 90\% yield in the stochastic simulations (markers) is plotted against the detachment rate $\tilde \delta_1$ ($\delta_1$ in units of $C \nu$) for systems with triangle- (red), square- (blue) and hexagon-shaped (green) building blocks for a fixed target structure size ${S = 98}$ (triangle-shaped building blocks) or ${S = 100}$ (square- and hexagon-shaped building blocks) \footnote{Note that in the case of triangle-shaped building blocks, there is no regular structure consisting of ${S = 100}$ particles since regular structures consist of $2 \, L^2$ (with an integer $L$) triangles; compare Fig.~\ref{fig:model}(a)). 
Hence we chose a structure size ${S = 98}$ (${L = 7}$), which is as close as possible to the structure size ${S = 100}$ for square and hexagonal monomers.}. Simulations were performed with particle numbers $N=1000 \, S$ (square- and hexagon-shaped building blocks) and $N=200 \, S$ (triangle-shaped building blocks).  
Dashed-dotted lines represent the prediction of the effective theory obtained by numerically integrating Eq.~\eqref{effective_theory_1st_order}. 
For sufficiently large detachment rates, the assembly time scales proportionally to $\tilde \delta_1^{\sigma-2}$, while the assembly time diverges in the limit of small detachment rates. 
The value of $\tilde \delta_1$ at which the assembly time attains its minimum, $T_{90}^{\text{min}}$, defines the optimal detachment rate $\tilde \delta_1^\text{opt}$.
The minimal assembly time $T_{90}^\text{min}$ (b) and the optimal detachment rate $\tilde \delta_1^\text{opt}$ (c) inferred from the stochastic simulations (markers and solid lines) and effective theory (dashed-dotted lines) are plotted against the size $S$ of the target structure.  
They both exhibit approximate power-law dependencies for all three morphologies of the building blocks. 
Tables show the scaling exponents $\theta_\text{sim}$ and $\phi_\text{sim}$ inferred from the last three data points of the stochastic simulation in comparison with their theoretical asymptotic values $\theta_\text{th}$ and $\phi_\text{th}$ derived by the mathematical analysis; see  Eq.~\eqref{eq:control_parameter_exponent} and Eq.~\eqref{eq:time_complexity_exponent}.}
\label{fig:scaling_laws}
\end{figure*}

\subsection{Scaling laws for the optimal detachment rate and the minimal assembly time}
\label{sec:scaling_delta_and_T90min}

The assembly time $T_{90}$ as a function of the detachment rate $\tilde \delta_1$ in the strong binding limit is displayed in Fig.~\ref{fig:scaling_laws}(a) for the different morphologies of the building blocks. 
The curves exhibit a characteristic shape: 
For small $\tilde \delta_1$, the assembly time shows a U-shape with a pronounced minimum, followed by a power law scaling ${T_{90} \sim \tilde \delta_1^{\sigma-2}}$ as $\tilde \delta_1$ increases.
This power law results from the fact that for large $\tilde \delta_1$, the formation of the nucleus is the rate-limiting step in the assembly process and that $\sigma-2$ unstable states must be traversed to form a nucleus [cf. Fig.~\ref{fig_supp:illustration_nucleation}]. 
In contrast, the regime of the U-shape is characterized by simultaneous structure assembly: Heuristically, if the detachment rate is close to its optimal value, the time scale of nucleation is slightly smaller than the time scale of growth of an entire structure. Therefore, the required number of nuclei can form initially within a short time interval while the concentration of monomers is still large. Since the rate of nucleation depends strongly on the monomer concentration, the assembly time is very sensitive to variations in $\tilde \delta_1$ in this regime, resulting in the typical U-shape of the $T_{90}-$curves with a pronounced minimum.

Remarkably, the minimum assembly times for the three different morphologies differ by almost three orders of magnitude.
Even more strikingly, the minimum assembly times $T_{90}^\text{min}$ at the optimal detachment rates $\tilde \delta_1^\text{opt} (S)$ behave differently as a function of the structure size $S$ [Fig.~\ref{fig:scaling_laws}(b)]: 
For triangle- and square-shaped building blocks, the minimum assembly time increases approximately as a power law in the structure size with exponent ${\theta \approx 1.2}$ (for ${S \gg 1}$), while for hexagon-shaped building blocks, it increases with a much smaller exponent ${\theta \approx 0.4}$.
Similarly, the optimal detachment rate shows approximate power law dependencies on the target structure size $S$ with characteristic exponents $\phi$ for each particle morphology [Fig.~\ref{fig:scaling_laws}(c)]. 
We will refer to the (asymptotic) exponents $\theta$ and $\phi$ as \textit{time complexity} and \textit{control parameter exponent}, respectively~\cite{Gartner2021}. 
Differences in the time complexity exponents imply that the disparities in the assembly times of the various monomer morphologies become ever more pronounced with the increasing size of the target structure. 
We thus conclude that for the self-assembly of large objects, the morphology of the building blocks plays a crucial role, as it decisively determines the time efficiency of their self-assembly process. 

Please note that for the minima depicted in Fig.~\ref{fig:scaling_laws} (a) to be not affected by higher order detachment rates, a minimal binding energy $E_B^{\text{irr}}/(k_\text{B} \, T) \approx \log{(T_{90}^{\text{min}} \, \delta_1^{\text{opt}})}$ between 10 and 12 is required as a sufficient condition. 
These values appear plausible, even though typical values for the binding energies between subunits, for example, in virus capsid assembly, are reported to be still a bit smaller \cite{Hagan2011, Perlmutter2015, Hagan2006a}. We thus expect that the minimal assembly time in typical biological and experimental systems is reasonably well described by the approximation that neglects higher order detachment rates. 

The assembly time can alternatively be characterized as a function of the total average binding energy of a completed structure, $E_{\text{tot}} = \tfrac{1}{2} S \,  z_{\text{bond}} \, E_B$, taking into account the total number $z_{\text{bond}}$ of bonds per particle. Using $E_B / (k_B \, T) = \log{(A/ \delta_1)}$ and $\tilde \delta_1 \approx S^{\phi_{\text{th}}}$, we find that the \textit{optimal} total binding energies of structures formed from triangle-, square-, and hexagon-shaped monomers are generally different but become identical if $\tilde A \approx S^{2.5}$ ($E_{\text{tot}}^{\text{opt}} \approx 3 \, S \, \log S $). Upon crossing this manifold, the total binding energies for the different monomer morphologies change their relative ordering. Thus, it is important to note that system-specific stability requirements can affect the choice of the optimal particle morphology. \\
 
\textit{Adjoined time complexity exponents.} 
The time complexity exponent $\theta$ above describes the scaling of the assembly time in units of the reactive time scale $(C\nu)^{-1}$. 
Thus, if the monomer concentration $C$ or binding rate $\nu$ are changed rather than the binding energy $E_B$ or temperature $T$, the change in time scale must be taken into account as well.
For example, assuming that binding energy and temperature are held constant, and only $C$ or $\nu$ are changed to optimize the control parameter $\tilde \delta_1$ for a given structure size $S$, the minimal assembly time (in physical units) scales as ${\sim S^{\theta+\phi}:=S^{\tilde \theta}}$. 
We call ${\tilde \theta=\theta+\phi}$ the \textit{adjoined} time complexity exponent. 
Table~\ref{tab:adjoined_exponents} displays the values of the adjoined exponents for the different particle morphologies.
The adjoined exponents are larger than the time complexity exponents, indicating that self-assembly in the strong binding limit is more efficient if $C$ and $\nu$ are kept constant (optimally at their maximum values), and the binding energy or temperature are varied (the product $C\nu$ then fixes the minimal achievable assembly time in a specific system). We will further deepen this discussion in Sec.~\ref{sec:interpretation_scaling_results}.
Also note that the adjoined exponents rank the particle morphologies differently concerning their efficiency, with triangular and hexagonal monomers having the smallest adjoined exponents in the limit of large structure sizes. 
\begin{table}[h]
\centering
\includegraphics[width=0.55\linewidth]{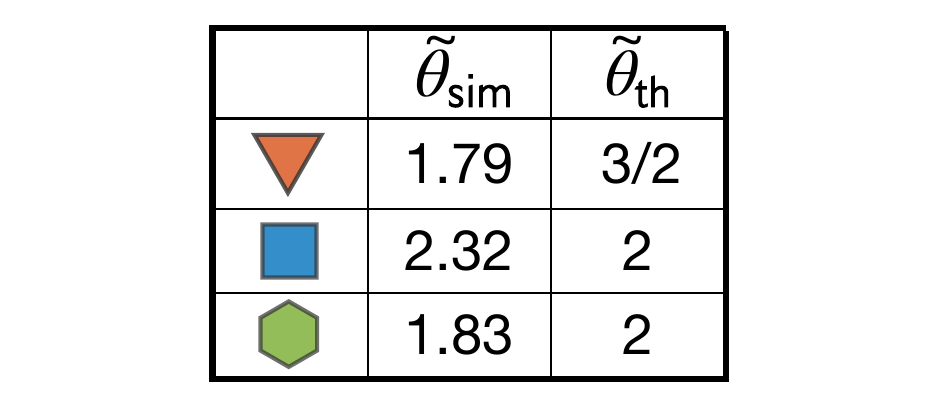}   
\caption{\textbf{Adjoined time complexity exponents}, $\tilde \theta_{\text{sim}}=\theta_{\text{sim}}+\phi_{\text{sim}}$, measured with the stochastic simulation (cf.~Fig.~\ref{fig:scaling_laws}; for $400\leq S \leq 1032$) together with their theoretical asymptotic estimates, $\tilde \theta_{\text{th}}$, for the different monomer morphologies.}
\label{tab:adjoined_exponents}
\end{table}

\section{Effective kinetic theory}
\label{sec:effective_model}

To understand how the kinetic parameters and the morphology of the building blocks affect the assembly process and, in particular, how the power law dependence of the assembly time on the structure size emerges, we formulate an effective kinetic theory \cite{krapivsky2010kinetic} for the concentrations $c_s(t)$ of clusters of size $s$ and the concentration of monomers ${m(t):= c_1 (t)}$.
To accomplish this, we leverage the fact that, in the strong binding limit, the assembly process typically follows a particular path that proceeds through a sequence of stable intermediate states, as shown in Fig.~\ref{fig:cluster_size_distribution} for square-shaped building blocks.
In general, the larger the number of unstable states that must be traversed to reach a new stable configuration, the less likely the particular path will be. 
Therefore, in formulating the effective theory, we can restrict ourselves to considering only the most likely assembly pathways that connect stable configurations via the smallest number of unstable intermediate states. The predominance of these assembly pathways could also be retraced with calculations of the free energy landscape of all cluster sizes \cite{jacobs2015rational,reinhardt2016effects}.

\paragraph{Nucleation process.} 

\begin{figure}[!t]
\centering
\includegraphics[width=0.80\linewidth]{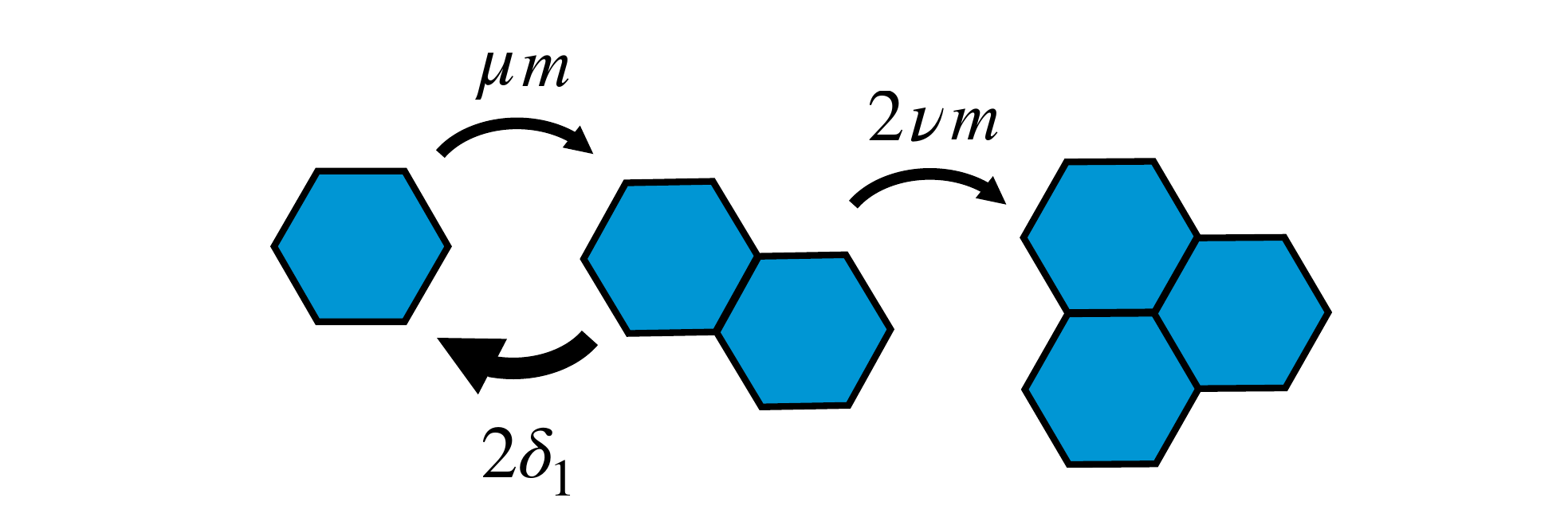}
\caption{
The \textbf{nucleation process} (illustrated for hexagon-shaped monomers) corresponds to a one-step Markov process with the nucleus as the absorbing state. 
The forward rates $\mu \, m$ and $2 \, \nu \, m$ correspond to dimer formation and the subsequent assembly step, respectively, where the numerical prefactor in the second step accounts for the multiplicity of the process. 
Note that the forward rates are proportional to the monomer concentration $m$, considering the monomers as an unlimited reservoir.  
The backward rate equals $2 \, \delta_1$ as either of the two monomers can detach. 
Hence, in the effective rate constant $\bar \mu$ for the nucleation process, Eq.~\eqref{eq:nucleation_rate}, the combinatorial prefactors of the forward and backward reactions cancel out exactly. 
The same holds for the more complex nucleation processes for particles with square and triangular morphology.  
}
\label{fig:illustration_nucleation}
\end{figure}

The assembly process starts with the formation of a stable nucleus consisting of $\sigma$ monomers [Fig.~\ref{fig:model}(a)]. 
Its formation can be viewed as a one-step Markov process with intermediate cluster states ${1 \leq s < \sigma}$ and the final nucleus as an absorbing state $\sigma$; see the illustration for hexagon-shaped monomers in Fig.~\ref{fig:illustration_nucleation}. 
In the limit where the backward rates are large compared to the forward rates, which here corresponds to ${\delta_1 \gg \nu \, m}$, the effective rate for the transition from a monomer to a nucleus is simply given by the ratio of the product of the forward and backward rates [cf.~App.~\ref{app:effective_model_irr_limit}]
\begin{equation}
\label{eq:nucleation_rate}
    r_{1\to \sigma}
    = 
    \mu \, 
    \left( 
    \frac{\nu}{\delta_1} 
    \right)^{\sigma-2}
    m^{\sigma}
    \equiv 
    \bar \mu \, 
    m^\sigma 
    \, .
\end{equation}
In a kinetic theory, this translates to an effective loss term $- \, \sigma \, \bar \mu \, m^\sigma$ in the rate equation for the monomer concentration $m$ with the effective rate for a reaction of order $\sigma$ given by ${\bar \mu = \mu\left( \nu / \delta_1 \right)^{\sigma-2}}$.
The corresponding gain term in the rate equation for the nuclei concentration  $c_\sigma$ is $\bar \mu \, m^\sigma$. 

\paragraph{Effective assembly process.} 

After the formation of the nucleus, a minimum number $\gamma$ of monomers must attach to the nucleus before the next stable cluster configuration is reached [Fig.~\ref{fig:model}(c)].
Thus, as in the nucleation process, one has a one-step Markov chain with the cluster of size ${\gamma + \sigma}$ representing an absorbing state. 
The corresponding effective rate, again under the assumption ${\delta_1 \gg \nu \, m}$, is given by [cf.~App.~\ref{app:effective_model_irr_limit}]
\begin{align}
\label{eq:rate_assembly_step_after_nucleation}
    r_{\sigma \to \sigma + \gamma} 
    =
    p_\sigma \, q_{\sigma} \,
    \nu 
    \left( \frac{\nu}{\delta_1} \right)^{\gamma -1} 
    m^\gamma
    \equiv p_\sigma \, q_{\sigma} \, \bar \nu \, m^\gamma
    \, .
\end{align}
Here $p_\sigma$ denotes the perimeter of the nucleus, accounting for the fact that the set of $\gamma$ monomers can attach at any point along the nucleus boundary. 
Furthermore, $q_{\sigma}$ is a numeric factor that accounts for the multiplicity of the seed-forming reaction pathways (e.g., for square-shaped monomers ${q_\sigma=1}$, and triangle-shaped monomers ${q_\sigma=4}$, cf.~App.~\ref{app:effective_model_irr_limit} and Fig.~\ref{supp_fig:multiplicity_factor_q_s}(a) therein). 
The effective kinetic parameter for this reaction of order $\gamma$ is given by ${\bar \nu = \nu \, (\nu/\delta_1)^{\gamma-1}}$. 

\begin{figure}[htb]
\centering
\includegraphics[width=0.9\linewidth]{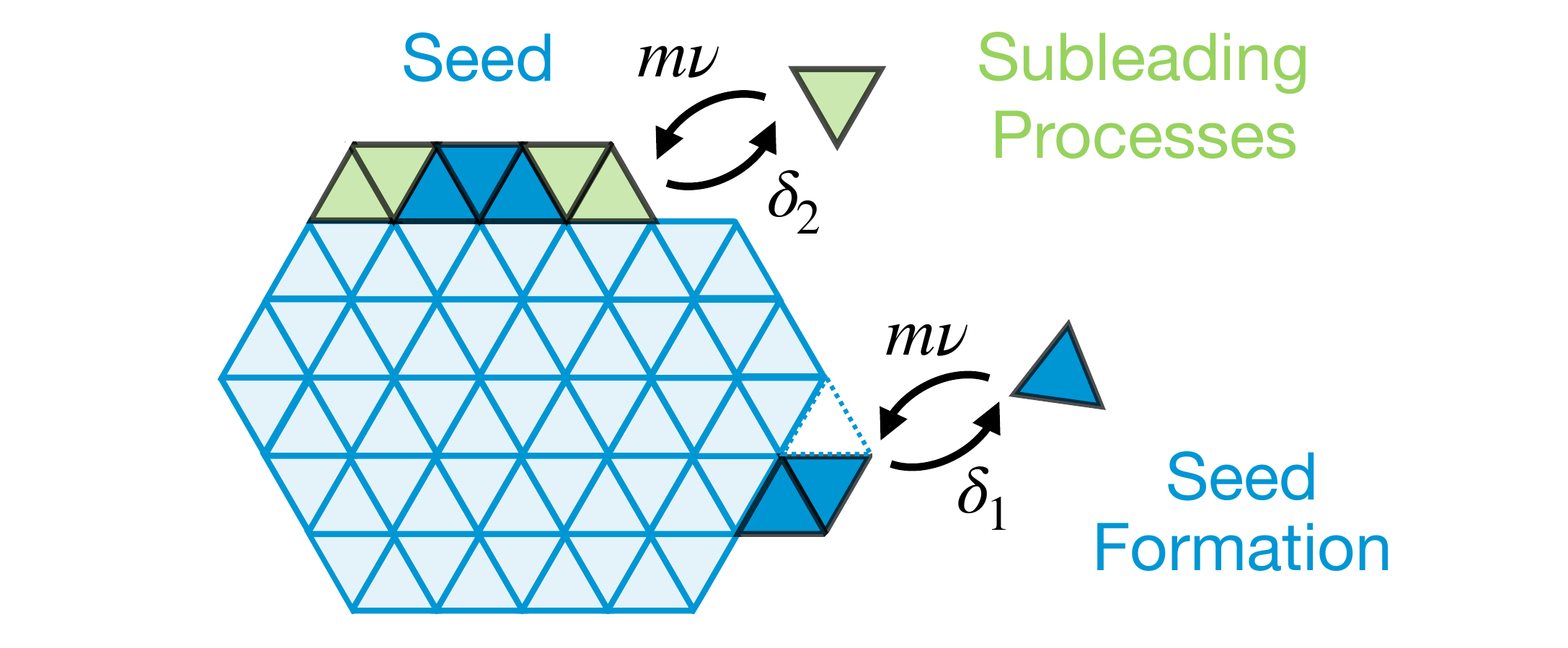}
\caption{
\textbf{Illustration of the assembly steps for large clusters} (illustrated for triangle-shaped monomers). Cluster growth after nucleation proceeds in two steps for assembly processes with attachment order ${\gamma >1}$ (here, ${\gamma = 3}$): The formation of a seed of size $\gamma$ at one edge of the cluster (rate limiting slow process with an effective rate $p_s \, q_s \, \bar \nu \, m^\gamma$) is followed by a sequence of subleading fast processes that complete the edge very quickly (`domino effect').  
}
\label{fig:assembly_steps_large_cluster}
\end{figure}

In general, all of the following assembly steps will depend on the morphology of the cluster and not just on its size $s$. 
However, in the spirit of a mean-field approximation, we disregard the specific cluster morphology and consider only typical (large) clusters, as illustrated in Fig.~\ref{fig:assembly_steps_large_cluster} for triangle-shaped monomers.

In the case of hexagon-shaped monomers, clusters can grow by the attachment of a single monomer without having to go through unstable configurations. 
Hence, the attachment order for hexagon-shaped monomers is ${\gamma=1}$ and clusters simply grow by monomer attachment at a rate ${r_{s \to s+1} = p_s \, q_s \,  \nu \, m}$ with ${q_s=1}$. 
We approximate the perimeter of the cluster as ${p_s = a \sqrt{s}}$ independent of the cluster morphology, where $a$ can be estimated by considering a typical cluster, or regarded as a parameter to be determined numerically by fitting to the results of stochastic simulations [cf.~App.~\ref{app:effective_model_irr_limit}].
Hence, in the effective rate equation for $c_s$ (with ${s>\sigma}$) there is a gain term of the form $f_s \, \nu \, m \, c_s$ with the combinatorial factor ${f_s = p_s \, q_s = a \, s^\omega}$ with ${\omega = 1/2}$ \footnote{Since cluster growth is by the addition of monomers, stochastic effects will result in some interface roughness (akin to Eden growth \cite{murray1960two})
leading to a larger surface-to-volume ratio than given by ${p_s = a \sqrt{s}}$ and hence to effective growth exponent ${\omega > 1/2}$. Since this only affects the system with hexagon-shaped monomers, we will not discuss this additional effect here further and simply approximate $\omega$ by $1/2$ for hexagon-shaped monomers.}.
We call $\omega$ the \textit{growth exponent} for cluster growth.

In contrast, for square- and triangle-shaped monomers, cluster growth proceeds in two steps [Fig.~\ref{fig:assembly_steps_large_cluster}]:
First, a leading order slow reaction involving $\gamma$ monomers must occur to arrive at a new stable cluster configuration. 
This \textit{seed formation process} occurs along one of the edges of the polygon at a rate given by Eq.~\eqref{eq:rate_assembly_step_after_nucleation}, where the nucleus size $\sigma$ is replaced by the cluster size $s$. 
For large $s$, the numeric factor $q_s$, which accounts for the degeneracy of the seed-forming reaction, is approximately a constant given by ${q_s=2}$ for square-shaped monomers and ${q_s=3}$ for triangles; see App.~\ref{app:effective_model_irr_limit} with Fig.~\ref{supp_fig:multiplicity_factor_q_s}(b).
Once such a seed is formed, it can further grow in a series of reactions of lower effective order (order 1 in the case of square-shaped monomers and order 2 in the case of triangle-shaped monomers). 
Since the corresponding rates are large compared to the rate of seed formation,  the leading order process is followed with high probability---in a kind of ``domino effect''---by a series of such lower-order processes until the entire edge of the polygon is filled up.
Subsequent growth then requires another seed formation process along any polygon edge.
Since the seed formation process is rate-limiting, the total rate ${r_{s \to s+l_s}}$ of filling up the edge of a polygon of length $l_s$ approximately equals the rate of seed formation ${r_{s \to s+ \gamma}^{\text{seed}}}$. 

To analyze systems with different monomer morphologies on equal footing, we now formulate the growth dynamics in terms of an effective rate at which the cluster grows by one unit. 
The corresponding \textit{effective monomer attachment rate} is given by ${\tilde r_{s \to s + 1} = l_s \, r_{s \to s + l_s}}$, where ${l_s \sim \sqrt{s}}$ is a typical edge length of the cluster. 
Thus, in an effective rate equation for the concentration $c_s$ of clusters of size $s$, the gain term has the form $f_s \, \bar \nu \, m^\gamma \, c_s$, where ${f_s = q_s \, p_s \, l_s \approx a \, s^\omega}$ with a growth exponent ${\omega = 1}$.
This takes into account that the multiplicity of the reactions in the growth process is proportional to the product of the typical perimeter ${p_s \sim \sqrt{s}}$, the multiplicity factor $q_s$ for the seed forming reaction, and the edge length ${l_s \sim \sqrt{s}}$ of a cluster. 
Similar to hexagon-shaped monomers, for large $s$, the prefactor $a$ is approximately a constant that depends only on the morphology of the building blocks \footnote{Unlike for hexagon-shaped building blocks, in the case of square and triangular building blocks, the prefactor $a$ additionally includes the multiplicity $q_s$ of the seed forming reaction pathways. For example, the multiplicity considers that the seed can typically grow in two directions unless it forms directly at the corner of an edge.}.
It can be determined numerically by fitting it to the results of the stochastic simulations or estimated by combinatorial arguments [cf. App. \ref{app:effective_model_irr_limit}].

Taken together, one obtains the following set of \textit{rate equations} for the effective kinetic model
\begin{subequations}
\label{effective_theory_1st_order}
\begin{align}
    \partial_t m (t)
    &= 
    -\sigma \, \bar \mu \, m^\sigma (t) 
    - 
    \bar\nu \, m^{\gamma} (t) \sum\nolimits_{s=\sigma}^{S-1} f_s \, c_{s} (t) \, , 
    \label{effective_theory_1st_order:a}  
    \\
    \partial_t c_{\sigma} (t)
    &= 
    \bar \mu \, m^\sigma (t)
    - 
    \bar\nu \,  m^{\gamma} (t) \, 
    f_\sigma \, c_{\sigma} (t) \, ,   
    \label{effective_theory_1st_order:b} 
    \\
    \label{effective_theory_1st_order:c}
    \partial_t c_{s} (t)
    &=  
    \bar\nu \, m^{\gamma} (t) \, 
    \big[ 
    f_{s-1} \, c_{s-1}(t) - f_s \, c_{s}(t) \big] 
    \, .
\end{align}
\end{subequations}
Here, to describe the cluster growth, we used the effective one-particle growth rates derived in the previous paragraph.
For the combinatorial factor $f_s$ we use the asymptotic expression for the entire domain ${\sigma \leq s < S}$, 
\begin{align}
\label{eq:growth_multilpicity}
    f_s = a\, s^\omega \, ,
\end{align} 
with the prefactor $a$ determined numerically for each monomer morphology [cf.~App.~\ref{app:effective_model_irr_limit}] \footnote{The combinatorial factors $f_s$ for clusters of small size $s$ typically deviate from the asymptotic form $a\, s^\omega$. 
Note, however, that for increasing target size $S$, the approximation becomes ever more accurate as the asymptotic form of $f_s$ applies to an increasingly large fraction of all assembly steps.}; for completed structures (absorbing state) ${s = S}$, we set ${f_S = 0}$. 
Equation~\eqref{effective_theory_1st_order:a} accounts for the loss of monomers due to nucleation events with the effective rate $\bar \mu$ and attachment of monomers to incomplete complexes with the effective rate $\bar \nu$.
Similarly, Eq.~\eqref{effective_theory_1st_order:b} describes the gain and loss of stable nuclei due to nucleation events and the initial growth of stable nuclei, respectively. 
Note that for triangle-shaped monomers, the first assembly step following nucleation has a larger attachment order ${\gamma=4}$ than all subsequent assembly steps where ${\gamma=3}$.
This is accounted for explicitly in the implementation of the effective model for triangle-shaped monomers [cf.~App.~\ref{app:effective_model_irr_limit}] as we found that it significantly improves the agreement with the results from the stochastic simulations.  
Finally, Eq.~\eqref{effective_theory_1st_order:c} specifies the cluster dynamics with a gain due to the growth of smaller complexes and loss due to the growth of complexes of the same size $s$.
As shown in Fig.~\ref{fig:scaling_laws}, we find excellent agreement between the effective kinetic theory and the stochastic simulations for all particle morphologies.

In conclusion, the effective kinetic theory identifies the three parameters $\sigma$ (\textit{nucleation size}), $\gamma$ (\textit{attachment order}), and $\omega$ (\textit{growth exponent}) as the major determinants of self-assembly kinetics. 
Since the above kinetic approach is general, we suppose that it can accurately describe the self-assembly of a broad class of systems, including, for example, three-dimensional structures.

\section{Scaling Theory}
\label{sec:scaling_theory} 

The numerically observed power laws for the optimal detachment rate $\tilde \delta_1^\text{opt}$ and the minimum assembly time $T_{90}^\text{min}$ as a function of the structure size $S$ suggest that the self-assembly dynamics exhibit some underlying scale invariance [Fig.~\ref{fig:scaling_laws}].
In this chapter, we analyze the kinetic theory for cluster growth in the strong binding limit, Eq.~\eqref{effective_theory_1st_order}, using several complementary approaches that offer crucial insights into the nature of the dynamics. 
We limit our presentation to discussing the main ideas, leaving the formal mathematical analysis to the appendices~\ref{app:onset_scaling} and \ref{app:general_scaling}.

\subsection{Onset criterion for finite yield}
\label{sec:onset_criterion}

In the strong binding limit, the yield curves obtained from the stochastic simulations exhibit scaling behavior (cf.~the data collapse in Fig.~\ref{fig:yield_scaling})
\begin{equation}
    Y(\tilde \delta_1, S) 
    = 
    \widehat Y(\tilde \delta_1/\tilde \delta_1^\text{opt}(S)) \, .
\end{equation}
This means that all detachment rate defining characteristic features of the yield curves, such as the threshold rate for finite yield $\tilde \delta_1^\text{on}$, must show the same scaling behavior with structure size $S$ as the optimal detachment rate $\tilde \delta_1^\text{opt}$.
Because of this equivalence in scaling behavior, it is sufficient to analyze the self-assembly dynamics for a vanishingly small yield. 
This considerably simplifies the rate equations Eq.~\eqref{effective_theory_1st_order}, which we will do first in the following. 
Using the asymptotic form for the  combinatorial factor $f_s$, the monomer dynamics [Eq.~\eqref{effective_theory_1st_order:a}] can be written as 
\begin{align}
\label{eq:monomer_dynamics}
    \partial_t m(t)
    =
    -\sigma \, \bar \mu \, m^\sigma
    - a \, \bar \nu \, \langle s^\omega \rangle \, m^\gamma  K 
    \, ,
\end{align}
where ${K:=\sum_{s=\sigma}^{S-1} c_s}$ is the total concentration of incomplete clusters, and $\langle s^\omega \rangle$ the $\omega$--moment of the cluster size distribution $c_s$. 
As long as the yield is zero, no clusters are completed, and thus, the number of incomplete clusters grows with each nucleation event.
\begin{align}
\label{eq:cluster_dynamics}
    \partial_t K (t)
    =
    \bar \mu \, m^\sigma 
    \, .
\end{align}
These equations for the monomers and incomplete clusters are not closed because they require information about the moments $\langle s^\omega \rangle$ of the cluster size distribution. 
However, in the limiting case of vanishingly low yield considered here, where no appreciable amount of finished structure is present, these equations can be greatly simplified by taking advantage of the fact that all bound monomers are contained in unfinished clusters,
\begin{align}
    C-m = K \, \langle s \rangle \, ,
\end{align}
where $\langle s \rangle$ is the average size of a cluster. 
For a growth exponent ${\omega =1}$ (valid for triangle- and square-shaped monomers), this immediately leads to a closed set of equations; otherwise, for general $\omega$, we use a mean-field-like ``factorization approximation'' ${\langle s^\omega \rangle \approx \langle s \rangle^\omega}$ to close the equations.

To determine when the system displays finite yield, one must specify under which conditions the cluster size distribution $c_s$ can develop a finite weight at fully formed clusters of size $S$.
The dominant factor in the dynamics of the cluster size distribution is drift due to (irreversible) cluster growth, which implies that (at the mean-field level) the largest cluster $s_\text{max}$ obeys the growth law,
\begin{align}
\label{eq:cluster_front_dynamics}
    \partial_t s_\text{max} (t)
    =
    a \, \bar \nu \, s_\text{max}^\omega \, m^\gamma
    \, ,
\end{align}
corresponding to cluster growth by forming a seed of size $\gamma$ at one of the cluster's edges, followed by a `domino effect' filling up the cluster edge as described in Sec.~\ref{sec:effective_model}.  

The set of equations, Eqs.~\eqref{eq:monomer_dynamics}--\eqref{eq:cluster_front_dynamics}, exhibits scale invariance, as shown in App.~\ref{app:onset_scaling}. 
In particular, one finds that the maximal cluster size obeys the scaling form
\begin{align}
\label{eq:scaling_max_cluster_dynamics}
    s_\text{max}(t) 
    = 
    \eta^y \, 
    \tilde s_\text{max} 
    \left( 
    a \, \bar \nu \, C^\gamma \, \eta^z t 
    \right)
\end{align}
with ${\eta = \frac{\bar \mu}{a \, \bar \nu} \, C^{\sigma - \gamma - 1}}$ and the scaling exponents
\begin{align}
    z = \frac{1-\omega}{2-\omega}
    \qquad  \text{and} \qquad
    y = - \frac{1}{2-\omega} \, .
\end{align}
The onset condition for finite yield, ${S=s_{\text{max}}(\infty)}$, can then be written as a power law using the definitions of the effective rates $\bar \mu$ and $\bar\nu$ in Eqs.~\eqref{eq:nucleation_rate} and \eqref{eq:rate_assembly_step_after_nucleation},
\begin{align}  \label{eq:scaling_phi}
    \tilde \delta_1^\text{on} 
    \sim S^\phi
    \, ,
\end{align}
with the \textit{control parameter exponent}
\begin{align}
\label{eq:control_parameter_exponent}
    \phi
    =
    \frac{2-\omega}{\sigma-\gamma-1}
    \, .
\end{align}
Furthermore, from Eq.~\eqref{eq:scaling_max_cluster_dynamics}, we read off that the typical time scale of the assembly process scales as
\begin{equation}
    \tau_{\text{assem}} 
    \sim 
    \left( a \, \bar \nu \, C^\gamma \, \eta^z \right)^{-1}
    \sim 
    \tilde \delta_1^\alpha  \, ,
\end{equation}
where ${\alpha = \gamma - 1 + z \, (\sigma - \gamma - 1)}$. 
By using the scaling of the onset detachment rate $\tilde \delta_1^\text{on}$ with target size $S$, Eq.~\eqref{eq:scaling_phi}, one thus obtains
\begin{equation}
\tau_{\text{assem}} \sim S^{\alpha \phi} := S^{\theta}
\end{equation}
with the \textit{time complexity exponent} given by
\begin{equation}   \label{eq:time_complexity_exponent}
    \theta 
    = 
    \frac{(1-\omega)\,\sigma+\gamma+2\,\omega-3}{ \sigma-\gamma-1} \, . 
\end{equation}
Strictly speaking, $\tau_{\text{assem}}$ corresponds to the time scale for forming the first complete structure starting from an existing nucleus. 
However, we are actually interested in the minimal assembly time $T_{90}^{\text{min}}$ required to achieve 90\% yield.
To see that $T_{90}^{\text{min}}$ scales identically as $\tau_{\text{assem}}$, we first determine what the rate-limiting step in the assembly process is by comparing $\tau_{\text{assem}}$ with the typical time required to form a nucleus,
\begin{equation}
    \tau_\text{nuc} = 
    \left( \bar \mu \, C^\sigma \right)^{-1} 
    \sim
    \tilde \delta_1^\beta 
    \, ,
\end{equation}
where $\beta = \sigma - 2$. 
For all the particle morphologies considered here, one has ${\beta > \alpha}$, \footnote{In general, ${\beta > \alpha}$ holds if ${\sigma>\gamma+1}$ and ${\omega< 3/2}$, meaning that this must be true in general.} implying that nucleation is the rate-limiting step of the self-assembly process.
Using the scaling equivalence ${\tilde \delta_1^\text{opt} \sim \tilde \delta_1^\text{on} \sim S^\phi}$ and that nucleation is rate-limiting, the minimal assembly time (for 90\% yield), $T_{90}^\text{min}$, can be approximated by the time required for $0.9 \, N/S$ nucleation events to happen, each of which occurs at an effective rate $\sim \bar \mu_\text{opt} \, C^\sigma$.
This results in the scaling law 
\begin{align}
\label{eq:scaling_minimal_assembly_time}    
    T_{90}^\text{min}
    \sim 
    \frac{C}{S} \cdot 
    \frac{1}{\bar \mu_\text{opt} C^{\sigma}}
    \sim 
    S^{\theta}
\end{align} 
with $\theta$ identical to Eq.~\eqref{eq:time_complexity_exponent}. 
In conclusion, the time to assemble the first structure and the minimal time to reach 90\% yield show the same scaling behavior with target size $S$.  

The theoretical estimates of the exponents for the three different particle morphologies are shown in the tables in Fig.~\ref{fig:scaling_laws}(b,c) compared with the values inferred from stochastic simulations. 
The theoretical exponents generally provide good estimates for the numerically determined exponents. 
The remaining differences can be attributed to the asymptotic scaling regime not being fully achieved for the simulated system sizes and the actual growth exponent $\omega$ deviating slightly from its estimated value \footnote{The latter is particularly pronounced in the system with hexagonal monomers, where, due to the absence of a `domino effect' that fills up the edges of a cluster after an initial seed has formed, the typical morphologies of the cluster exhibit some interface roughness during growth. Note that the assembly time obtained from integrating the effective model in which we used ${\omega=1/2}$ coincides with the analytic estimate ${\theta_\text{th}=1/2}$ for the system with hexagonal building blocks. Furthermore, using a value of $\omega$ slightly larger than $1/2$ in Eq.~\eqref{eq:time_complexity_exponent} (which would be plausible due to the roughening effect) consistently results in a value of $\theta$ smaller than $1/2$ as observed from the simulation data. For example, the measured value ${\theta\approx 0.4}$ is obtained with ${\omega=0.6}$ according to Eq.~\eqref{eq:time_complexity_exponent}.}.  

For a given $\omega$, the smallest possible value for the time complexity exponent according to Eq.~\eqref{eq:time_complexity_exponent} is obtained for ${\gamma = 1}$, giving ${\theta = 1-\omega}$. 
In contrast, if $\gamma>1$, $\theta$ decreases steadily with increasing
nucleation size $\sigma$, asymptotically reaching the minimum ${1-\omega}$ in
the limit ${\sigma \to \infty}$. 
Realistically, however, this limit cannot be reached because, for very large $\sigma$, the description by our effective model breaks down since $\delta_1^{\text{opt}}$ can no longer be assumed to be large compared to the reaction rate $C \nu$. 
In summary, this suggests that a minimum attachment order ${\gamma = 1}$ is key to achieving a minimal time complexity exponent $\theta$ and thus maximal time efficiency for the self-assembly of large structures in the strong binding limit.

\subsection{Scale invariance of the self-assembly dynamics}
\label{sec:general_scaling_theory}

The theory in the strong binding limit can be shown to exhibit general scale invariance, regardless of the assumption of a small yield.
We will outline the main arguments leading to this scale invariance below while leaving the mathematical details to App.~\ref{app:general_scaling}.
This scale invariance explains the collapse of the yield curves observed in Fig.~\ref{fig:yield_scaling} and has further essential consequences, as discussed in this subsection.

We use a hydrodynamic approximation to see that the self-assembly dynamics shows scale invariance and write Eq.~\eqref{effective_theory_1st_order} as an effective advection equation for the cluster size distribution $c(s)$, as proposed by previous work~\cite{Zlotnick1999a, Endres2002, Morozov2009}. 
We consider $c(s)$ as continuous function of a real variable ${s \in[ \sigma, S ]}$ and Taylor expand the first term in Eq.~\eqref{effective_theory_1st_order:c} to first order in $\partial_s$ to obtain the advection equation
\begin{equation}
    \partial_t c (s,t) 
    = - a \,\bar\nu \, m^{\gamma} \partial_s [s^{\omega} c(s)] \, .
\end{equation}
This equation must be completed with the appropriate boundary conditions: 
\begin{equation}   \label{eq:advection_eq_boundary_condition}
{a \bar\nu \, m^{\gamma} (t) \, \sigma^{\omega} c(\sigma,t) = \bar\mu \, m^\sigma} (t)
\end{equation}
accounting for the influx of complexes due to nucleation, and an absorbing boundary condition ${c(S,t) = 0}$ for the complexes that get completed. 
In the equation for the monomer concentration, Eq.~\eqref{effective_theory_1st_order:a}, we neglect the loss term due to nucleation since monomer loss due to attachment processes plays a much more significant role whenever there is significant growth of the structures (if there is substantial yield, roughly $S/\sigma$ times as many monomers are consumed by attachment reactions as compared to nucleation reactions). 
Approximating the sum in Eq.~\eqref{effective_theory_1st_order:a} by an integral, we thus obtain for the concentration of the monomers
\begin{equation}  
\label{eq:advection_eq_monomers}
    \partial_t m (t) 
    =  
    - a \, \bar\nu \, m^{\gamma} 
    \int_{\sigma}^{S}  s^{\omega} c(s) \ ds 
    \, ,  
\end{equation}
with initial condition $m(0)=C$.
As the last step, we approximate the lower limit of the integral $\sigma$ by a small target-size-dependent value $\epsilon \, S$, i.e., we neglect a fraction ${(\epsilon S - \sigma)/S \approx \epsilon}$ of all assembly steps, where $\epsilon$ can in principle be arbitrarily small. 
This step is helpful as it allows us to perform a variable transformation in the integral, thus making the integral, including its boundaries, independent of $S$. If $\omega>1$, which can be case for the self-assembly of three-dimensional structures [cf.~App.~Sec.~\ref{app:sec:3D_structures}], the approximation of the lower integral boundary can become inaccurate because the growth of small clusters from size $\sigma$ to $\epsilon \, S$ defines the predominant time scale in the growth process of a cluster. In this case, we therefore approximate the upper boundary of the integral instead, see App.~Sec.~\ref{app:sec:3D_structures} for details. \\ 
With these approximations \footnote{Note that some of the approximations used here actually become exact in the limit of large $S$. For example, neglecting the diffusion term and higher order terms in the advection equation is irrelevant in the limit ${S \to \infty}$ as their contribution becomes negligible against the leading order advection term (see Eqs.~\eqref{eq:advection_diffusion} and \eqref{eq:jump_moments}, showing that the advective contribution to cluster growth equals the square of the diffusive contribution). Similarly, neglecting the nucleation term in Eq.~\eqref{eq:advection_eq_monomers} becomes increasingly irrelevant as $S$ increases.} it can be shown with a scaling ansatz that the monomer concentration and complex concentration as functions of both $s, t$, the parameter $\delta_1$, and the target structure size $S$ obey the scaling forms 
\begin{equation}
\label{eq:scaling_form_m}
    m(t, \delta_1, S) 
    = 
    C \, \widehat m(S^{-\theta}  \, \tilde t, \, S^{-\phi}\, \tilde \delta_1)
\end{equation}
and 
\begin{equation}
\label{eq:scaling_form_c}
	c(s,t,\delta_1,S) 
	= 
	C S^{-2} \, \widehat c(S^{-1} s, \, S^{-\theta} \, \tilde t, \, S^{-\phi} \, \tilde \delta_1) \, ,
\end{equation}
with ${\tilde t = C\nu \, t}$ and $\phi$ and $\theta$ identical to Eq.~\eqref{eq:control_parameter_exponent} and Eq.~\eqref{eq:time_complexity_exponent} as obtained in Sec.~\ref{sec:onset_criterion}. 
Moreover, the condition to obtain a yield $Y$ translates directly into a condition on the   scaling functions $\widehat m$ and $\widehat c$: 
\begin{equation}   \label{scale_less_yield_condition}
    \widehat m (S^{-\theta} \tilde t, \, S^{-\phi} \, \tilde \delta_1) + 
    \int_{\epsilon}^{1} x \, \widehat c(x, S^{-\theta} \tilde t, \, S^{-\phi} \, \tilde \delta_1) \, dx 
    = 1-Y \, ,
\end{equation}
where we used that the yield being $Y$ is equivalent to the number of resources that remain in the system, i.e., that have not yet been absorbed by the right boundary, equals $(1\,{-}\, Y)\, C$. This shows that the yield obeys the scaling form
\begin{equation}  \label{eq:yield_scaling_form}
Y(t,\delta_1,S) = \widehat Y(S^{-\theta} \tilde t, S^{-\phi} \tilde \delta_1) \, ,
\end{equation}
implying that all detachment rate-defining characteristics of the yield curve scale identically ($\sim S^{\phi}$) as functions of the structure size. 
For example, the threshold rate to obtain a fixed yield $Y$, $\delta_1^{\text{on}}(Y)$, scales identically as the optimal rate $\delta_1^{\text{opt}}(Z)$ to obtain a yield $Z$ in the minimal amount of time, explaining the collapse of the yield curves in Fig.~\ref{fig:yield_scaling}.
Furthermore, by solving Eq.~\eqref{eq:yield_scaling_form} for the time argument, we find that the time $T_Y$ required to achieve a yield $Y \in (0,1)$ as a function of $\delta_1$ obeys
\begin{equation} 
\label{eq:T90_scaling_form}
	T_Y(\delta_1,S) 
	=  
	(C\,\nu)^{-1} \, S^{\theta} \, 
	\widehat T_Y(S^{-\phi} \, \tilde \delta_1) \, ,
\end{equation}
whenever the function has a finite value. 
This shows that the same scaling behavior that we observed for the optimal detachment rate and minimal assembly time in Fig.~\ref{fig:scaling_laws} (b) and (c), in fact, applies identically to the entire $T_Y$ curve. 

\textit{Robustness of $T_{90}^{\text{min}}$.} The fact that the $T_{90}-$curves in Fig. \ref{fig:scaling_laws}(a) will maintain their shape when increasing $S$, allows us to uniquely characterize the sensitivity of the minimal assembly time as a function of the control parameter. Figure \ref{fig:scaling_laws}(a) shows that the morphology strongly influences how sensitively the minimal assembly time depends on $\tilde \delta_1$: For example, for the assembly time to differ from the minimal assembly time by at most one order
of magnitude, the parameter $\tilde \delta_1$ may only vary within a relative range of 40\% in the case of triangular monomers, but 340\% for hexagonal monomers. These values correspond to respective ranges of the binding energies of $\pm 0.18 \, k_B T$ and $\pm 1.0 \, k_B T$. This suggests that hexagonal monomers not only self-assemble more efficiently, they also require less fine-tuning of the control parameter to achieve their maximal efficiency. In general, we expect that the nucleation size $\sigma$ crucially determines the sensitivity of the minimal assembly time, since $\sigma$ impacts both the depth of the U-shape and the scaling of $T_{90}$ in the limit of large $\tilde \delta_1$ [cf. Fig. \ref{fig:scaling_laws}(a)]. Small nucleation sizes therefore seem favorable to reduce the sensitivity.   \\     

\textit{Dimerization barrier.}
Note that, due to the boundary condition, Eq.~\eqref{eq:advection_eq_boundary_condition}, the scaling functions Eqs. \eqref{eq:scaling_form_m} and \eqref{eq:scaling_form_c} still depend on the ratio $\tilde \mu := \mu/\nu$ between the dimerization and the binding rate (dimerization barrier). In biological systems, due to cooperative or allosteric binding effects, the dimerization rate $\mu$ can be significantly smaller than the rate of attachment $\nu$ of monomers to larger clusters \cite{zlotnick2011virus,packianathan2010conformational,kushner1969}, resulting in $\tilde \mu \ll 1$. In App.~Sec.~\ref{scaling_theory_dimerization_barrier}, we show that, in an analogous way, scaling functions for the parameter $\tilde \mu := \mu/\nu$ can be obtained. 
A fixed value of $\tilde \mu$ does not affect the scaling behavior of the assembly time in dependence on the target structure size. However, we find that for square- and triangle-shaped monomers (generally for systems with $\gamma>1$), the minimal assembly time decreases proportionally with $\tilde \mu$ [cf.~App.~Sec.~\ref{scaling_theory_dimerization_barrier}]. Furthermore, one can also determine the time complexity exponent for a scenario in which the assembly process is controlled entirely through $\tilde \mu$ (while $\tilde \delta_1$ is kept constant). For this scenario, we obtain the same time complexity exponent $1-\omega$ as when changing $\tilde \delta_1$ in the case of hexagonal monomers (cf.~App.~Sec.~\ref{scaling_theory_dimerization_barrier} for a further discussion). However, in experiments, it is presumably considerably simpler to control $\tilde \delta_1$ (via temperature, monomer concentration, etc.) rather than $\tilde \mu$, which requires control over allosteric effects and hence a particular molecular design of the building blocks.

\subsection{Heuristic onset criterion for high yield}
\label{sec:heuristic_onset_criterion}

The results from the scaling analysis of the rate equations can also be understood heuristically based on the following scaling arguments for the onset of high yield.
To achieve a high yield of target structures of size $S$, in total, about $S$ times as many attachments (cluster growth) events as nucleation events must occur
\begin{align}
    \nu_\text{tot} 
    = 
    S \cdot \mu_\text{tot} \, ,
\end{align}
or, in other words, during the assembly process, any structure formed by a nucleation event must also have a significant probability of completion. 
The total nucleation rate is given by [cf.~Eq.~\eqref{eq:cluster_dynamics}] ${\mu_\text{tot} = \bar \mu \, m^\sigma}$, and the total growth rate can be estimated as [cf.~Eq.~\eqref{eq:monomer_dynamics}]
\begin{align}
    \nu_\text{tot}
    =
    a \, \bar \nu \, \langle s^\omega \rangle \, m^\gamma \, K \, .
\end{align}
where ${\langle s^\omega \rangle := \int_{\sigma}^{S} s^{\omega}\, ds / (S-\sigma) \approx S^{\omega}/(\omega+1)}$ is the average growth rate of a cluster. 
Note that during the assembly process, most clusters undergo all assembly steps to obtain a high yield, so it is reasonable to consider $\langle s^\omega \rangle$ as proportional to $S^\omega$, as calculated here. 
Finally, since a finite fraction of the $K$ clusters formed should become finished structures, one has ${K \sim C/S}$. 
Using the scaling of the effective kinetic parameters $\bar \mu$ and $\bar \nu$ with the detachment rate one finds
\begin{align}
\label{attachment_nucleation_balance}
    S 
    = 
    \frac{\nu_{\text{tot}}}{\mu_{\text{tot}}} 
    \sim 
    \tilde \delta_1^{\sigma-\gamma-1} \, S^{\omega-1} \, ,
\end{align}
where we assumed that the monomer density is of the order of $C$ and hence independent of the structure size $S$.
Therefore, to achieve high yield, the detachment rate must scale as ${\tilde \delta_1^\text{full} \sim S^\phi}$, where $\phi$ is identical to Eq.~\eqref{eq:control_parameter_exponent}. 
Thus, both $\tilde \delta_1^\text{full}$ and $\tilde \delta_1^\text{on}$ [cf.~Sec.~\ref{sec:onset_criterion}] scale in the same way with the structure size, providing further confirmation of the scaling law for yield curves.  

\subsection{Slow nucleation limit}
\label{sec:slow_nucleation_limit}

Interestingly, the assembly time $T_{90}$ can be calculated analytically when considering the extreme case of very slow nucleation that applies when ${\tilde \delta_1 \gg \tilde \delta_1^\text{opt}}$. 
To see this, one needs to realize that in this regime, the concentration of unfinished structures is low at any time, so one can assume that each cluster is completed before the next one is nucleated.
Each nucleation event then reduces the number of monomers by $S$ units; hence the evolution of the monomer concentration obeys ${\partial_t m = -S \, \bar\mu \, m^{\sigma}}$. 
Separating variables and integrating $m$ from $C$ to $(1-Y) \, C$, where $Y \in (0,1)$ is the desired yield, and $t$ from 0 to $T_Y$, one obtains
\begin{align}
\label{Ty_slow_nucleation}
   T_Y
   = Q_Y \, \frac{C}{S} 
   \frac{1}{\bar \mu \, C^{\sigma}}  \,
\end{align}
with
\begin{equation}
\label{qy_slow_nucleation}
	Q_Y
	:= \frac{(1-Y)^{1-\sigma}-1}{\sigma-1} \, ,
\end{equation}
and $\bar \mu\sim \delta_1^{\sigma-2}$ [cf.~Eq.~\eqref{eq:nucleation_rate}].
From a comparison with the numerical data [see Fig.~\ref{fig:T50_scaling}(a) in App.~\ref{app:invariance_to_yield_threshold}], we find that this analytic result matches the assembly time exactly in the regime where it scales as $T_{90} \sim \tilde \delta_1^{\sigma-2}$. 

Note that $T_Y$ has the same functional form as Eq.~\eqref{eq:scaling_minimal_assembly_time}, confirming the asymptotic scaling of $T_{90}$ [Fig.~\ref{fig:scaling_laws}] and the validity of the general scaling law for the assembly time given by Eq.~\eqref{eq:T90_scaling_form}.  
Furthermore, the analytic result, Eq.~\eqref{Ty_slow_nucleation}, reveals a strong dependence of the time efficiency in the strong binding limit on the nucleation size $\sigma$: 
Not only does $\sigma$ determine the dependence on the detachment rate (${T_{90} \sim \tilde \delta_1^{\sigma-2}}$), but it also strongly affects the prefactor $Q_Y$.
For example, to achieve 90\% yield, one has ${Q_{0.9} = 10^5}$ for triangle-shaped monomers but only ${Q_{0.9} = 333}$ for square-shaped monomers. 
Hence, this explains why square-shaped monomers assemble more efficiently than triangle-shaped monomers in the strong binding limit [see Fig.~\ref{fig:scaling_laws}], although both systems exhibit the same time complexity exponent ${\theta=1}$. 
Note also that by decreasing the demanded yield $Y$ from $0.9$ to $0.5$, the assembly time for triangular monomers in the limit of large $\tilde \delta_1$ decreases by a factor of ${Q_{0.9}/Q_{0.5} = 3225}$. In contrast, for square and hexagonal monomers, it decreases only by a factor of $143$ and $33$, respectively. 
Thus, we expect that differences in the assembly time between the different particle morphologies tend to become less pronounced at a lower yield (note, however, that the yield $Y$ does not affect the scaling of $T_{90}$ with the target structure size $S$).
Or, to put it another way, increasing the yield in systems with large $\sigma$ is more expensive in terms of time efficiency than in systems with small $\sigma$.

\subsection{Interpretation of the scaling results}
\label{sec:interpretation_scaling_results}

By exploiting the scale invariance of the dynamic equations in Secs.~\ref{sec:onset_criterion} and \ref{sec:general_scaling_theory}, we showed that the assembly time and detachment rate in the strong binding limit exhibit a characteristic scaling behavior as functions of the target structure size.
This scaling behavior is characterized by the exponents $\theta$ (time complexity exponent) [Eq.~\eqref{eq:time_complexity_exponent}] or $\tilde \theta$ (adjoined time complexity exponent) [cf.~Sec.~\ref{sec:scaling_delta_and_T90min}] and $\phi$ (control parameter exponent) [Eq.~\eqref{eq:control_parameter_exponent}], which themselves are functions of the three structural parameters $\sigma, \gamma$ and $\omega$ determined by the morphology of the building blocks. 
The different time complexity exponents associated with distinct monomer morphologies lead to considerable differences in the corresponding assembly times for large target structure sizes. 
How can we understand intuitively why monomers with different morphologies assemble at different speeds? To answer this question, first of all, we have to distinguish two experimental scenarios: In the first scenario, we assume that the monomer concentration and attachment rate are constant and the detachment rate $\delta_1$ is changed (for example, by changing binding energy or temperature) to tune $\tilde \delta_1 = \delta_1/(C\nu)$. Conversely, in the second scenario, the monomer concentration (or attachment rate) is varied while $\delta_1$ is kept constant. The scaling of the assembly time in the second scenario is therefore described by the adjoined time complexity exponents, see Sec. \ref{sec:scaling_delta_and_T90min}. We start by discussing the first scenario, in which the reactive time scale $C\nu$ remains constant and $\delta_1$ is varied. 

Here, the heuristic argument from Sec.~\ref{sec:heuristic_onset_criterion} provides a useful interpretation for why different particle morphologies assemble at vastly different speeds. 
Intuitively speaking, monomers with certain types of shapes assemble faster than others because their particular morphologies allow for a reduction of the nucleation rate (by decreasing $\delta_1$) while the overall cluster growth rate remains unaffected by changing $\delta_1$. 
Hexagon-shaped monomers are an example of such a system:
To form a nucleus of size ${\sigma=3}$ the system has to pass through an unstable state which makes the rate of nucleus formation sensitive to variations in the detachment rate $\delta_1$; 
see Fig.~\ref{fig:illustration_nucleation}. 
In contrast, once a nucleus has formed, monomers can always attach in a way so that they immediately have two binding partners, whereby the growth process is independent of $\delta_1$ 
[in Refs.~\cite{jacobs2015rational, reinhardt2016effects}, this is reflected by the finding that the free-energy gradient of large clusters is generally much larger if the subunits have higher coordination numbers].
Since a high yield of the assembly process in the strong binding limit requires the ratio between the overall nucleation and cluster growth rate to be small [cf.~Eq.~\ref{attachment_nucleation_balance}], the process can thus be controlled efficiently by tuning $\delta_1$. 
This means that only a comparatively small change in the nucleation time scale (upon changing $\delta_1$) is necessary for the ratio to decline and the yield to grow.  

In contrast, if even only a tiny fraction of the growth events following nucleation requires a transition through an unstable state (e.g., for square- and triangle-shaped monomers), these slow steps will eventually dominate the time scale of cluster growth, which thus also becomes dependent on $\delta_1$. 
Therefore, the nucleation rate must now be decreased more significantly compared to the case in which the growth rate is independent of $\delta_1$ to achieve the same ratio between nucleation and growth rate [Eq.~\eqref{attachment_nucleation_balance}]. Thus, with nucleation being the rate-limiting step [cf.~section~\ref{sec:onset_criterion}], this makes the overall process less time efficient.  

Hence, as exemplified by the system with hexagon-shaped monomers, a minimal attachment order ${\gamma = 1}$ is key to achieving maximal time efficiency in the scenario where the binding energy or temperature are varied.
In section~\ref{sec:summary_and_applications}, we will discuss several ways particle morphologies can be designed to assemble with ${\gamma=1}$ and, generally, how this can be applied to increase the efficiency of artificial self-assembly processes.   

But what happens if instead of the detachment rate $\delta_1$ the monomer concentration $C$ (or binding rate $\nu$) is varied to tune the ratio $\tilde \delta_1$? In this case, with increasing target structure size, the monomer concentration must be further reduced to enhance $\tilde \delta_1$. This necessarily increases the time scale of all binding reactions and, thus, decreases time efficiency. Since the nucleation size $\sigma$ is smallest for hexagonal monomers, the monomer concentration must be reduced even more than for square- and triangular building blocks to achieve the required reduction in the effective nucleation rate, as expressed by the magnitude of the respective control parameter exponents $\phi$ [cf.~Fig.~\ref{fig:scaling_laws}(c)]. Table~\ref{tab:adjoined_exponents} displays the adjoined time complexity exponents $\tilde \theta=\theta+\phi$, which describe the scaling of the assembly time with the target structure size for this scenario.  
Hence, assuming that the binding energy per bond for triangular and hexagonal monomers is the same and the monomer concentration is optimized, triangular monomers will self-assemble into large structures faster than hexagonal monomers. 
Irrespective of the morphology of the monomers, however, the first scenario (with the monomer concentration being constant and chosen as large as possible) is more efficient than the second (since $\theta<\tilde \theta$).
In other words, for hexagonal monomers to self-assemble significantly faster than square or triangular monomers, the binding energy per bond needs to be considerably lower for hexagon-shaped monomers (note, however, that the total binding energy per particle might still be larger, see Sec. \ref{sec:scaling_delta_and_T90min}). Hence, in our discussion in Sec.~\ref{sec:summary_and_applications}, we generally relate to the first scenario.
In the next section, we discuss a slightly different system where hexagonal particles self-assemble more efficiently, even in both experimental scenarios.

\subsection{Dependence on the morphology of the target structure}
\label{sec:tubular_structures}

\begin{figure}[tb]
\centering
\includegraphics[width=0.85\linewidth]{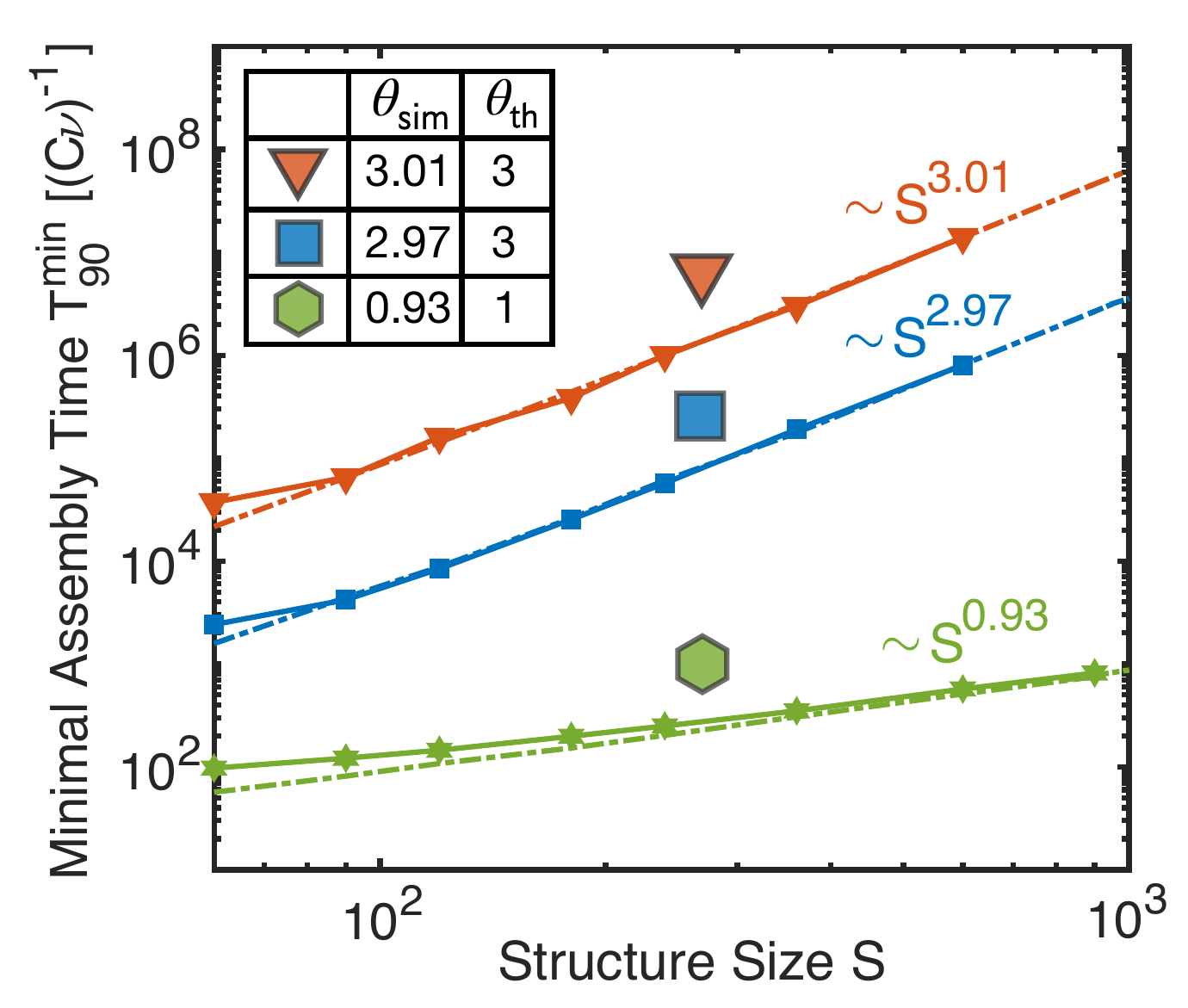}
\caption{
\textbf{Time efficiency of the self-assembly of tubes.} The minimal assembly time $T_{90}^{\text{min}}$ for the self-assembly of two-dimensional tubes (cylinders) inferred from the stochastic simulations (markers and solid lines) and effective theory (dashed-dotted lines) [cf.~App.~Sec.~\ref{app:sec:growth_rate_tube-structures}] is plotted against the size S of the target structure. The tubes have a fixed width (circumference) of six monomers and a variable height of $S/6$ monomers (with periodic boundary conditions in both dimensions). Simulations were performed with particle numbers $N=500 S$. The minimal assembly time was determined by averaging the assembly times over 10 independent runs for each different value of $\delta_1$, and subsequently taking the minimum of the averages. The table shows the time complexity exponent $\theta_{\text{sim}}$ inferred from the last three data points of the stochastic simulation in comparison with the theoretical value $\theta_{\text{th}}$ according to Eq.~\eqref{eq:time_complexity_exponent} with growth exponent $\omega=0$. Predictions of the effective model agree excellently with the simulation data. The fact that the exponents are very different from those in Fig.~\ref{fig:scaling_laws}(b) shows that also the morphology of the target structure crucially influences self-assembly efficiency.}
\label{fig:tubular_structures}
\end{figure}

We have discussed in detail how the assembly time depends on the morphology of the building blocks. Here we  briefly demonstrate that the assembly time also depends on the morphology of the target structure. Specifically, so far, we have considered the formation of bulk structures that extend symmetrically into both dimensions. An important application of self-assembly in nanotechnology is the formation of elongated tubes (cylinders)~\cite{bekyarova2005applications,de2013carbon}. Therefore, we have simulated the self-assembly of tubes with a fixed circumference of 6 monomers and a variable length of $S/6$ monomers as an example of structures that extend unequally in both dimensions. Again, we assumed periodic boundary conditions in both dimensions so that structures stop growing once they reach the target size $S$ (the case of unlimited cluster growth is analyzed in App.~Sec.~\ref{app:sec:unlimited_growth}, showing that the scaling properties are indeed identical). What time complexity exponent do we expect for the self-assembly of tubes? The key difference to the growth of symmetric target structures lies in the fact that once the tubes close along the short dimension, they keep growing at a constant growth rate $g_s$ since the perimeter of the boundary remains constant. The growth rate $g_s$ can easily be calculated with the same formalism as in Sec.~\ref{sec:effective_model} as we show explicitly in App.~Sec.~\ref{app:sec:growth_rate_tube-structures}. In particular, the growth of large clusters is characterized by a growth exponent $\omega=0$. Hence, Eq.~\eqref{eq:time_complexity_exponent} yields the time complexity exponents $\theta=3$ for square- and triangle-shaped monomers and $\theta=1$ for hexagonal monomers, respectively. The time complexity exponents measured from stochastic simulations precisely coincide with the theoretical values; see table in Fig.~\ref{fig:tubular_structures}. Furthermore, the assembly time is predicted very accurately by the effective model, with the growth rates modified accordingly [cf.~App.~Sec.~\ref{app:sec:growth_rate_tube-structures}]; see Fig.~\ref{fig:tubular_structures}. Due to the large discrepancies of the time complexity exponents, hexagonal monomers self-assemble into large tubes significantly more efficiently than square or triangular monomers. Furthermore, the adjoined time complexity exponents [cf.~Sec.~\ref{sec:scaling_delta_and_T90min}] $\tilde \theta = 3 / 4 / 5$ for hexagonal, triangular, and square monomers, respectively, are quite different in the case of tubes. This suggests that hexagon-shaped monomers are favorable also in the scenario where the monomer concentration is varied to find the optimal control parameter value.  

This demonstrates that the time complexity also depends considerably on the morphology of the target structure. Our scaling approach, however, is very general and allows us to readily predict the scaling exponents if the growth exponent $\omega$ is correctly identified for the respective target structures. Our results suggest that, also in the case of two-dimensional tubes, self-assembly is significantly more efficient if hexagonal rather than square or triangular monomers are used.

\subsection{Robustness of the scaling behavior}

To demonstrate the relevance and robustness of the scaling results, we investigated various differing scenarios and modifications of the model. The following briefly summarizes these investigations and their findings. Detailed discussions and explanations of the following analyses can be found in App. Secs. \ref{app:role_of_cluster_cluster_interactions} to \ref{app:sec:heterogeneous_systems}.  

\textit{Cluster interactions.} In the above model and analysis, we made the simplifying assumption that clusters do not interact with each other and grow only by attachment of single monomers (ideality assumption). 
In App.~Sec.~\ref{app:role_of_cluster_cluster_interactions}, we discuss in detail whether and how cluster interactions could affect the assembly time and how this affects our results. 
We find that the potential role of cluster interactions depends strongly on the morphology of the particles as well. 
In any case, however, the assembly time is not or only slightly affected by cluster interactions, showing that our results derived under the ideality assumption are robust. 

\textit{Lower yield thresholds.} Furthermore, in App.~Sec.~\ref{app:invariance_to_yield_threshold}, we show that our scaling results are invariant to the threshold for the yield we use to define the assembly time. Specifically, we verify that the time $T_{50}$ required to achieve 50\% yield scales identically as the time $T_{90}$ to achieve 90\% yield.

\textit{Seeded systems.}
In App.~Sec.~\ref{app:sec:seeding} we show that our scaling results are also robust to varying initial conditions. Specifically, we show that the scaling behavior is the same if we consider a `seeded' system in which we initially put in a number $\sim N/S$ of stable nuclei (seeds), as is often done in experiments \cite{Evans2017,kulkarni2015nanotechnology}.  
Generally, we find that the minimal assembly time increases significantly due to seeding if $\gamma>1$ (square- and triangle-shaped monomers)
 but remains constant if $\gamma=1$ (hexagonal monomers) [cf.~App.~Sec.~\ref{app:sec:seeding} and Fig.~\ref{app:fig:seeding}(b)]. 
In any case, however, we find that the scaling of the assembly time as a function of target structure size remains invariant if the system is seeded initially with a number $\sim N/S$ of nuclei [cf.~Fig.~\ref{app:fig:seeding}(c)]. This can be easily explained with our scaling analysis since the new initial condition $c_{\sigma}(t=0)\sim S^{-1}$ is consistent with the scale invariance described by the scaling form for the cluster size density, $c(s,t,S)\sim S^{-2}\widehat c (S^{-1}s, S^{-\theta} \tilde t)$ [cf.~Eq.~\eqref{eq:scaling_form_c}]; see App.~Sec.~\ref{app:sec:seeding} for details. Hence, all our scaling results remain valid for seeded systems.

\textit{Annealing.}
Annealing is another frequently used experimental technique in which temperature continuously decreases during self-assembly. The idea is to compensate for the decreasing attachment rate resulting from monomer consumption with a simultaneous reduction of the detachment rate. 
In this way, the ratio between the frequencies of detachment events (rate $\delta_1$) and attachment events (rate $\nu \, m$) can be kept at a constant level during self-assembly.  
In App.~Sec.~\ref{app:sec:annealing}, we study the behavior of self-assembly under a perfect annealing protocol, in which the ratio between attachment and detachment frequencies remains exactly constant during the process. We find that annealing enhances the time efficiency of self-assembly, but the time complexity exponent remains the same. In the framework of the effective theory, annealing is described with a time-dependent detachment rate $\delta_1(t)\sim m(t)$. Since the monomer concentration does not scale with $S$ [cf.~Eq.~\eqref{eq:scaling_form_m}], the scaling analysis can be performed in precisely the same way, and the invariance of the scaling exponents becomes evident.

\textit{Unlimited cluster growth.}
Furthermore, in App.~Sec.~\ref{app:sec:unlimited_growth}, we study the scaling properties of systems in which cluster growth is not self-limiting (i.e., there are no defined target structures), but clusters can grow indefinitely. In nature and nanotechnology, there are plentiful examples of such kinds of systems where structure growth is in principal unlimited \cite{alberts2017molecular,Hagan2021}. Describing infinite cluster growth with an effective model as in Sec.~\ref{sec:general_scaling_theory} only requires us to replace the upper limit of cluster sizes in the integral in Eq.~\eqref{eq:advection_eq_monomers} by infinity and to discard the absorbing boundary at $s=S$. Clearly, these modifications do not affect the scaling analysis. We define the yield in systems with unlimited cluster growth simply as the fraction of resources bound in clusters of a size $s\geq S$. Thereby, Eq.~\eqref{scale_less_yield_condition} immediately shows the scale invariance of this yield metric [Eq.~\eqref{eq:yield_scaling_form}] and the corresponding assembly time [Eq.~\eqref{eq:T90_scaling_form}]. Alternative definitions of the assembly time are possible as well [cf.~App.~Sec.~\ref{app:sec:unlimited_growth}]. 
Hence, our scaling results transfer directly to systems with unlimited cluster growth \footnote{We remark that the precondition for the applicability of our theory is that the growth rate of large clusters consistently scales as $\tilde r_{s\to s+1}\sim s^{\omega}$ with some arbitrary growth exponent $\omega$}. In App.~Sec.~\ref{app:sec:unlimited_growth}, we furthermore characterize the resulting final cluster size distribution in systems with unlimited cluster growth.

\textit{Heterogeneous systems.}
In the main text, we have focused exclusively on homogeneous systems in which all particles are identical. In nature and nanotechnology, there are plentiful examples of heterogeneous systems that self-assemble from distinct species of constituents \cite{ke2012three,wei2012complex,alberts2017molecular}. Heterogeneous designs can furthermore reduce assembly errors \cite{hayakawa2022geometrically} and provide a viable way to limit the growth of structures. Hence, heterogeneous systems represent a suitable testing ground to verify our scaling results experimentally. Therefore, it is important to show that our results apply likewise to heterogeneous systems. To test the validity of our results for heterogeneous systems, in App.~Sec.~\ref{app:sec:heterogen_identical_rates} we first consider the case of ideal heterogeneous systems, where distinct species only bind with their specific neighbors (see Fig.~\ref{fig:heterogeneous_sketch}), but the reaction rates and concentrations of all species are the same. We show mathematically that the ideal heterogeneous system decouples into $S$ identical homogeneous systems (one for each species). Thus, the ideal heterogeneous system is equivalent to a homogeneous system provided particle numbers are sufficiently large to suppress fluctuations between species' concentrations \cite{gartner2020stochastic}. Departing from this idealized case, in App.~Sec.~\ref{app:sec:heterogen_random_rates} and \ref{app:sec:heterogen_explicit_boundaries} we also consider systems with random heterogeneous binding rates as well as clusters that have explicit boundaries (rather than periodic boundary conditions). In both cases, self-assembly requires slightly longer, but the time complexity exponent is identical as in the ideal heterogeneous or homogeneous case. We furthermore study scenarios in which four neighboring species are provided in significantly larger concentrations [App.~Sec.~\ref{app:sec:heterogen_concnetrations}], or the respective detachment rates of these same four species are strongly reduced [App.~Sec.~\ref{app:sec:heterogen_detachment_rates}]. Both scenarios enhance the rate of nucleation of these specified four species and thus affect the assembly dynamics in similar terms as seeding discussed above. In particular, the minimal assembly time drastically increases as a result of these counter-productive measures [cf.~App.~Fig.~\ref{fig:heterogen_detachment_and_concentrations}]. In contrast, a viable way to improve time efficiency is by enhancing the cluster growth rate (rather than the nucleation rate). In heterogeneous systems, this can particularly well be done through hierarchical self-assembly as we discuss in detail in the next section (see also App.~Sec.~\ref{app:sec:heterogen_hierarchical_assembly}). Taken together, we find that heterogeneous systems behave very similarly to homogeneous systems and our results explain and describe their properties in an analogous way.

\textit{Higher order reversibility.}
Besides the assumption of an ideal  self-assembly process, the second basic assumption we made in our analysis is that the binding energy is large, and hence all higher order detachment rates $\delta_2$, $\delta_3$, ... vanish (strong binding limit). 
In a follow-up paper, we will relax this assumption and investigate how higher-order detachment processes affect the assembly dynamics in cases where the binding energy is low. In this context, we will show that our effective theory for the strong binding limit can be generalized to incorporate the effect of higher-order detachment processes. We demonstrate that the minimum assembly time and its scaling behavior remain widely unaffected by higher-order detachment processes in a broad parameter range of structure sizes and binding energies. Hence, our scaling results for the strong binding limit remain valid and useful even beyond the regime where the binding energy is very large and all higher-order detachment processes are strictly negligible.

\section{Summary and Applications}
\label{sec:summary_and_applications}

In the following, we will provide a concise overview of our findings and demonstrate how these insights contribute to a better understanding of biological self-assembly processes.
Furthermore, we will explore their potential applications in optimizing the efficiency of artificial self-assembly processes. 

\subsection{Summary}

In this work, we considered the strong binding limit of self-assembly in well-mixed systems, which applies if the binding energy is large so that only the detachment of particles with a single bond from a cluster is relevant to the dynamics. 
On the other hand, detachment processes of particles with multiple bonds (higher-order detachment processes) can be neglected on the relevant time scales.  
We found that, in this regime, the morphology of the building blocks crucially determines the dynamic properties of the self-assembly processes, most importantly, their time efficiency. 
Three structural parameters essentially determine the self-assembly dynamics: 
The nucleation size and attachment order, which describe the effective order of reactions by which clusters nucleate and grow, respectively, and the growth exponent, which determines how the growth rate scales with the structure size.
Importantly, we showed that the kinetic rate equations exhibit an inherent scale invariance that allowed us to derive scaling laws for the minimum required self-assembly time (`time complexity exponent') and the optimal control parameter value (`control parameter exponent') in dependence of the target structure size $S$. 
Both exponents are determined in terms of the three structural parameters alone.
The time complexity exponent describes an important constraint for the time efficiency of self-assembly processes for large structure sizes. 
From its dependence on the structural parameters, it can be seen that particles with different morphologies can self-assemble into large structures at vastly different speeds. 

Specifically, we found that hexagon-shaped monomers, which have an attachment order ${\gamma=1}$, can self-assemble particularly fast and exhibit a low time complexity exponent. 
An attachment order ${\gamma=1}$ means that clusters grow after nucleation by attachment of individual monomers (without having to go through a set of unstable states).
This results in fast cluster growth and efficient self-assembly provided nucleation is sufficiently retarded. 
On the other hand, if the attachment order is larger than one (as in the case of square- and triangle-shaped monomers), i.e., if the growth processes are effectively few-particle reactions, a large nucleation size is favorable for the time complexity exponent to remain small.
A large nucleation size, however, makes the time efficiency sensitive to variations in the control parameter and thus necessitates enhanced fine-tuning [cf.~Eq.~\eqref{Ty_slow_nucleation}]. 
Furthermore, the self-assembly time increases more strongly as a function of the target yield [cf.~Eq.~\eqref{qy_slow_nucleation}]. 

We thus conclude that the particles' morphology strongly impacts the time efficiency of self-assembly processes in the strong binding limit. In particular, particles with an attachment order $\gamma=1$ (like hexagon-shaped monomers) were found to self-assemble very efficiently. 
What do these results imply for nanotechnology and for our understanding of biology? 
In the following, we will outline some ideas in this regard.
 
\subsection{Increasing self-assembly efficiency by exploiting morphology effects}
\begin{figure}[b]
\centering
\includegraphics[width=0.95\linewidth]{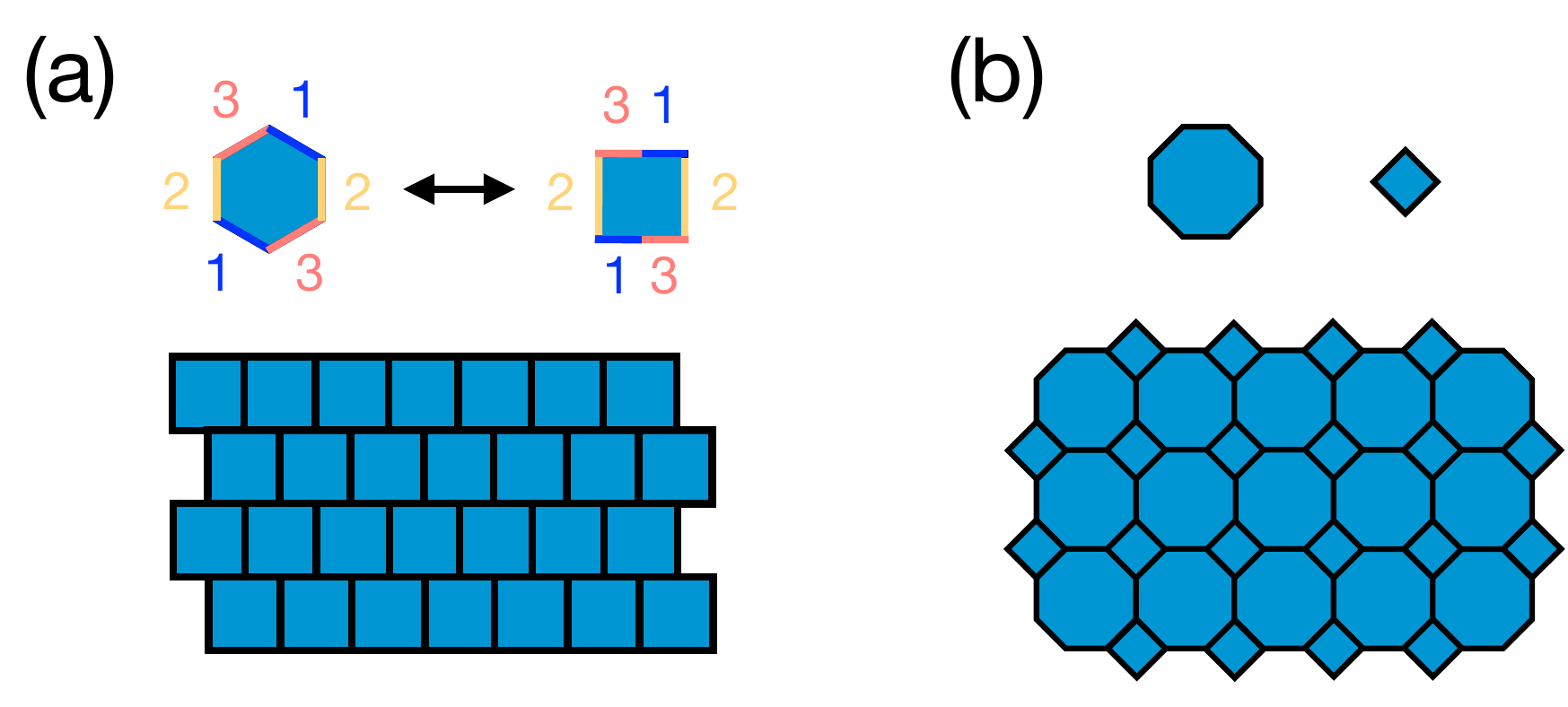}
\caption{
\textbf{Examples for morphologies that self-assemble efficiently in the strong binding limit.} Squares with six instead of four binding sites form structures topologically equivalent to those formed by hexagons [cf.~Fig.~\ref{fig:model}] (i.e., the squares in the finished structure can be smoothly deformed into hexagons without breaking any bonds). The same color and number indicate matching bonds. Hence,  self-assembly of the squares is described by the same kinetic parameters ($\sigma=3$, $\gamma=1$ and $\omega\approx 1/2$) as for hexagons and exhibits the same minimal time complexity exponent $\theta\approx 1/2$. (b) Using two or more different types of constituents (here, octagons and squares) offers additional possibilities for designing efficient assembly schemes. For the example illustrated, again ${\sigma=3}$, ${\gamma=1}$ and ${\omega\approx 1/2}$ and hence ${\theta\approx 1/2}$ as for hexagons.}
\label{fig:optimal_morphologies_examples}
\end{figure}

Self-assembly processes should be designed with specific characteristics to achieve both speed and resource efficiency. 
Our research indicates that particles that assemble with an attachment order ${\gamma=1}$, such as hexagon-shaped monomers, possess a minimal time complexity exponent. 
This implies that these systems show rapid and efficient self-assembly of large structures.

More generally, one may now ask which other particle morphologies can be identified have an attachment order ${\gamma=1}$ and thus exhibit a minimal time complexity exponent.
To give an idea for an answer, we first note that it is actually not only the shape of the monomers that determines the kinetically relevant parameters (like $\sigma$, $\gamma$, and $\omega$)  but that the concept of morphology also includes the number and positions of their binding sites. 
For example, the same kinetically relevant parameters as for hexagonal monomers can also be obtained with square-shaped monomers, provided each monomer has six instead of four binding sites, and the rows assemble `on gap' as shown in Fig.~\ref{fig:optimal_morphologies_examples}(a). 
The morphology of these squares is \textit{topologically equivalent} to that of hexagons, meaning that by deforming the building blocks while maintaining the adjacency relations with their neighbors, the resulting lattice of squares can be deformed into that of hexagons. 
Topologically equivalent morphologies thus have the same parameters $\sigma$, $\gamma$, and $\omega$. 
Hence, morphologies that are topologically equivalent to that of hexagons have the same minimal time complexity exponent. 

A vast number of additional possibilities to achieve ${\gamma = 1}$ arises if two or more different kinds of building blocks are used to form the structures. 
One such example, in which structures are built from squares and octagons, is depicted in Fig.~\ref{fig:optimal_morphologies_examples}(b), but the number of possibilities here is sheer unlimited.
Using particles with such morphologies can optimize the time- and resource efficiency of their self-assembly processes.   

\begin{figure}[!t]
\centering
\includegraphics[width=0.9\linewidth]{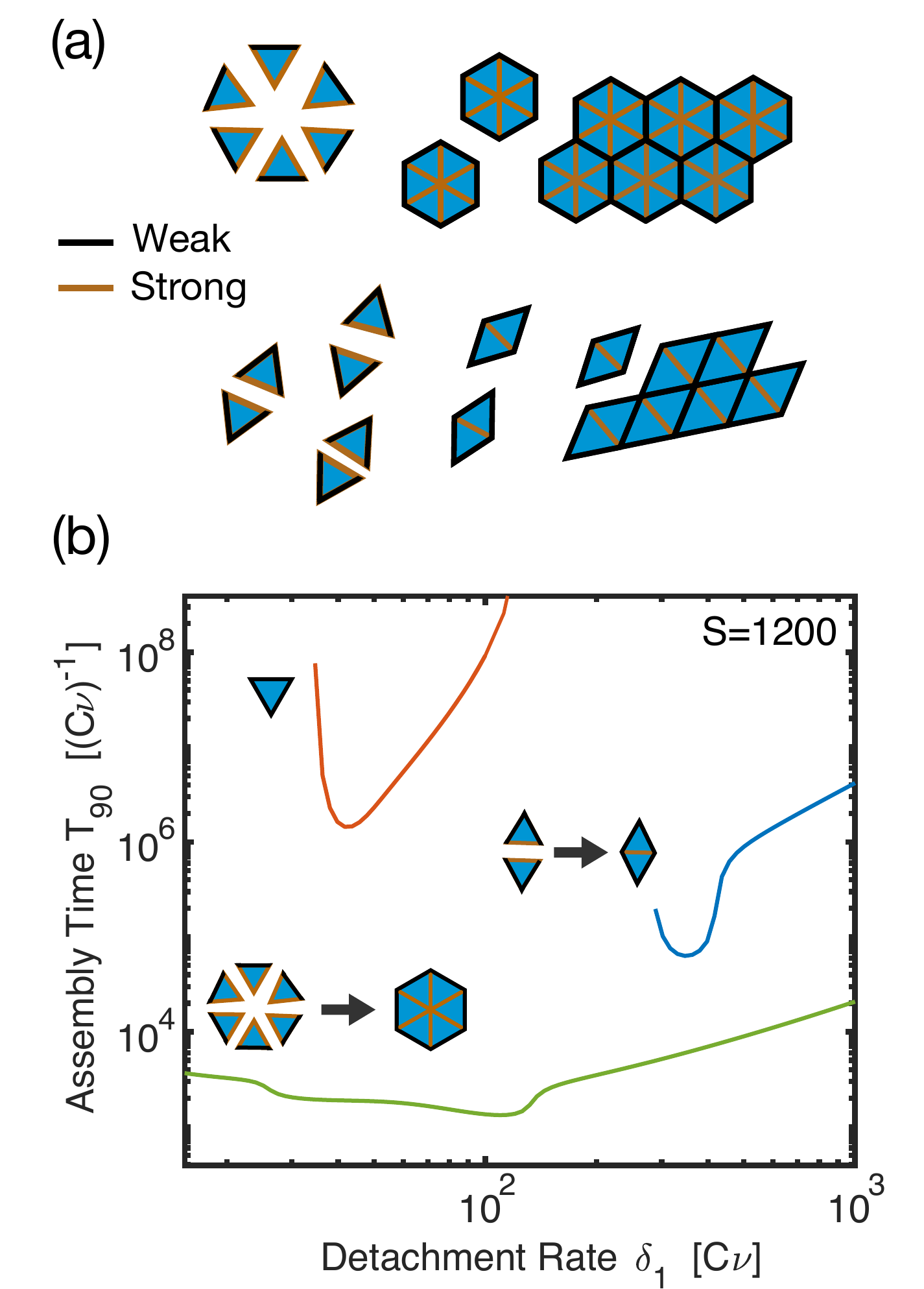}
\caption{
\textbf{Hierarchical self-assembly as a way to design optimized assembly schemes.}
Hierarchically organizing the assembly process can increase self-assembly efficiency significantly by optimizing the morphology of intermediate assembly products.
\textbf{(a)} To induce triangle-shaped monomers to assemble hierarchically, bonds between edges indicated in brown are assumed to be more stable than bonds between black edges. 
As a result of this variation of the binding strengths, monomers first form hexamers (hexagons) or dimers (rhomboids), respectively, which subsequently assemble into the structures more efficiently than the original triangles, cf.~Fig.~\ref{fig:scaling_laws}. 
\textbf{(b)} The assembly time $T_{90}$ is plotted against the detachment rate $\tilde \delta_1$ for the system with triangle-shaped monomers in comparison with the two hierarchical assembly schemes illustrated in (a) for a target structure of size ${S=1200}$. 
Simulations were performed with the effective model (see App.~\ref{app:effective_model_irr_limit}), assuming a constant ratio $r_h=10^{-4}$ between the detachment rates of strong and weak bonds. 
Since, according to our findings, the self-assembly of particles with a hexagonal morphology is remarkably efficient, the assembly time can thus be reduced by two to three orders of magnitude over a broad range of the control parameter $\tilde \delta_1$. 
}
\label{fig:hierarchical_assembly}
\end{figure}

\textit{Hierarchical self-assembly.} 
Furthermore, the morphology effect can be used to optimize self-assembly experiments by designing highly time-efficient hierarchical self-assembly schemes.
For instance, Fig.~\ref{fig:hierarchical_assembly} shows two possibilities to significantly enhance the self-assembly efficiency of triangle-shaped monomers by inducing them to form higher-order constituents with a more favorable morphology through a hierarchical assembly step.
Specifically, this is achieved by making certain bonds more stable than others [cf.~Fig.~\ref{fig:hierarchical_assembly}(a)], enabling the monomers to first form squares or hexagons, which subsequently assemble into the final structures much more efficiently than the original triangles.
We have investigated this hierarchical scenario by modifying our effective kinetic theory [cf.~App.~\ref{app:effective_model_irr_limit}] and found that both a huge reduction in assembly time and increased robustness to parameter variations can be achieved.
Specifically, for structures of size ${S = 1200}$, a reduction of the minimum assembly time by $3$ orders of magnitude can be achieved, while at the same time, the assembly time only weakly depends on the detachment rate (increasing robustness); see Fig.~\ref{fig:hierarchical_assembly}(b).
Please note that the triangle-shaped building blocks do not assemble hierarchically per se but that they must be made to do so by discriminating the bond strengths along their edges. 
Therefore, hierarchical self-assembly will provide a simple way to benefit from this morphology effect, thereby significantly increasing the efficiency and robustness of artificial self-assembly.

\subsection{Virus and carboxysome capsid assembly}

Our results might furthermore contribute to a better understanding of the kinetics of biological self-assembly processes. 
For example, several icosahedral virus capsids self-assemble from triangle-shaped capsid proteins~\cite{masson2012viralzone,kushner1969}, 
but the mechanisms underlying their self-assembly are still largely unknown. 
Our results suggest that a hierarchical step might be a possible strategy to allow large capsids to assemble time-efficiently. 
However, the icosahedral symmetry of the spherical virus capsids poses an additional challenge to the self-assembly process:
At precisely 12 sites, the capsid proteins must form pentamers, while everywhere else, they form hexamers. Only in this way closed capsids can form, whose geometry self-limits their own growth \cite{Hagan2021}. 
It has therefore been suggested that large virus capsids can only assemble in the presence of a scaffold~\cite{Li2018} that coordinates their formation. 
Thus our results might not apply directly in this case as a consequence of the scaffolding. 
Otherwise, assuming that icosahedral shells assembled without a scaffold by forming a pentameric nucleus with nucleation size ${\sigma = 5}$, Eq.~\eqref{eq:time_complexity_exponent} would predict a time complexity exponent of ${\theta = 2}$, implying that self-assembly of large icosahedral capsids would be even significantly more inefficient than self-assembly of the planar triangle-shaped system (${\theta = 1}$). 
Hence, besides structural aspects, enhancement of time efficiency could be another essential reason for the requirement of a scaffold in forming large icosahedral viruses.   

In contrast, alpha-carboxysomes, which also form large icosahedral structures, were shown to assemble also in the absence of a scaffold and are composed of proteins with hexameric and pentameric quaternary structure (CsoS1ABC and CsoS4AB)~\cite{Kerfeld2016}. 
Hence, carboxysome capsids represent the icosahedral analog of the hexagonal particle system studied here, being characterized by the same parameters $\sigma$, $\gamma$ and $\omega$. 
This particular morphology of the constituents could be an essential factor for such huge structures to assemble efficiently even in the absence of a scaffold. 
Similar hierarchical assembly pathways, in which triangle-shaped capsid proteins first form pentameric and hexameric constituents, have also been confirmed for certain classes of smaller viruses~\cite{baschek2012stochastic}, like Picorna viruses~\cite{li2012vitro}, the Brome Mosaic Virus~\cite{flasinski1997structure}, Human Papillomavirus~\cite{hanslip2006assembly} or Cowpea Chlorotic Mottle Virus~\cite{willits2003effects}, all of which are also known to assemble without a scaffold. 
These insights might be particularly relevant for experiments that try to mimic the self-assembly of artificial capsids for biotechnological applications~\cite{Sigl2021}. 
Taken together, this suggests that natural self-assembly processes have been optimized concerning their time efficiency by exploiting the dependence on the morphology of the building blocks.
It is plausible that these adaptations occurred during evolution since, according to our analysis, relatively small modifications (like alterations of individual bond strengths) could be enough to reduce the assembly time and increase the robustness of the self-assembly process significantly.

\subsection{Conclusion and Outlook}

Our study suggests that understanding the kinetic properties of self-organizing processes is crucial for properly understanding these phenomena in biology and for the experimental realization of efficient self-organizing systems.
Specifying structural determinants and conditions under which high yield is achieved in self-assembly is probably insufficient because the time it takes could be extremely long. 

Specifically, we have demonstrated that the time required to achieve high yield for large target structures can vary by several orders of magnitude depending on the morphology of the building blocks.
Therefore, in this work, we described a general mathematical framework that allows us to analyze and understand the kinetics of self-assembly processes in well-mixed systems better. 
In particular, we found that the morphology of the constituents is an essential kinetic determinant that strongly impacts self-assembly efficiency. 
We discussed how these insights could be used in nanotechnology to significantly enhance the efficiency and robustness of artificial self-assembly.   
Furthermore, we showed how this might impact our understanding of biological self-assembly phenomena. 

In this paper, we considered the strong binding limit of self-assembly, which applies if the binding energy between the constituents is large. 
In a subsequent paper, we will extend our analysis to systems with low binding energy and investigate how higher-order detachment processes affect the assembly dynamics. 

In our model and analysis, we assumed a well-mixed self-assembly system in which monomers from a dilute solution bind to sparse clusters. 
As discussed in the context of virus capsid assembly, however, some systems first form dense disordered aggregates (e.g., via phase separation or by binding to a scaffold like DNA) \cite{Li2018,ianiro2019liquid,fang2020two}. The disordered aggregates subsequently transit to an ordered state. Our well-mixed model does not describe such a scenario, and hence our time complexity results might not apply in this case. Importantly, such a spatial accumulation could significantly enhance self-assembly efficiency by increasing the local concentration of resources. Unspecific binding and local aggregation of the monomers might therefore crucially influence the time efficiency in spatial systems and could undermine the role of the monomer morphology. An interesting question to ask is whether the transition from the disordered to the ordered state can be described by a similar formalism as discussed in this manuscript, and, independent of that, what the time complexity of such spatial systems will be. In particular, it would be interesting to see how the time complexity of spatial systems compares to the time complexity of the well-mixed systems considered here. 

Furthermore, in the current model, we disregard any possibilities of misbinding or assembly errors due to defects. 
The results of grand canonical ensemble simulations \cite{jacobs2015rational,reinhardt2016effects} suggest that systems with fast cluster growth tend to be more prone to defects and assembly errors. It would therefore be interesting to add reasonable error rates to the dynamic model and see whether this leads to additional constraints regarding the optimal morphology of the subunits. Finally, it would be interesting to see whether our theory can be extended to the case of an open system, i.e., when the total number of particles is not fixed but increases through a steady influx of monomers~\cite{boettcher2015role}, and how this affects the kinetic behavior of the system.

\begin{acknowledgments}
We thank Severin Angerpointner and Richard Swiderski for stimulating discussions and proofreading the manuscript. 
This research was financially supported by the Deutsche Forschungsgemeinschaft (DFG, German Research Foundation) under Germany’s Excellence Strategy – EXC-2094–390783311 and under – Project-ID 364653263 – TRR 235 as well as by the Chan-Zuckerberg-Initiative (CZI). FMG was supported by a DFG fellowship through the Graduate School of Quantitative Biosciences Munich (QBM).
\end{acknowledgments}

\appendix

\section{Stochastic simulation}
\label{app_stochastic_simulation}

The simulations were performed using Gillespie's stochastic algorithm \cite{Gillespie2007}. In the simulation, complexes are represented as boolean arrays of size $S$, which contain ones at sites occupied by a particle and zeros otherwise. For simplicity, one can imagine the structures as two-dimensional arrays. However, internally, all data structures are represented as one-dimensional memory sequences, and two-dimensional subscript indices must be converted to linear indices. The neighboring- or adjacency relations between the sites in the two-dimensional array thereby determine the morphology and are defined as follows: In the square system, site $(i,j)$ is adjacent to sites $(i\pm 1 \mod L, j)$ and $(i, j\pm 1 \mod L)$, where $L=\sqrt{S}$ is the linear extension of the target structure and by taking the modulo periodic boundary conditions are implied. Accordingly, in the hexagonal system, site $(i,j)$ is adjacent to the six sites $(i\pm 1 \mod L, j)$, $(i, j\pm 1 \mod L)$ and $(i \pm 1 \mod L, j\mp 1 \mod L)$ and in the triangle-shaped  system, site $(i,j)$ has three neighbors $(i\pm 1 \mod L, j)$ and $(i, j+1\mod L)$ if $j$ is even or $(i, j-1\mod L)$ if $j$ is odd (see Fig.~\ref{fig:model}(a) for the logic behind these definitions).  

When a dimer forms, such a boolean array is reserved for the complex, and two arbitrary neighboring sites are chosen and set to 1 to represent the initial dimer. Due to the periodic boundary conditions, which two neighboring positions are chosen is irrelevant because the structure is translationally invariant. Subsequently, each unoccupied site in the complex with at least one occupied neighbor is occupied by a monomer at rate $\nu n$, where $n$ is the number of monomers in the system (monomer attachment). The reaction rate $\nu$ is typically set to 1. Similarly, an occupied site becomes empty again with rate $\delta_i$, where $i$ denotes the number of occupied neighboring sites (monomer detachment). Each attachment (detachment) event decreases (increases) the number $n$ of monomers in the system. In this way, the simulation respects all possible configurations of clusters that can emerge. By counting the number $M$ of complete structures, i.e., structures with $S$ occupied sites, the yield is calculated as $\text{yield}=\frac{M S}{N}$.  

It is important to optimize the code for efficiency because, since the detachment rate $\delta_1$ is typically much larger than the reaction rate $N\nu$, many Gillespie steps are generally needed until a yield of 90\% is reached even for intermediate particle numbers $N$ (typically, we simulated the system with N between 100S and 1000S so that a maximum number of 100 to 1000 target structures is built). In particular, the simulation of the triangle-shaped system is computationally expensive due to the comparatively large number of intermediate steps between two stable configurations with the triangle-shaped morphology and the longer time spans required to be simulated to reach 90\% yield. Partly, several billion Gillespie steps were needed for a single run to complete. The computational cost of the simulation strongly increases with the size of the target structure because both the detachment rate $\delta_1$ and the required simulation time $T_{90}$ increase with $S$. Furthermore, the particle number $N$ should typically be increased with $S$ to keep the number of assembled structures constant. 
Hence, to simulate also large system sizes up to a size of $S=1000$, the efficiency of the simulation is crucial. By associating additional data structures with the complexes that allow us to choose and update attachment and detachment events efficiently, our simulation written in C++ was able to perform more than one million Gillespie steps per second on a 3,1 GHz CPU. The C++ code of the simulation is available online.

\section{Effective kinetic theory in the strong binding limit} 
\label{app:effective_model_irr_limit}

In this appendix, we show how, in the strong binding limit, the self-assembly dynamics can be approximately described by effective rate equations for the cluster size distribution $c_s(t)$, cf.~Eq.~\eqref{effective_theory_1st_order}.
Our analysis assumes that the dynamics can be reduced to the most probable self-assembly paths, i.e., those reaction paths that traverse the minimum number of unstable configurations to move from one stable cluster configuration to the next.

\begin{figure*}[htb]
\centering
\includegraphics[width=0.9\linewidth]{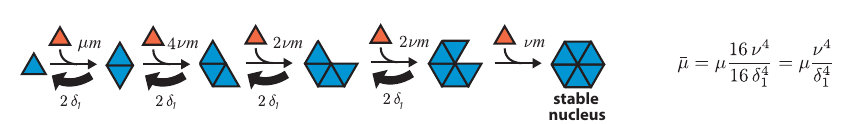}
\caption{
\textbf{Illustration of the nucleation process for triangle-shaped monomers.} 
Starting from a monomer, the nucleus is formed by subsequent attachment of monomers with growth rates ${g_i \sim \nu \, m}$ indicated in the graph. 
The detachment of monomers counteracts this with the shrinkage (detachment) rate ${d_i = 2 \delta_1}$. 
The prefactors account for the degeneracy of attachment or detachment reactions leading to equivalent configurations.
In the limiting case of large detachment rates, the effective nucleation rate $\bar \mu$ is given by the product of the forward rates divided by the product of the reverse rates.
}
\label{fig_supp:illustration_nucleation}
\end{figure*}

\subsubsection*{Effective nucleation process}

The self-assembly process starts with the formation of a stable nucleus, which corresponds to the formation of a cluster of $\sigma$ monomers, as illustrated in Fig.~\ref{fig_supp:illustration_nucleation} for triangle-shaped monomers.
This stochastic process is a one-step Markov chain with $\sigma$ states and the stable nucleus as the absorbing state.
The effective forward (growth) rates $g_i$, and backward (shrinkage) rates $d_i$ are indicated in Fig.~\ref{fig_supp:illustration_nucleation} for triangle-shaped and in Fig.~\ref{fig:illustration_nucleation} for hexagon-shaped monomers. 
Let $p_i (t)$ be the probability for the Markov chain to be in a state with $i$ monomers.
Then the set of master equations for all states ${1 \leq i < \sigma}$ read
\begin{align}
\label{eq:master_bulk}
    \partial_t p_i (t) 
    =
    g_{i-1} \, p_{i-1} 
    - d_i \, p_i 
    - g_i \, p_i
    + d_{i+1} p_{i+1} 
    \, ,
\end{align}
where ${g_0=0}$ and ${d_1=0}$ since ${i=1}$ is the initial state.
For the absorbing state, one has
\begin{align}
\label{eq:master_absorbing_state}
    \partial_t p_\sigma 
    = g_{\sigma -1 } \, p_{\sigma -1} \, .
\end{align}
In the limiting case, where the decay rates $d_i$ are large compared to the growth rates $g_i$, Eq.~\eqref{eq:master_bulk} can be assumed to be stationary (${\partial_t p_i = 0}$), leading to
\begin{align}
    p_{\sigma - 1} = 
    \frac{g_1 \, g_2 \ldots g_{\sigma-2}}
    {d_2 \, d_3 \ldots d_{\sigma-1}}
    \, p_1 \, .
\end{align}
This results in an effective equation for the nucleation process ${1 \to \sigma}$
\begin{align}
    \partial_t p_\sigma (t) 
    = 
    r_{1 \to \sigma} \, p_1 (t) 
\end{align}
with the effective nucleation rate given by the product of the forward (growth) and backward (shrinkage) rates
\begin{align}  \label{eq:rate_parameter_general}
    r_{1 \to \sigma} 
    = 
    \frac{g_1 \, g_2 \ldots g_{\sigma-1}}
    {d_2 \, d_3 \ldots d_{\sigma-1}}
    \, .
\end{align}
For the three different monomer morphologies studied here, we find that the numerical prefactors cancel (compare Fig.~\ref{fig_supp:illustration_nucleation}), and one obtains 
\begin{equation}  \label{eq:effective_nucleation}
    r_{1 \to \sigma} 
    = 
    \bar \mu \, m^\sigma \, ,
\end{equation}
with the effective kinetic parameter $\bar \mu$ for the nucleation process of order $\sigma$ given by 
\begin{align}
    \bar \mu = \mu\left(\frac{\nu}{\delta_1}\right)^{\sigma-2} \, .
\end{align}
Henceforth we will refer to this kinetic parameter simply as an \textit{effective nucleation rate}.

\subsubsection*{Effective assembly processes after nucleation}

\begin{figure}[htb]
\centering
\includegraphics[width=0.9\linewidth]{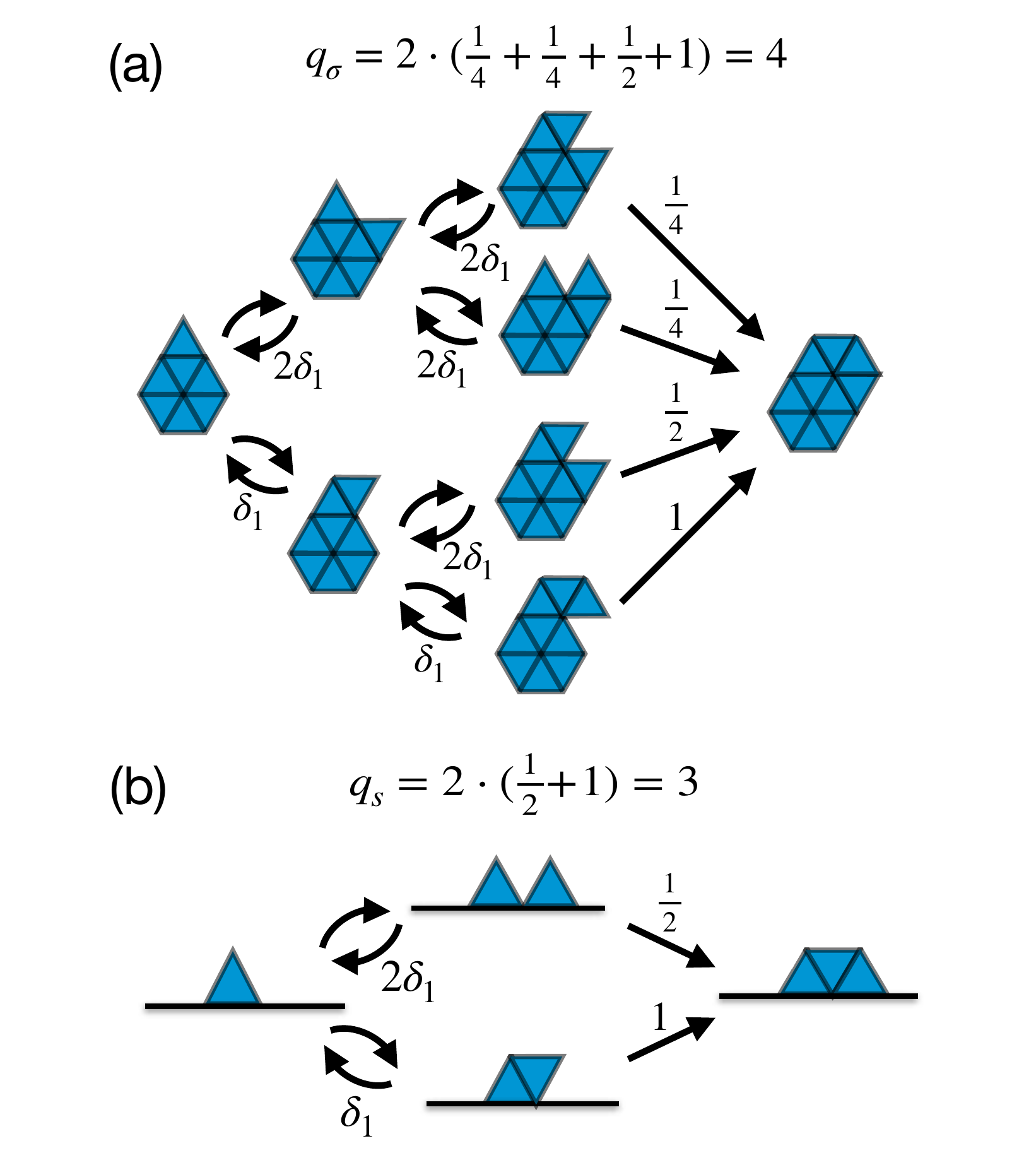}
\caption{\textbf{Derivation of the multiplicity factor $q_s$ for seed formation} for (a) nuclei and (b) larger clusters with triangle-shaped monomers. The diagram depicts all equivalent assembly pathways by which a seed can be completed after the first seed-forming monomer has attached. The rate for all forward steps in the diagram is $\nu m$ while the rates for the backward steps are given by $\delta_1$ or $2\delta_1$, depending on whether there are one or two monomers with a single bond. The total weight of each path (indicated above the rightmost arrows) is calculated by multiplying all backward rates along the respective path, i.e., by applying Eq.~\eqref{eq:rate_parameter_general} to the path. The multiplicity factor $q_s$ corresponds to the sum of the weights of all contributing paths. Note that $q_s$ has an additional factor of 2 because one must additionally account for the analogous pathways in which the seed forms to the left-hand side of the first monomer instead of to the right.}
\label{supp_fig:multiplicity_factor_q_s}
\end{figure}

\begin{figure*}[!t]
\centering
\includegraphics[width=0.9\linewidth]{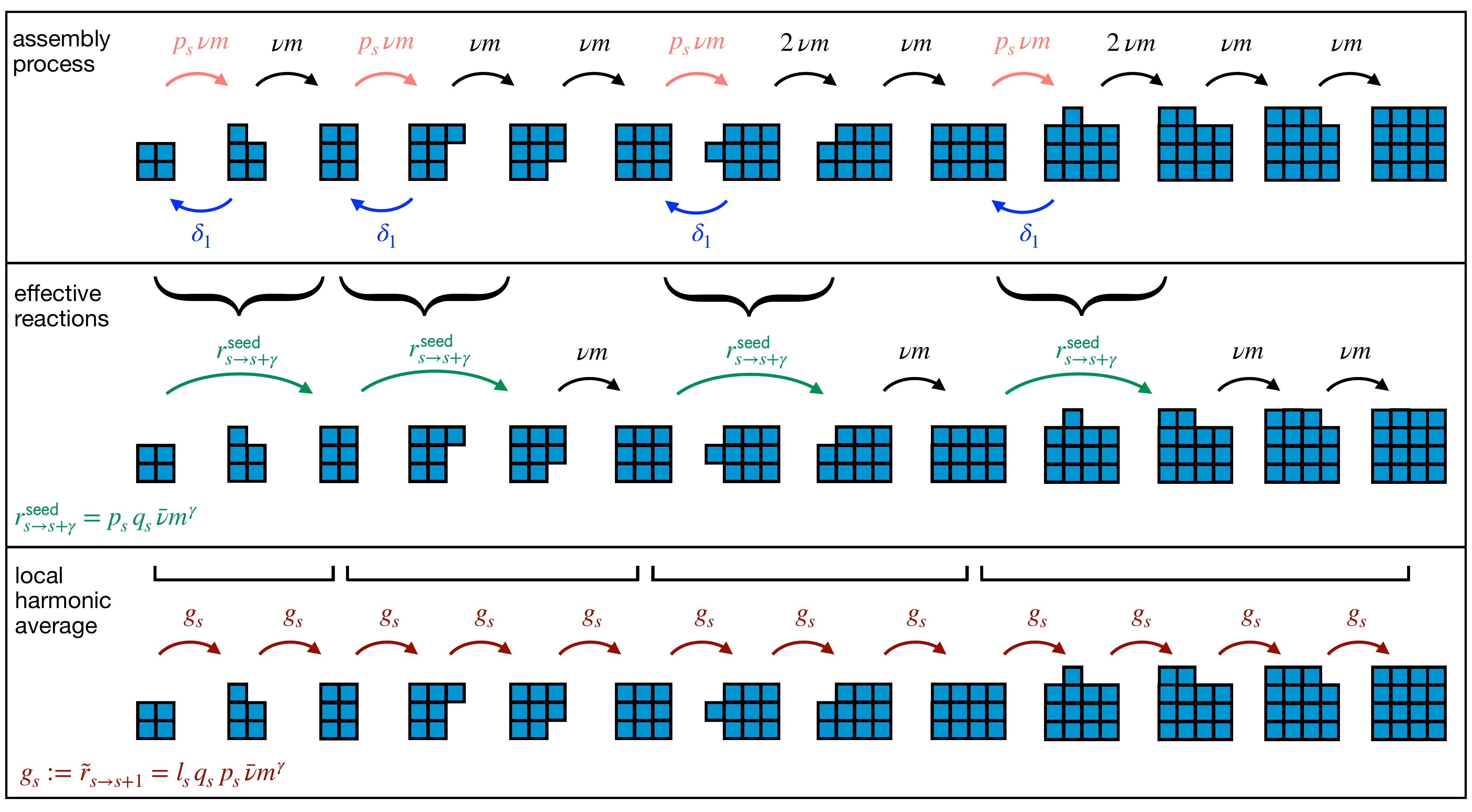}
\caption{
\textbf{Effective assembly process in the strong binding limit}, illustrated for square-shaped monomers. 
\textit{Top panel:} 
Typical assembly path leading from a stable nucleus of size ${\sigma = 4}$ to a structure of size ${s=16}$, with the corresponding transition rates between the states, indicated in the graph (assuming $\delta_n=0$ for $n\leq 2$). 
The prefactor $p_s$ denotes the average perimeter (number of binding sites for the seed) of a cluster of size $s$. 
The effective theory is derived in two steps illustrated by the following panels: 
\textit{Middle panel:} 
First, we describe transitions passing through unstable states as effective few-particle-reactions of order $\gamma$ (here, for square-shaped monomers, ${\gamma = 2}$) with rate $r_{s\to s+\gamma}^{\text{seed}}$. 
This is possible since the backward rate $\delta_1$ is large compared to the forward rates (cf. App. \ref{app:effective_model_irr_limit}). This way, we `integrate out' the fast detachment rate $\delta_1$ from the process. 
\textit{Bottom panel:} 
In the second step, we introduce one-step forward rates $g_s$, which are continuous functions of the structure size $s$. These average one-step rates are calculated by dividing the number of particles added per row by the total time it takes to add the entire row. In other words, this corresponds to taking a moving harmonic average of the rates of the leading order reactions and the subsequent subleading order reactions as indicated by the square brackets.
Since the leading order reactions are significantly slower than the subleading order reactions, the latter can be assumed as infinitely fast. 
In this way, we derived a 1-step Poisson process with transition rates that vary continuously as a function of the cluster size. 
This enables a convenient mathematical analysis and allows us to describe systems with different particle morphologies on equal footing. 
}
\label{supp_fig:derivation_1st_order_theory}
\end{figure*}

As discussed in Sec.~\ref{sec:effective_model}, there are typically two different types of processes by which clusters grow: the formation of a seed of size $\gamma$ and the subsequent growth of the seed by a series of lower-order reactions. We refer to these processes as leading and subleading processes, respectively.
For illustration, refer to Fig.~\ref{supp_fig:derivation_1st_order_theory}, where a representative example of an assembly process for square-shaped building blocks in the early stages of cluster growth is shown.

A \textit{leading-order} (seed formation) process requiring the sequential attachment of $\gamma$ monomers typically occurs for cluster configuration with `completed (or smooth)' edges, i.e., the rectangles shown in Fig.~\ref{supp_fig:derivation_1st_order_theory}.  
Using the same reasoning in terms of a Markov chain as for the nucleation process, the effective rate for seed formation reads 
\begin{align}
\label{eq:app:effective_seed_formation_rate}
    r_{s \to s + \gamma} 
    =
    p_s \, q_s \, \nu \, \left( \frac{\nu}{\delta_1} \right)^{\gamma -1} \, m^\gamma 
    \equiv 
    p_s \, q_s \, \bar \nu \, m^\gamma
    \, , 
\end{align}
where $p_s$ denotes the perimeter, or rather the number of available binding sites, of a cluster of size $s$. 
The additional factor $q_s$ is a combinatorial factor that indicates the number of equivalent ways through which the remaining ${\gamma-1}$ monomers can complete the seed after the first monomer has been attached. 
For example, in the case of square-shaped monomers, the second particle forming the seed can attach to either side of the first one unless the first particle attaches directly to a corner. Hence, for large clusters, one has $q_s=2$. 
For smaller clusters, however, the average value of $q$ is smaller; in the case of a nucleus, it is ${q_\sigma = 1}$. For triangle-shaped monomers, the equivalent assembly pathways by which a seed can be formed are illustrated in Fig.~\ref{supp_fig:multiplicity_factor_q_s} both for small clusters (nuclei) and larger clusters. Each reaction pathway contributes with a weight that is calculated by Eq.~\eqref{eq:rate_parameter_general} and the sum of these weighted pathways (times a factor of 2 since the seed can grow to both sides) yields the combinatorial factors $q_\sigma=4$ and $q_s=3$, respectively.    
The value of the perimeter $p_s$ will generally depend on the cluster's shape. For simplicity, we generally assume that clusters have a regular shape. For example, in the case of square-shaped monomers, assuming that a typical cluster has a quadratic shape, we obtain ${p_s = 4 \sqrt{s}}$.
However, only the scaling of the multiplicity factor ${p_s q_s \sim \sqrt{s}}$ for typical large clusters is relevant for the following analysis.

These \textit{slow} processes of leading order (seed formation along the edges) are usually followed by a sequence of \textit{fast} processes of lower order, referred to as a `domino effect', cf.~Fig.~\ref{fig:model}(c) and Fig.~\ref{fig:assembly_steps_large_cluster}. 
These processes typically fill up the entire row that was initiated by the seed and continue until the process can only proceed by another reaction of leading order $\gamma$. 
In the case of square-shaped monomers, these lower order reactions have order $1$ and happen at a rate $2 \, \nu \, m$ or $\nu \, m$, depending on whether they can extend the seed in both directions or only in one [cf.~Fig.~\ref{supp_fig:derivation_1st_order_theory}]. 
Analogously, in the case of triangle-shaped monomers, the effective order of these reactions is two, and the effective rates are given by $4 \, \nu \, (\nu/\delta_1) \, m^2$ or $2 \, \nu \, (\nu/\delta_1) \, m^2$, respectively, with an additional factor of 2 accounting for the two distinct but equivalent ways of adding two monomers to the seed. 
Since we assume ${\delta_1 \gg \nu m}$ (strong binding limit), the corresponding reaction rate for both morphologies of the building blocks is much larger than that of the leading order reaction [cf. Eq.~\ref{eq:app:effective_seed_formation_rate}], which therefore defines the rate-limiting steps.

The total number of monomers attaching by a leading order process and the subsequent domino effect is ${l_s \sim  \sqrt{s}}$, where $l_s$ is the typical edge length of a cluster of size $s$ \footnote{There are either no or only a small number of successive lower-order processes for the assembly steps immediately following nucleation. Hence, the growth rates for small clusters behave a bit differently, as we discuss below, specifically for triangle-shaped building blocks. In the other cases, however, good agreement with the simulation data can be obtained even if we do not treat the initial assembly steps separately and assume the same growth rate for them as for larger clusters.}.  
Since the seed formation process is rate-limiting, the effective total rate between two leading order events approximately equals the rate of seed formation
\begin{align} \label{eq:app:neglect_domino_effect_time_scale}
    r_{s \to s+ l_s} 
    \approx  
    r_{s \to s+ \gamma}^{\text{seed}} \, .
\end{align}
The alternating sequence of leading and sub-leading order processes is illustrated in the middle panel of Fig.~\ref{supp_fig:derivation_1st_order_theory}.
To arrive at a set of rate equations that have the same form for all the different morphologies of the monomers, we reduce them to an effective average growth rate $g_s$ at which a cluster grows by one monomer unit
\begin{align}  \label{eq:app:growth_rate}
    g_s := \tilde r_{s \to s + 1}
    =
    l_s \cdot p_s \, q_s \, \bar \nu \, m^\gamma 
    \, .
\end{align}
Assuming that both the typical perimeter $p_s$ and edge length $l_s$ of a cluster scale as $\sqrt{s}$ for $s\gg 1$, one finds for large clusters 
\begin{align}
    g_s
    =
    f_s \, \bar \nu \, m^\gamma 
    \, ,
\end{align}
with ${f_s = l_s \, p_s \, q_s := a \, s}$. The factor $a$ is a numerical prefactor that can be estimated (see below) by considering a typical large cluster or determined by fitting the assembly time to data from the stochastic simulation for a fixed target structure size $S$.

For assembly processes where the attachment order is ${\gamma = 1}$, like in the case of hexagon-shaped monomers, there are no subleading processes, and hence the combinatorial factor $f_s$ scales as ${f_s \sim p_s \sim \sqrt{s}}$. 

Taking the results for all monomer morphologies together, we assume that the asymptotic form of the combinatorial factor can be written as
\begin{align}    
\label{f_s_definition}
    f_s = a \, s^\omega \, ,
\end{align}
where we call $\omega$ the growth exponent, which is ${\omega = 1/2}$ for ${\gamma = 1}$ (hexagonal building blocks) and ${\omega = 1}$ for ${\gamma > 1}$ (triangular and square building blocks).

\subsubsection*{Numerical prefactor of the multiplicity factor}

The numerical prefactor $a$ in the multiplicity factor $f_s$ of the growth rate, Eq.~\eqref{f_s_definition}, affects the minimal assembly time and the optimal detachment rate. Still, it does not affect their scaling, as shown in the next section.  
We determined the corresponding values of $a$ by comparing the optimal replacement rate predicted by theory with stochastic simulation for target structures of size ${S=100}$; see Table~\ref{tab:numeric_prefactors}, where the fitted values of $a$ are compared with their theoretical estimates.
Alternatively, the parameter $a$ can be estimated by assuming typical cluster morphologies, as we briefly discuss in the following.

\setlength{\tabcolsep}{0.3cm}
\renewcommand{\arraystretch}{1.4}
\begin{table}[]
    \centering
    \begin{tabular}{|c|c|c|c|}
\hline
         &  square  &  triangle  &  hexagon \\ 
\hline
   $q_s$   &     2     &      3      &   1  \\
\hline
$a_\text{th}$  &  8   &     6      &   -  \\
\hline
$a_\text{fit}$  &  5.3   &     4      &   2.3  \\
\hline
\end{tabular}
\caption{Table showing the multiplicity factors $q_s$ for large cluster sizes $s$, as well as the theoretically estimated and fitted prefactors $a_{\text{th}}$ and $a_{\text{fit}}$, respectively, for the different particle morphologies. The theoretical values $a_{\text{th}}$ were estimated assuming a large cluster in the form of a square (square-shaped monomers) or a regular hexagon (triangle-shaped monomers). Small clusters typically have smaller values for $q_s$ and $a$, which explains why the best fit is obtained with a smaller value $a_{\text{fit}}$ compared to $a_{\text{th}}$. Clusters assembling from hexagon-shaped monomers have a rough interface, whereby estimating the prefactor $a$ (and the growth exponent $\omega$) in this case is more involved.}
\label{tab:numeric_prefactors}
\end{table}

For systems with square-shaped monomers, assuming that typical clusters have the shape of a square, the cluster boundary is ${p_s = 4\sqrt{s}}$, and the number of particles attached per attachment sequence corresponds to one-quarter of the boundary, ${l_s = \sqrt{s}}$, implying ${a=8}$ and $\omega=1$ for large clusters.  
The best fit was obtained with a slightly smaller prefactor of ${a=5.3}$, probably because small clusters have a smaller value ${a \approx 4}$ since $q_s \approx 1$ if $s$ is small. 
\begin{figure}[!t]
\centering
\includegraphics[width=0.9\linewidth]{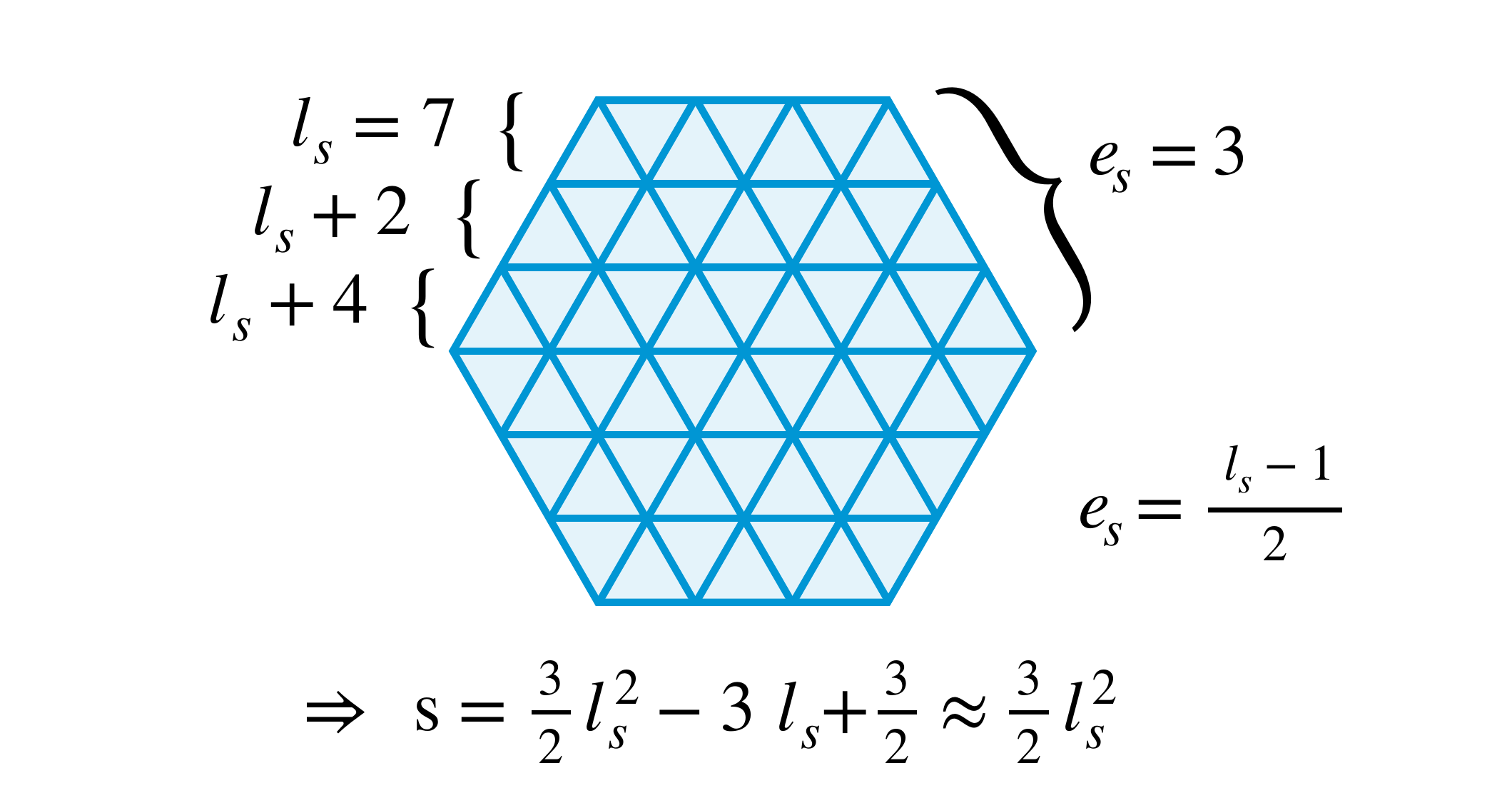}
\caption{\textbf{Regular hexagonal cluster} consisting of $s=54$ triangular building blocks to illustrate the relation between $e_s$ (number of binding sites per edge), $l_s$ (number of particles forming an edge) and the cluster size $s$. The cluster size $s$ equates to $s=2\sum_{k=0}^{e_s-1} (l_s+2\,k)$, which can be calculated to yield the relation shown in the figure. In the limit of large $s$, one thus finds $s\approx \tfrac{3}{2}l_s^2$. }
\label{fig:trigonal_ns_ls}
\end{figure}

In the case of triangle-shaped monomers, clusters typically have a hexagonal morphology. 
Assuming a cluster has a regular hexagonal shape, its edge length $e_s$ is given by ${e_s = \sqrt{s/6}}$, cf. Fig.~\ref{fig:trigonal_ns_ls}. 
The perimeter is ${p_s = 6 \, e_s}$ and the total number of particles attached in an attachment sequence (domino effect) is ${l_s=2 \, e_s + 1}$ (because an edge of length $e$ consists of $2 \, e_s + 1$ triangles, see Fig.~\ref{fig:trigonal_ns_ls}). 
Thus, ${p_s \cdot l_s \approx 2 \, s}$ and together with ${q_s = 3}$ we obtain ${a \approx 6}$ for large clusters. 
In contrast, for small clusters, where the seed tends to form at the corner of an edge, one has ${q_s \approx 1.5}$ and thus approximately ${a \approx 3}$. 
The best fit between the estimated values for large and small clusters was obtained for ${a = 4}$.   

There are no lower-order reactions for systems with hexagon-shaped monomers because ${\gamma=1}$. 
Consequently, the clusters typically have a rough interface. 
This makes the calculation of $a$ (and the growth exponent $\omega$) more involved than in the other two cases, and we refrain from making an estimate.
The best numerical fit we obtained with a relatively small prefactor ${a=2.3}$, which probably accounts for the fact that, to leading order, monomers only bind to sites at which they have two neighbors, which amounts to only a fraction of all boundary sites. 

\subsubsection*{Assembly following nucleation for triangle-shaped monomers}

For hexagon-shaped and square-shaped monomers, we have used the asymptotic expression ${f_s = a \, s^\omega}$ with the above numerically estimated values to compare the results of the effective kinetic theory with the stochastic simulations.
However, the system with triangle-shaped monomers is special because the first binding reaction after the formation of the nucleus has a higher attachment order ${\gamma=4}$ than all subsequent binding reactions, which have ${\gamma=3}$ [cf.~Fig.~\ref{fig:model}(a)].
As it turned out, the fit to the stochastic simulation is significantly improved (in the region around the minimum at $\tilde \delta_1^{\text{opt}}$) if the higher order of the initial attachment reaction is taken into account. 
Specifically, by accounting for all equivalent assembly pathways that lead from the nucleus to the next stable configuration, the effective rate of the initial attachment reaction is found to be
\begin{equation}
    r_{\sigma \to \sigma+4} 
    = 
    24 \, \bar \nu' c_{\sigma} m^4   
\end{equation}
with 
\begin{align}
    \bar\nu' = \nu\left(\frac{\nu}{\delta_1}\right)^{3} \, .
\end{align}
 Consequently, the corresponding effective attachment rate per monomer for ${\sigma \leq s \leq \sigma+3}$ is given by  
\begin{equation}
    g_s = \tilde r_{s \to s + 1}
     = 
     96 \, \bar \nu' c_{s} m^4  \, .
\end{equation}
We modified the effective theory Eq.~\eqref{effective_theory_1st_order} for the system with triangle-shaped monomers accordingly to account for the higher order of the initial attachment reaction.

\subsubsection*{Numerical solution of the kinetic rate equations}

To numerically solve the resulting kinetic rate equations given by Eq.~\eqref{effective_theory_1st_order}, we used Matlab's ode15s, a semi-implicit solver designed for solving stiff problems. 
Typically, semi-implicit ODE solvers allow one to provide the Jacobian matrix or a sparsity pattern of the Jacobian for the specific problem. 
The sparsity pattern indicates only the nonzero elements of the Jacobian without specifying them. 
For the given problem, Eq.~\eqref{effective_theory_1st_order}, the sparsity pattern is easily found by observing that the temporal evolution of $c_s$ depends only on itself, on $c_{s-1}$ and the monomer concentration ${m=c_1}$. 
Hence, the sparsity pattern is given by a ${S\times S}$ matrix with ones only in its diagonal, its subdiagonal, as well as its first column and first row, while all other entries are zero. 
By providing the sparsity pattern for the Jacobian, the performance of the solver could be accelerated significantly, especially for large structure sizes $S$. In this way, Eq.~\eqref{effective_theory_1st_order} can be solved quickly despite the very small absolute and relative error tolerance of $10^{-12}$ that we used. 
The speed could be further increased by providing an analytic Jacobian, which was unnecessary in this case.

\subsubsection*{Growth rate for tube-like structures}
\label{app:sec:growth_rate_tube-structures}

In Sec.~\ref{sec:tubular_structures} we investigate the growth of two-dimensional tubes (cylinders) with a constant width (circumference) of 6 monomers and a variable height of $S/6$ monomers. Hence, in contrast to the case of symmetric target structures, the tubes have one dimension (the width) that is significantly shorter than the other one (the height). 
Thus, as soon as the structures close in the short dimension, they continue growing with a constant growth rate since the structure's perimeter remains constant. To adopt the effective model accordingly, we determine the tubes' constant growth rate $g_s$. To this end, we note that, in the case of square and hexagonal building blocks, the number of possible binding sites $p_s$ for the first monomer to form a new seed is $p_s=2 \cdot 6$ since the monomer can attach either to the upper or the lower edge of the tube. In contrast, in the case of triangular monomers, $p_s=2\cdot 3$, since a row of width $l_s=6$ monomers only has 3 binding sites. The rate of seed formation is then given by Eq.~\eqref{eq:app:effective_seed_formation_rate} with the multiplicity factors $q_s$ shown in Table~\ref{tab:numeric_prefactors}. Due to the domino effect following seed formation, $l_s=6$ monomers in total are added as the result of a single seed-forming event in the case of square or triangular monomers ($l_s=1$ for hexagonal monomers due to $\gamma=1$). Hence, with Eq.~\eqref{eq:app:growth_rate}, we obtain the constant effective growth rate $g_s=a \bar \nu m^{\gamma}$, with $a=144/108/12$ for square, triangular and hexagonal monomers, respectively. 
For simplicity, we assume that up to size $s=30$, clusters grow with the same (size dependent) growth rate $g_s$ as symmetric structures and for $s>30$ with the constant growth rate derived above. The effective model modified this way accurately matches the simulation data for tube-like structures, as shown in Fig. \ref{fig:tubular_structures}.

\subsubsection*{Interaction of oligomers}

Equation~\eqref{effective_theory_1st_order} describes the basic reaction kinetics of reversible self-assembly. 
The basic theory can easily be extended by accounting for additional effects. 
For example, in App.~ \ref{app:role_of_cluster_cluster_interactions}, we investigate how reactions between larger oligomers might affect the assembly time [see Fig.~\ref{fig:non_ideal_assembly}(b)]. 
To estimate this effect with the help of the effective theory, we additionally accounted for the reactions of any two clusters of size ${i \geq \sigma}$ and ${j \geq \sigma}$ with ${i+j \leq S}$ to a cluster of size ${i+j}$ at rate $\nu$:
\begin{equation}
    r_{ (i,j) \to i+j} 
    = \nu \, c_{i} \, c_{j}  
    \, .
\end{equation}
To this end, we augmented the right-hand side of Eq.~\eqref{effective_theory_1st_order} by the following terms:
\begin{equation}
    \partial_t c_s 
    = ... \, + \frac{1}{2} \, \nu \sum\limits_{\substack{i,j\geq \sigma \\ i+j=s}} c_{i} c_{j} - \nu c_{s} \sum\limits_{\substack{i\geq \sigma \\i+s \leq S}} c_{i}   \, ,
\end{equation}
for all ${\sigma \leq s \leq S}$, where the factor $1/2$ in front of the first sum avoids double counting and serves as a stoichiometric factor in the case ${i=j}$. 

\subsubsection*{Hierarchical self-assembly system}

Furthermore, we used the effective theory to simulate the hierarchical self-assembly scenario in Sec.~\ref{sec:summary_and_applications} [cf.~Fig.~\ref{fig:hierarchical_assembly}].
To this end, we have formally distinguished between weak and strong bonds of the monomers. 
This causes the monomers to first form either dimeric rhombuses or hexameric hexagons, as shown in Fig.\ref{fig:hierarchical_assembly}. 

In the first case, we simulated the irreversible dimerization of two (triangle-shaped) monomers via strong bonds into rhomboids at rate ${\mu=\nu}$. 
In the second case, we considered the formation of hexagons by explicitly simulating the nucleation process of triangle-shaped monomers depicted in Appendix Fig.~\ref{fig_supp:illustration_nucleation} with detachment rate ${\delta_{\text{str}} = 10^{-4} \cdot \delta_1}$. 
The rhomboids and hexagons, once formed, are stable and react in the usual way as described by Eq.~\eqref{effective_theory_1st_order} mediated by the weak bonds. 
To simulate the assembly of these higher-order particles, we used Eq.~\eqref{effective_theory_1st_order} together with the morphological parameters for square- or hexagon-shaped particles, respectively, and the corresponding detachment rate for the weak bonds.  

Interactions between (triangle-shaped) monomers and incomplete hexagons with larger oligomers are entirely neglected, which is justified by the time interval between the reactions mediated by strong and weak bonds. 
In the hexagonal case, the rate of detachment $\delta_{\text{str}}$ of the strong bonds must be neither too large nor too small relative to $\delta_1$: 
In both cases, the formation of the hexagons would be slow, leading to an increase in the total build-up time. We found that a ratio of ${ r_h := \delta_\text{str} / \delta_1 \approx 10^{-4}}$ between the detachment rates of the strong and weak bonds minimizes about $T_{90}$.

\section{Scaling theory for yield onset}
\label{app:onset_scaling}

Here we analyze the solutions of the rate equations in the strong binding limit, Eq.~\eqref{effective_theory_1st_order}, in a parameter regime where the assembly process has a low yield. The solutions are shown to exhibit scale invariance. Using this scale invariance, we can derive a scaling law for the parameter condition for yield onset, i.e., the condition to obtain a nonzero yield, in dependence on the target structure size. 
The same scaling law also describes how the optimal parameter value scales as a function of the target structure size, which we will show rigorously in section \ref{app:general_scaling}. 

\subsubsection*{Effective dynamics of monomers and clusters}

Using the asymptotic form $f_s = a s^\omega$ for the combinatorial factor, one can rewrite the sum on the right-hand side of the rate equations, Eq.~\eqref{effective_theory_1st_order:a}, governing the evolution of the concentration of monomers, as
\begin{equation}        
\label{sum_transformation}
    \sum_{s=\sigma}^{S} s^{\omega} c_{s} 
    = 
    K \sum_{s=\sigma}^{S} s^{\omega} \frac{c_{s}}{K} 
    = K \, \langle s^{\omega} \rangle \, .
\end{equation}
Here ${K:= \sum\nolimits_{s=\sigma}^{S} c_s}$ denotes the total concentration of all complexes above the nucleation size and $\langle s^{\omega} \rangle$ defines the $\omega$-moment of the distribution of cluster sizes. 
For ${\omega=1}$, which holds for square- and triangle-shaped monomers, $\langle s^{\omega} \rangle$ reduces to the average cluster size $\langle s \rangle$.
As long as the yield is zero, the average cluster size equals the total number of bound monomers divided by the total number of clusters: ${\langle s \rangle = (C-m)/K}$.
For general $\omega$, there is no such exact identity, but to make progress, we approximate the $\omega$-- moment by ${\langle s^{\omega} \rangle \approx \langle s \rangle^{\omega}}$.
Since $\omega$ is usually close to 1 (e.g., for the hexagon-shaped monomers ${\omega \approx 1/2}$), we expect the approximation to be generally sufficiently accurate. 
Hence, as long as the yield is zero, the dynamics of the monomer concentration are governed by [cf.~Eq.~(\ref{effective_theory_1st_order}a)]
\begin{equation}     
\label{dynamics_m}
    \partial_t m(t) 
    = 
    - a \bar\nu \, m^{\gamma} 
    \left( C-m \right)^{\omega}
    K^{1-\omega} 
    - \sigma \, \bar \mu \, m^{\sigma} 
    \, .
\end{equation}
Moreover, the total concentration of complexes $K$ increases by nucleation events and thus follows the equation
\begin{equation}    
\label{dynamics_K}
    \partial_t K(t)
    = 
    \bar \mu \, m^{\sigma}  
    \, .
\end{equation}
Given initial conditions for both the concentrations of monomers $m$ and complexes $K$, Eq.~\eqref{dynamics_m} and Eq.~\eqref{dynamics_K} form a closed system of differential equations from which the number of particles that remain in the monomer pool and the number of complexes can be calculated.

\subsubsection*{Continuum limit of cluster assembly}

However, to determine the condition under which nonzero yield is achieved in the self-assembly process, we need to know how the distribution of cluster sizes evolves: Nonzero yield is achieved if the outermost front of the cluster size distribution eventually reaches the target structure size $S$, i.e., ${\lim_{t\to \infty} c_S(t) \neq 0}$.
To determine a condition for the onset of the yield, we rewrite Eq.~(\ref{effective_theory_1st_order}c) in a continuum limit by considering ${c(s)=c_s}$ as a continuous function of the cluster size $s$. Since the cluster size $s$ is typically large, we approximate ${(s-1)^{\omega} \approx s^{\omega}}$ and then use a second order Taylor expansion ${c_{s-1} \approx c(s) - \partial_s c(s) + \frac{1}{2} \partial_s^2 c(s)}$ to arrive at 
\begin{equation}   
\label{eq:advection_diffusion}
    \partial_t c (s,t) 
    =  
    - a \, \bar\nu s^{\omega} m^{\gamma} \partial_s c (s,t)
    + \tfrac{1}{2} a \, \bar\nu s^{\omega} m^{\gamma} \partial_s^2 c (s,t) \, .
\end{equation}
This constitutes an advection-diffusion equation for the time-dependent cluster size distribution $c(s,t)$ with coefficients depending on the cluster size $s$ and the instantaneous monomer concentration $m(t)$. 
The drift velocity reads ${v(s)= a \, \bar\nu s^{\omega} m^{\gamma}}$ and the diffusion coefficient ${D(s) = \tfrac{1}{2} a \, \bar\nu s^{\omega} m^{\gamma}}$.
A similar approach of approximating the system of rate equations by a continuous advection-diffusion equation has been used previously to describe virus capsid assembly~\cite{Zlotnick1999a, Endres2002, Morozov2009}. 

Equation~\eqref{eq:advection_diffusion}
can be interpreted as a Fokker-Planck equation for $c(s,t)$.
The corresponding jump moments for small time increments $\Delta t$ are given by (note that $D(s) = \frac12 v(s)$)
\begin{align}   \label{eq:jump_moments}
\begin{split}
    \langle \Delta s \rangle 
    &= v(s) \, \Delta t \, ,\\
    \langle (\Delta s)^2 \rangle 
    &= v(s) \, \Delta t \, ,
\end{split}
\end{align}
where the average is taken over the cluster size distribution. Hence the Fano factor (variance-to-mean ratio) ${F = 1}$, i.e., fluctuation effects are fairly large.
Here, as we are interested in the asymptotic scaling for large target sizes $S$, we neglect the diffusive contribution against the advective contribution. 
However, as it turns out, and is expected from the value of the Fano factor, the diffusive contribution is one of the main reasons why the yield and assembly time for small structure sizes deviate from their asymptotic scaling.  
Considering only the advective contribution, i.e., the average time evolution of the cluster distribution, and assuming that several complexes nucleate at ${s=\sigma}$ at time ${t=0}$, the leading edge of the density profile evolves according to 
\begin{equation}   
\label{dynamics_s}
    \partial_t s(t) 
    = 
    v(s(t))
    =
    a \, \bar\nu \, s(t)^{\omega} \,  m(t)^{\gamma} 
    \, ,
\end{equation}
with initial condition ${s(0)=\sigma}$. 
In this advective limit, the onset condition for the existence of a finite yield reads ${\lim_{t \to \infty} s(t) = S}$, i.e., to obtain a finite yield, the leading edge of the cluster size distribution must eventually reach the target structure size $S$. 

In summary, Eqs.~(\ref{dynamics_m},~\ref{dynamics_K},~\ref{dynamics_s}) constitute a closed set of equations that allows us to compute the onset of yield by the condition ${S=s(\infty)}$.

\subsubsection*{Analytic solution}

We were able to find a closed analytic solution only for systems with square-shaped monomers (${\sigma = 4}$, ${\omega=1}$, and ${\gamma=2}$) using elementary calculus. 
 Specifically, the complex concentration $K$ drops out of Eq.~\eqref{dynamics_m} if $\omega=1$, which can therefore be solved for the monomer concentration $m$. Using $m$ in Eq.~\eqref{dynamics_s} then gives the solution
\begin{align}
    s(\infty)
    =
    \sigma \, \frac{1+\sqrt{1-4\,\tilde \eta}}{1-\sqrt{1-4\,\tilde \eta}}
\end{align}
with ${\tilde \eta := \sigma \, \bar  \mu \, C /(a \bar \nu)}$, which, for large $S$ (and ${\eta\ll 1}$), implies
\begin{align}
    \frac{\delta_1^\text{on}}{C\nu} 
    \sim 
    \frac{\mu}{\nu} \, S \, .
\end{align}

\subsubsection*{Scaling analysis}

To determine the asymptotic scaling exponents for general cluster morphologies, however, finding an explicit solution to the above equations is not necessary.  
It is sufficient to demonstrate that in the limit of large target, structure sizes $S$, the equations obey a scale invariance, which can be exploited to determine the exponents without finding explicit solutions.  

To reveal this scale invariance, we first note that in the limit of large structure sizes, the nucleation term can be neglected in the dynamics of the monomer concentration Eq.~(\ref{dynamics_m}) since nucleation events are rare compared to the attachment of monomers to existing clusters. The rate equations can then be written in dimensionless form by measuring concentrations in units of $C$ and time in units of the effective time scale $ (a \, \bar\nu \, C^{\gamma})^{-1}$. 
Specifically, with the variable transformations ${m \to m \, C}$, ${K \to K \, C}$, and ${t \to t/(a \, \bar \nu \, C^{\gamma})}$ one obtains
\begin{subequations}
\label{mKs_system}
\begin{align}     
    \partial_t m 
    &= -{m}^{\gamma}(1-m)^{\omega}K^{1-\omega} \, , 
    \\
    \partial_t K 
    &= \eta \, {m}^{\sigma}  \, ,
    \\
    \partial_t s 
    &= s^{\omega} m^{\gamma} \, ,
\end{align} 
\end{subequations}
where we have introduced the dimensionless parameter 
\begin{align}
\label{eq:eta_parameter}
    \eta := 
    \frac{\bar \mu}{a \, \bar \nu} \, C^{\sigma-\gamma-1} \, .
\end{align}
Using a scaling ansatz 
\begin{subequations}
\begin{align}
    m 
    &= \tilde m(\eta^z t) 
    \, , \\ 
    K 
    &= \eta^{x} \tilde K(\eta^z t) 
    \, \\
    s 
    &= \eta^{y} \tilde s(\eta^z t) \, ,
\end{align}
\end{subequations}
with the dynamical exponent ${z=( 1{-}\omega)/(2{-}\omega)}$, and the scaling exponents ${x=(2{-}\omega)^{-1}}$ and $ {y=-(2{-}\omega)^{-1}}$ for the cluster density and the size $s$ of the largest clusters, one can easily verify that the resulting set of equations for the scaling functions becomes independent of the parameter $\eta$:
\begin{subequations}
\label{mKS_system_scaleless}
\begin{align}   
    \partial_{\tilde t} \, \tilde m (\tilde t) 
    &= -{\tilde m}^{\gamma}(1-\tilde m)^{\omega} \tilde K^{1-\omega} \, , 
    \\
    \partial_{\tilde t} \, 
     \tilde K (\tilde t)  
    &= {\tilde m}^{\sigma} \, , 
    \\
    \partial_{\tilde t} \, \tilde s (\tilde t) 
    &= \tilde s^{\omega} \tilde m^{\gamma} \, ,
\end{align} 
\end{subequations}
where ${\tilde t = \eta^z \, t}$.
Therefore, the scaling functions $\tilde m$, $\tilde K$ and $\tilde s$ still depend on the parameters $\sigma$, $\gamma$ and $\omega$ characterizing the morphology, but are independent of $\eta$.
The fact that the dynamics obey scale invariance can now be exploited to write the onset condition for a finite yield as
\begin{align}
    S 
    = 
    s(\infty) 
    = \eta^y \, \tilde s(\infty) 
    \sim \eta^{y} \, ,
\end{align}
since $\tilde s(\infty)$ is just a number that is independent of $\eta$. We conclude that there must be a scaling relation
\begin{equation}   
\label{scaling_eta}
    \eta^{-1} 
    \sim 
    S^{2-\omega} \, .
\end{equation}
which, by using the definition of the dimensionless parameter $\eta$ [Eq.~\eqref{eq:eta_parameter}] and the definitions for $\bar\mu$ and $\bar\nu$ from Eqs.~\eqref{eq:nucleation_rate} and \eqref{eq:rate_assembly_step_after_nucleation},
can be written as a scaling relation for the detachment rate
\begin{equation}  
\label{eq:supp:scaling_delta_opt}
    \frac{\delta_1^\text{opt}}{C\nu} 
    \sim 
    \frac{\delta_1^\text{on}}{C\nu} 
    \sim 
    \left( \frac{\mu}{\nu} \right)^{\frac{1}{\sigma-\gamma-1}}
    S^\phi
    \, ,
\end{equation}
with the control parameter exponent 
\begin{align}  \label{eq:supp:definition_phi}
    \phi 
    = 
    \frac{2-\omega}{\sigma-\gamma-1}
    \, .
\end{align}
This relation is consistent with the result obtained by the closed analytical solution above for the system with square-shaped monomers. 

Since nucleation is the time-limiting step, the assembly time $T_{90}$ can be estimated as the time required for $0.9 \, C/S$ nucleation events to happen.
Denoting the effective nucleation rate resulting from ${\delta_1 = \delta_1^\text{opt}}$ by $\bar \mu^\text{opt}$, one hence obtains
\begin{align}    
\label{scaling_Tmin}
    T^\text{min}_{90} 
    &\sim 
    \frac{C/S}{\bar \mu^\text{opt}\,C^{\sigma}} 
    \sim
    (C\nu)^{-1} 
    \left(
    \frac{\mu}{\nu}
    \right)^{\frac{\gamma-1}{\sigma-\gamma-1}} 
    S^\theta
    \, ,
\end{align}
with the asymptotic time complexity exponent given by
\begin{align}  \label{eq:supp:definition_theta}
    \theta 
    = 
    \frac{(1-\omega)\,\sigma+\gamma+2\,\omega-3}{\sigma-\gamma-1} 
    \, .
\end{align}
Alternatively, as described in the main text, the time complexity exponent can also be read off directly from Eq.~\eqref{mKS_system_scaleless}. Specifically, by reintroducing the proper dimensions, the time scale of the dynamics described by Eq.~\eqref{mKS_system_scaleless} is found to obey
\begin{equation}
\tau_{\text{assem}} \sim \left( a \, \bar \nu \, C^\gamma \, \eta^z \right)^{-1}
    \sim 
    \delta_1^\alpha  \, ,
\end{equation}
with ${\alpha = \gamma - 1 + z \, (\sigma - \gamma - 1)}$. Using Eqs.~\eqref{eq:supp:scaling_delta_opt} and \eqref{eq:supp:definition_phi}, one thus obtains
\begin{equation}
\tau_{\text{assem}} \sim S^{\phi \alpha} := S^{\theta} \,
\end{equation}
with $\theta$ given by Eq.~\eqref{eq:supp:definition_theta}.

\section{General scaling symmetry}
\label{app:general_scaling} 

A similar scaling symmetry that we found for Eq.~\eqref{mKs_system} can be shown to hold approximately even more generally, namely for the entire dynamic system described by the rate equations, Eq.~\eqref{effective_theory_1st_order}. 
Note that in Eq.~\eqref{mKs_system}, we assumed that complexes can grow indefinitely, and therefore the derived scaling symmetry holds only as long as the yield of completed structures is zero.  
In contrast, a scaling symmetry for the entire system would have even more far-reaching consequences. 
It would prove, among other things, our basic assumption underlying the analysis in the previous section that the threshold rate for nonzero yield scales identically as the optimal rate to achieve 90\% yield in the shortest time. 

To see how a scaling symmetry also holds for the entire system, Eq.~\eqref{effective_theory_1st_order}, we again use a continuous (hydrodynamic) approximation and write the dynamic system as a partial differential equation. 
To this end, we interpret the density $c(s)$ as a continuous function of a real variable $s$ over the interval $[\sigma, S ]$. 
Expanding the first term in Eq.~\eqref{effective_theory_1st_order}c to first order in $\partial_s$, one obtains the advection equation, 
\begin{align}
    \partial_t c (s,t)
    &= 
    \big(
    s^{\omega} c(s)
    -\partial_s [s^{\omega} c(s)]
    -s^{\omega} c(s) 
    \big) \,
    a \, \bar\nu \, m(t)^{\gamma} 
    \nonumber \\
    &= 
    -a \, \bar\nu \, m(t)^{\gamma} \partial_s [s^{\omega} c(s)] \, ,
\end{align}
where ${m:=c_1}$ denotes the concentration of monomers.
This can also be written as a continuity equation ${\partial_t c (s,t) = -\partial_s J(s) }$ with the flux (of cluster sizes) given by ${J(s) =a \, \bar\nu \, m^\gamma  s^\omega c(s)}$.
Note that this first-order expansion, where we neglect the diffusion term and higher-order contributions, becomes exact in the limit ${S \to \infty}$, where advection becomes the dominant driving force for the transfer of cluster mass. 
The equation for the concentration of nuclei,
Eq.~(\ref{effective_theory_1st_order}b), translates into a boundary condition at ${s=\sigma}$, where the the influx of complexes ${J_\text{in} = \bar \mu m^{\sigma}}$ must match the outflux towards larger complexes ${J(\sigma) = a \, \bar \nu \, m^{\gamma} \sigma^{\omega} c(\sigma)}$. 
At ${s=S}$, we set an absorbing boundary condition, which removes completed structures from the system. 
Finally, in the equation for the monomers, Eq.~(\ref{effective_theory_1st_order}a), the sum is replaced by an integral,
${\sum_{\sigma}^{S-1} s^{\omega}c(s) 
    \to \int_{\sigma}^{S} s^{\omega}c(s) \, ds}$ 
and we neglect the first term describing the loss of monomers by nucleation versus the second term describing the loss by attachment of monomers to clusters.
The latter approximation is accurate whenever there is significant growth of the structures and $S$ is sufficiently large so that, in total, there are many more attachment events than nucleation events. 

Taken together, this gives the following set of hydrodynamic equations
\begin{subequations}
\begin{align}
\label{app:eq:monomer_conc}
    \partial_t m (t) 
    &= 
    -a \, \bar\nu \, m(t)^{\gamma} \int_{\sigma}^{S}  s^{\omega} c(s,t) \ ds \\
    \partial_t c (s,t)
    &= 
    -a \, \bar\nu \, m(t)^{\gamma} \partial_s (s^{\omega} c(s,t)) \, ,
\end{align}
with boundary conditions
\begin{align}
    J(\sigma,t)
    &= a\, \bar\nu \, m(t)^{\gamma} \sigma^{\omega} c(\sigma,t) 
    = \bar\mu \, m(t)^{\sigma}  \,
    \\
    c(S,t) &= 0 
    \, .
\end{align}
and initial conditions
\begin{align}
    m(0) &= C \, ,
    \\
    c(s,0) &= 0 
    \, .
\end{align}
\end{subequations}
Next, we use the variables transformations ${m \to  m \, C}$, ${c \to c \, C}$, ${t \to \tau \, /(\nu C)}$, and ${s \to x \, S}$, to convert the equations into dimensionless form. In these dimensionless units, the effective attachment and nucleation rate are given by $\bar \nu = \tilde \delta_1^{1-\gamma}$ and $\bar \mu = \tilde \mu \, \tilde \delta_1^{2-\sigma}$, with $\tilde \mu := \mu/\nu$. Furthermore, we approximate the system by replacing the position of the lower boundary $\sigma/S$ (in dimensionless units) with some small and constant $\epsilon$. 
Hence, instead of a process in which clusters grow from a size $\sigma$ to $S$, we consider one in which clusters only grow from size $\epsilon S$ to $S$, thereby neglecting a fraction ${(\epsilon S - \sigma)/S \approx \epsilon}$ of all assembly steps. If $\omega>1$, as we will discuss in detail in App.~Sec.~\ref{app:sec:3D_structures}, this approximation can become inaccurate because the growth of small clusters from size $\sigma$ to $\epsilon \, S$ defines the predominant time scale in the growth process of a cluster. In this case, we will therefore use a different approximation; see App.~Sec.~\ref{app:sec:3D_structures} for details. \\ 
In dimensionless form, with the lower boundary of the integral approximated by the constant, the system becomes  
\begin{subequations}
\label{app:eq:general_scaling_system}
\begin{align}  \label{app:eq:monomer_conc_dimensionless}
    &\partial_{\tau} m (\tau)
    = 
    - a \, \tilde \delta_1^{1-\gamma} \, m(\tau)^\gamma \,  S^{1+\omega} 
    \int_{\epsilon}^{1} 
    x^{\omega} c(x,\tau) \, dx \, 
    \\
    &\partial_{\tau} c (x,\tau) 
    = - a \, \tilde \delta_1^{1-\gamma} \, m(\tau)^\gamma \, S^{\omega-1} \partial_x(x^{\omega} c(x,\tau)) \, ,
\end{align}
with the boundary conditions
\begin{align}
\label{app:eq:general_scaling_left_boundary}
    J(\epsilon,\tau) 
    &=
    a \, \tilde \delta_1^{1-\gamma} \, S^\omega \epsilon^\omega m(\tau)^\gamma c(\epsilon,\tau) 
    = \tilde \mu \, \tilde \delta_1^{2-\sigma} \,m^{\sigma}
    \\
    \label{app:eq:general_scaling_right_boundary}
    c(1,\tau) &= 0 
    \, ,
\end{align}
and initial conditions
\begin{align}
\label{app:eq:general_scaling_initial_m}
    m(0) &= 1 \, ,
    \\
    \label{app:eq:general_scaling_initial_c}
    c(x,0) &= 0 
    \, .
\end{align}
\end{subequations}
Next, we perform the following general scaling ansatz, $\tilde \delta_1 \to S^{\phi} \, \widehat \delta_1$; $\tau \to S^{\theta} \, \widehat \tau$; $m \to S^{\chi} \, \widehat m$ and $c \to S^{\xi} \, \widehat c$. To show that the system exhibits scale invariance, the exponents $\phi, \theta, \phi$ and $\xi$ must be determined in a way so that the system of the new scaling variables (those with the hat) is independent of $S$. By equating powers of $S$ on both sides of the equations after the scale transformation, this results in a set of linear equations for the exponents which must be simultaneously fulfilled. Since Eqs.~\eqref{app:eq:general_scaling_right_boundary} and \eqref{app:eq:general_scaling_initial_c} equate to zero, they pose no additional constraint on the exponents and the initial condition Eq.~\eqref{app:eq:general_scaling_initial_m} immediately implies that $\chi = 0$. The remaining three linear equations then become
\begin{subequations}
\begin{align}
    -\theta &= \phi \, (1-\gamma) + 1+\omega + \xi \, ,  \label{eq:scale_trafo1}  \\
    \xi-\theta &= \phi \, (1-\gamma) + \omega-1 + \xi \, , \label{eq:scale_trafo2} \\
    \phi \, (2-\sigma) &= \phi \, (1-\gamma) + \omega + \xi \, .  \label{eq:scale_trafo3}
\end{align}
\end{subequations}
Subtracting the second from the first equation yields $\xi=-2$ and solving the third equation then gives 
\begin{align}
\label{app:eq:control_parameter_exp_general_scaling}
    \phi = \frac{2-\omega}{\sigma-\gamma-1} \, .
\end{align}
The first or second equation can then be solved for $\theta$, which gives the unique solution
\begin{align}   \label{app:eq:time_complexity_exp_general_scaling}
    \theta = \frac{(1-\omega)\sigma+\gamma+2\omega-3}{\sigma-\gamma-1} \, ,
\end{align}
consistent with the exponents derived in the previous section [cf.~Eqs.~\eqref{eq:supp:definition_phi} and \eqref{eq:supp:definition_theta}]. 
This scaling symmetry implies that it is sufficient to understand the behavior of the scale-free system (for the variables with the hat), whereupon the functions $m(t)$, $c(s,t)$, etc., for arbitrary (but sufficiently large) structure size $S$ are obtained by appropriate rescaling. 
Specifically, the monomer concentration and complex size distribution obey the scaling forms
\begin{subequations}
\begin{align}  \label{app:eq:scaling_form_m}
m(t,\tilde \delta_1,S) 
	&= 
	C \, \widehat m(S^{-\theta} \, \tau, \, S^{-\phi} \, \tilde \delta_1) \, , \\
 \label{app:eq:scaling_form_c}
    c(s,t,\tilde \delta_1,S) 
	&= 
	C S^{-2} \ \widehat c(x, \, S^{-\theta} \, \tau, \, S^{-\phi} \, \tilde \delta_1) \, . 
\end{align}
\end{subequations}
To relate the exponent $\theta$ with the time complexity exponent, however, it is necessary that the condition to realize a yield $Y$, translates into a scaleless condition depending only on the variables $\widehat m$ and $\widehat c$. This can be seen by noting that the yield being $Y$ is equivalent to the amount of resources that remain in the system (i.e., that have not been absorbed by the absorbing boundary) being equal to ${1-Y}$: ${m+\int_{\sigma}^{S} s \, c(s) \, ds = 1-Y}$.
Hence, performing the same variable- and scale transformations, this condition translates into the scale-free form ${\widehat m + \int_{\epsilon}^{1} x \, \widehat c \, dx = 1-Y}$.
Due to the scale invariance of the governing equation [Eq.~(\ref{app:eq:general_scaling_system})], as well as the yield condition, the exponents $\phi$ and $\theta$ can be identified with the parameter- and time complexity exponents measured in the simulations.

The general scaling symmetry implies that the same scaling behavior that was found to hold for the optimal value of the control parameter and the minimal assembly time, in fact, applies identically for the entire functions $T_Y(\tilde \delta_1)$, denoting the time required to achieve a yield ${Y \in (0,1)}$ as a function of $\tilde \delta_1$, whenever this has a finite value: 
\begin{equation}
T_Y(\tilde \delta_1) = S^{\theta} \, \widehat T_Y(S^{-\phi} \, \tilde \delta_1) \, ,
\end{equation}
In particular, this shows that for the scaling behavior of $T_Y$, it is irrelevant which value we choose for the yield threshold $Y$ (in the main text, we exclusively used ${Y = 0.9}$). 
Finally, the scaling symmetry also explains the collapse of the yield curves in Fig.~\ref{fig:yield_scaling} and, in particular, it shows that the threshold parameter values $\tilde \delta_1^{\text{on,Y}}$ scale identically as the optimal parameter values $\tilde \delta_1^{\text{opt,Z}}$, independently of the demanded values $Y$ and $Z$ for the yield.

\subsection{Scaling theory for the dimerization barrier $\tilde \mu$}
\label{scaling_theory_dimerization_barrier}

With the scaling ansatz in the last section, we have transformed the system into one that is independent of $S$, giving us the scaling forms Eqs.~\eqref{app:eq:scaling_form_m} and \eqref{app:eq:scaling_form_m}. However, through the boundary condition Eq.~\eqref{app:eq:general_scaling_left_boundary}, the solution still depends on the ratio $\tilde \mu = \mu/\nu$ between the dimerization and the attachment rate. For the simulations in the main text, we set $\mu=\nu$, corresponding to $\tilde \mu=1$. A ratio $\tilde \mu<1$ corresponds to the case where the dimerization rate (or any other subnucleation binding reaction rate) is smaller than the typical attachment rate $\nu$, e.g., as a result of cooperative binding effects. With a similar scaling ansatz as before, we can also eliminate the parameter $\tilde \mu$ from the system. Specifically, transforming the control parameter as $\tilde \delta_1 \to {\tilde \mu}^{\phi'} \, \tilde \delta_1$ and time as $\tau \to {\tilde \mu}^{\theta'} \, \tau$, the system can be shown to become independent of $\tilde \mu$ if 
\begin{equation}
\phi' = \frac{1}{\sigma-\gamma-1} \qquad \text{and} \qquad \theta' = \frac{\gamma-1}{\sigma-\gamma-1} \, .
\end{equation}
We call $\phi'$ and $\theta'$ the complementary control parameter- and time complexity exponent, respectively.
The corresponding general scaling forms for $m$ and $c$ thereby read:
\begin{subequations}
\begin{align}
m(t,\tilde \delta_1,S,\tilde \mu) 
	&= 
	C \, \widehat m(S^{-\theta} {\tilde \mu}^{-\theta'} \tau, \, S^{-\phi} {\tilde \mu}^{-\phi'} \tilde \delta_1) \, , \\
    c(s,t,\tilde \delta_1,S, \tilde \mu) 
	&= 
	C S^{-2} \ \widehat c(x, \,  S^{-\theta} {\tilde \mu}^{-\theta'} \tau,  \, S^{-\phi} {\tilde \mu}^{-\phi'} \tilde \delta_1) \, . 
\end{align}
\end{subequations}
and for the assembly time, we obtain
\begin{equation}  \label{app:eq:complementary_scaling_T_Y}
T_Y(\tilde \delta_1) = S^{\theta} {\tilde \mu}^{\theta'} \, \widehat T_Y(S^{-\phi} {\tilde \mu}^{-\phi'} \, \tilde \delta_1) \, .
\end{equation}
This implies that whether or not the minimal assembly time (as a function of $\tilde \delta_1$) can be reduced by decreasing $\tilde \mu$ depends on the morphology of the building blocks: For hexagonal building blocks, we have $\theta'=0$, so $T_{90}^{\text{min}}$ can no further be reduced. In contrast, for square- and triangle-shaped monomers, we have $\theta'=1$, and so the minimal assembly time decreases proportionally with $\tilde \mu$ within a certain range of $\tilde \mu$ for which our effective theory gives an accurate description.  \\
Note that, if the complementary exponents are nonzero, the efficiency of the simulation can be increased significantly by reducing $\tilde \mu$: First, reducing $\tilde \mu$ (for $\theta'>0$) diminishes the assembly time, and thus the simulation time. In addition, if $\phi'>0$, the optimal parameter value decreases as well. Hence, fewer simulation steps are needed to simulate one time unit if $\delta_1$ is close to the respective optimal value. This offers the possibility to speed up simulations that would normally be extremely inefficient (if the goal is to determine the minimal assembly time or its scaling properties). For example, we use this trick for systems in Sec.~\ref{app:sec:heterogeneous_systems} that would otherwise take extremely long to simulate. With a reduced value of $\tilde \mu$, however, we are able to determine their scaling properties with reasonable simulation effort. Plotting the assembly time in units of $(C\nu)^{-1} \tilde \mu^{\theta'}$ and the control parameter in units of $(C\nu) \tilde \mu^{\phi'}$ allows one to compare simulation data obtained with different values of $\tilde \mu$.  \\ 

Reducing $\tilde \mu$ decreases the nucleation rate relative to the growth rate of clusters. Thus, it has the same effect on the assembly dynamics as increasing the detachment rate $\tilde \delta_1$. Hence, we can also ask how the assembly time scales with the structure size if instead of $\tilde \delta_1$ we change only $\tilde \mu$ to control the assembly process. In terms of our scaling approach, this corresponds to switching the roles of $\tilde \delta_1$ and $\tilde \mu$. Specifically, we now use the scaling ansatz $\tilde \mu \to S^{\phi_{\mu}} {\tilde \delta_1}^{\phi_{\mu}'} \, \widehat \mu$; \, $\tau \to S^{\theta_{\mu}}  {\tilde \delta_1}^{\theta_{\mu}'} \, \widehat \tau$; \, $c\to S^{\xi_{\mu}} \, \widehat c$ on the system Eq.~\eqref{app:eq:general_scaling_system}. This yields two sets of linear equations for the exponents of $S$ and $\tilde \delta_1$ on both sides of the equations. Solving the linear systems analogously, as shown above, we obtain the exponents
\begin{equation}
\phi_{\mu}  = \omega-2\, ,   \qquad \theta_{\mu} = 1-\omega \, , \qquad \xi_{\mu} = -2  \, , 
\end{equation}
and complementary exponents
\begin{equation}
\phi_{\mu}' = \sigma-\gamma-1 \, ,  \qquad \quad \theta_{\mu}'= \gamma-1   \, .
\end{equation}
For the scaling of the assembly time, this implies 
\begin{equation}
T_Y(\tilde \mu) = S^{\theta_{\mu}}  {\tilde \delta_1}^{\theta_{\mu}'} \, \widehat T_Y(S^{-\phi_{\mu}} \, \tilde \delta_1^{-\phi_{\mu}'} \, \tilde \mu) \, .
\end{equation}
By tuning the dimerization barrier, we find that one achieves the same minimal time complexity exponent $1-\omega$ as by controlling $\delta_1$ for the case $\gamma=1$. Indeed, both these scenarios are essentially the same since either of the two scenarios (decreasing $\tilde \mu$ or increasing $\tilde \delta_1$ in the case $\gamma=1$) reduces the speed of nucleation while leaving the speed of cluster growth unaffected. Note, however, that the time complexity exponent in the $\tilde \delta_1$-scenario exists only if $\sigma>\gamma+1$, whereas for the $\tilde \mu$-scenario, $\theta_{\mu}$ is independent of $\sigma$ and $\gamma$. This implies that controlling $\tilde \mu$ allows to achieve high resource and time efficiency in the case where $\tilde \delta_1$ is very small so that single bonds can already be considered stable on the relevant time scales. The corresponding nucleation size in this case is $\sigma=2$ and the attachment order $\gamma=1$. Assuming a growth exponent $\omega=1/2$ for two-dimensional structures would thus yield a time complexity exponent $\theta_{\mu}= 1/2$. However, controlling the parameter $\tilde \mu$ in an experiment is presumably tricky since it relies on allosteric or cooperative binding effects. In contrast, the parameter $\tilde \delta_1$ can be controlled more effectively, e.g., by changing the temperature or the monomer concentration, among other possibilities (cf.~Sec.~\ref{sec:model_description}).

\subsection{Large growth exponents: 3D structures}
\label{app:sec:3D_structures}
\begin{figure*}[!t]
\centering
\includegraphics[width=1.0\linewidth]{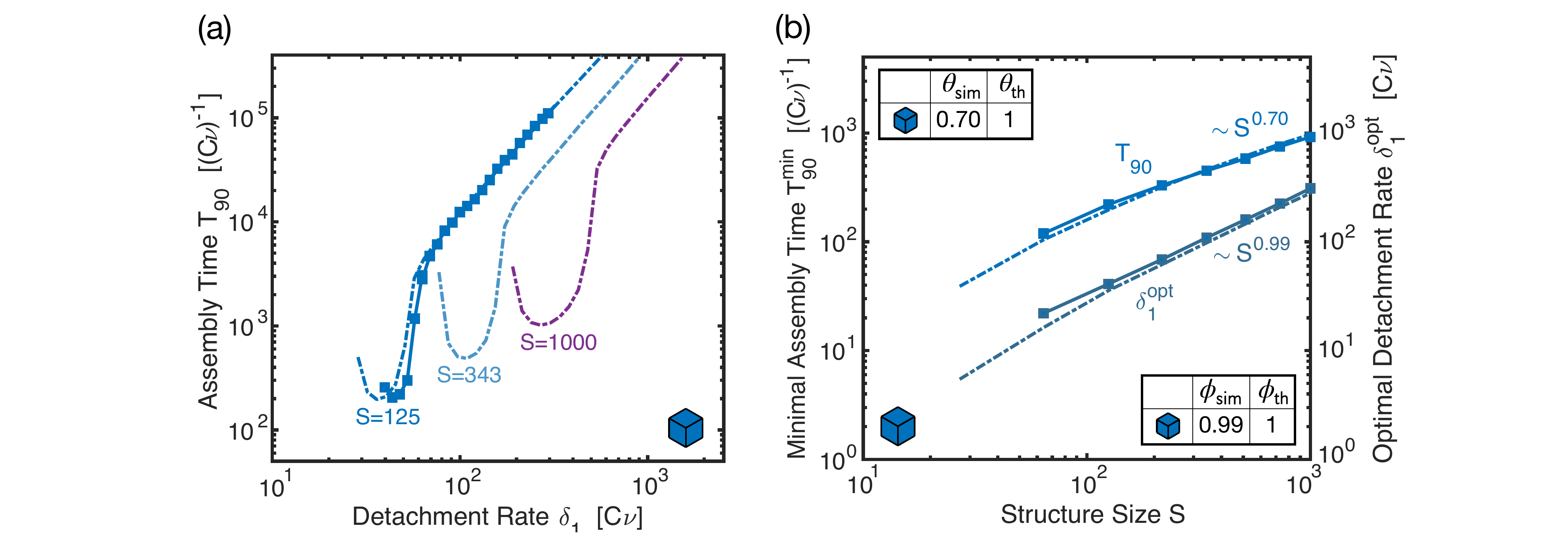}
\caption{\textbf{Assembly time and optimal detachment parameter for three-dimensional structures.} (a) The assembly time $T_{90}$ for the self-assembly of cube-shaped monomers into three-dimensional cubic structures is plotted against the detachment rate $\delta_1$. Markers show the results of stochastic simulations while dash-dotted lines show the prediction of the effective model for three different sizes of the target structure. Stochastic simulations were performed with particle numbers $N=500\, S$ and averaged over 10 independent runs. The effecitve model was integrated with parameters $\sigma=4, \, \gamma=2, \, \omega=4/3, \, a=15$ and $\mu=2\, \nu$; see App.~Sec.~\ref{app:sec:3D_structures}. (b) The minimal assembly time $T_{90}^\text{min}$ and the optimal detachment rate $\tilde \delta_1^\text{opt}$ inferred from the stochastic simulations (markers and solid lines) and effective theory (dashed-dotted lines) are plotted against the size $S$ of the target structure. The tables show their respective scaling exponents inferred from the last three data points of the stochastic simulation in comparison with their theoretical values according to Eq.~\eqref{app:eq:exponents_3D_systems}.}
\label{fig:3D_structures}
\end{figure*}
This section investigates the self-assembly of structures characterized by growth exponents $\omega>1$. In particular, three-dimensional structures can have growth exponents $\omega>1$. To see this, we consider as an example the self-assembly of cube-shaped monomers into a three-dimensional cube-shaped structure of edge length $L=S^{1/3}$. This example corresponds to the three-dimensional generalization of the square-shaped monomer morphology discussed in the main text: As in the two-dimensional case, $\sigma=4$ cube-shaped monomers form a stable nucleus. Subsequently, clusters grow by adding one face at a time in a domino effect that starts after a leading order attachment event involving $\gamma=2$ monomers has occurred. The face of a cube-shaped cluster of size $s$ comprises $l_s = s^{2/3}$ monomers and the total perimeter is thus given by $p_s = 6 \, s^{2/3}$. Therefore, Eq.~\eqref{eq:app:growth_rate} ff. implies that the growth exponent is $\omega=4/3$ in this case.   \\
For $\omega>1$, the approximation of the lower boundary of the integral in Eq.~\eqref{app:eq:monomer_conc_dimensionless}, which we used to derive the scale invariance, is no longer fully accurate, as we will elaborate in the following. Consequently, our previous results for the time complexity and control parameter exponent [Eq.~\eqref{app:eq:time_complexity_exp_general_scaling} and \eqref{app:eq:control_parameter_exp_general_scaling}] no longer accurately describe the scaling behavior for growth exponents larger than one. 
We recall that, to derive the scale invariance, we approximated the lower boundary $\sigma$ of the integral in the equation for the monomer concentration, Eq.~\eqref{app:eq:monomer_conc_dimensionless}, by a small target-size dependent value $\epsilon S$. In this way, a small fraction $\epsilon$ of all assembly steps is neglected. 
This approximation is accurate as long as $\omega\leq 1$ but becomes problematic when $\omega>1$. 
The reason is that, despite neglecting only a small fraction of the assembly steps, 
the time scale for the growth of a cluster from size $\sigma$ to $\epsilon S$, 
\[
T_{\sigma\to \epsilon S} = \int_{\sigma}^{\epsilon S} \frac{1}{m^{\gamma} \bar \nu S^{\omega}} \, ds \sim s^{1-\omega} |_{\epsilon S}^{\sigma} \approx \sigma^{1-\omega}
\]
 still exceeds the time scale for the growth from size $\epsilon S$ to the final size $S$, $T_{\epsilon S \to S} \sim (\epsilon S)^{1-\omega}$. Hence, neglecting the slow initial assembly steps significantly underestimates the overall time scale of the assembly process. \\
 A better approximation in the case where $\omega>1$ is obtained by approximating the upper boundary of the integral instead. Specifically, since large clusters grow very fast on average, we truncate the growth process at a fixed cluster size $K$ and assume that all subsequent growth steps are infinitely fast. On average, therefore, each assembly step from size $\sigma$ to $K$ consumes $(S-\sigma)/(K-\sigma)\approx S/K$ monomers. This stoichiometric factor needs to be included as a prefactor to the integral to guarantee that the total number of monomers is conserved. The equation for the monomer concentration replacing Eq.~\eqref{app:eq:monomer_conc} is thus given by 
\begin{equation}
\label{eq:app:monomer_concentration_3D_structures}
\partial_t m (t) = -a \, \bar\nu \, m(t)^{\gamma} \tfrac{S}{K} \int_{\sigma}^{K}  s^{\omega} c(s,t) \ ds \\ \, .
\end{equation}
Using the same scaling ansatz as in Sec.~\ref{app:general_scaling}, results in the linear system
\begin{subequations}
\begin{align}
    -\theta &= \phi \, (1-\gamma) + 1 + \xi \, ,  \label{eq:scale_trafo1}  \\
    \xi-\theta &= \phi \, (1-\gamma) + \xi \, , \label{eq:scale_trafo2} \\
    \phi \, (2-\sigma) &= \phi \, (1-\gamma) + \xi \, ,  \label{eq:scale_trafo3}
\end{align}
\end{subequations}
which is readily solved to give 
\begin{equation}    \label{app:eq:exponents_3D_systems}
\phi= \frac{1}{\sigma-\gamma-1} \, , \qquad \theta = \frac{\gamma-1}{\sigma-\gamma-1} \, .
\end{equation}
This suggests that if $\omega$ exceeds 1, the time complexity and control parameter exponent remain constant as functions of $\omega$. Note that, for $\omega=1$, Eqs.~\eqref{app:eq:control_parameter_exp_general_scaling} and \eqref{app:eq:time_complexity_exp_general_scaling} yield the same values for the exponents as Eq.~\eqref{app:eq:exponents_3D_systems}.
Comparing with the results from stochastic simulations [cf.~Fig.~\ref{fig:3D_structures}], we find that the control parameter exponent accurately matches the theoretical value $\phi_{\text{th}}=1$ predicted by Eq.~\eqref{app:eq:exponents_3D_systems}, while the time complexity exponent is slightly overestimated by the theoretical value $\theta_{\text{th}}=1$. The estimates are, however, significantly better than those obtained with Eqs.~\eqref{app:eq:control_parameter_exp_general_scaling} and \eqref{app:eq:time_complexity_exp_general_scaling}, which would yield $\phi_{\text{th}}=2/3$ and $\theta_{\text{th}}=1/3$.    
The overestimation of the time complexity exponent results from the fact that we average the rate of monomer consumption over the entire growth process of a cluster (accounting for the constant prefactor $S/K$ in Eq.~\eqref{eq:app:monomer_concentration_3D_structures}). This approximation particularly affects the kinetics in the regime of the U-shape, where clusters grow simultaneously and, thus, the precise temporal dynamics of the monomer concentration are important. 
In contrast, in the regime characterized by the power-law scaling, clusters tend to nucleate and grow consecutively, and the approximation is therefore quite accurate. Indeed, we find that the assembly time simulated with the effective model at a constant, large value of $\tilde \delta_1$ (e.g. $\tilde \delta_1=10^4$) scales with large target structure sizes as $\sim S^{-1.00}$ -- fully in accordance with the theoretical exponents $\phi_{\text{th}}=\theta_{\text{th}}=1$ and the power-law scaling $T_{90}\sim S^{\sigma-2}$.
We note from Fig.~\ref{fig:3D_structures}(a) that the form of the U-shape broadens when $S$ increases. Therefore, there can only be an approximate scaling relation but no exact one for the entire $T_{90}$-curve as in the case where $\omega \leq 1$. 
 
Three-dimensional structures with higher coordination numbers that self-assemble with attachment order $\gamma=1$ most likely have a growth exponent $\omega\approx 2/3$ (since there cannot be a domino effect if $\gamma=1$). In this case, Eq.~\eqref{app:eq:time_complexity_exp_general_scaling} suggests a time complexity exponent of $\theta_{\text{th}}=1/3$. This value is smaller than the value obtained for $\gamma\geq 2$ according to Eq.~\eqref{app:eq:exponents_3D_systems} for any realistic nucleation size $\sigma$. Hence, also in the case of three-dimensional structures, self-assembly is fastest if the attachment order is small. Structures with higher coordination numbers, therefore, tend to assemble faster. This is consistent with the results in reference \cite{reinhardt2016effects}, showing that high coordination structures typically have steeper free energy gradients. 

 \textit{Effective model for 3D structures.} The prefactors for the effective model used for the plots in Fig.~\ref{fig:3D_structures} are derived as follows. Once the first monomer has attached to the face of a three-dimensional structure, the second monomer forming the seed can attach in $q_s=4$ equivalent ways ($q_s$ binding sites) in the case of a large cluster and $q_s=2$ ways in the case of a small cluster. Hence, with the total perimeter surface given by $p_s=6\, s^{2/3}$, Eq.~\eqref{eq:app:growth_rate} predicts a value for the multiplicity factor $a$ between 12 and 24. Indeed, a good fit with the stochastic simulation is obtained with $a=15$ in Fig.~\ref{fig:3D_structures}. Furthermore, we note that in the nucleation process, the multiplicity for a dimer to grow into a trimer is 8, while the detachment rate of both the trimer and the dimer is $2\, \delta_1$. Hence, we obtain an overall prefactor of the effective nucleation rate of 2 (for 2D structures, the prefactors for the nucleation processes canceled to 1 for all three monomer morphologies, cf.~Fig.~\ref{fig_supp:illustration_nucleation}). We account for the additional prefactor in the nucleation process by setting $\mu=2\, \nu$ [cf.~Eq.~\eqref{eq:nucleation_rate}].
 To further improve the fit, we explicitly set the growth rates for the first four post-nucleation assembly steps to $g_s = 64 \, \bar \nu \, m^{\gamma}$. This improves the fit since the nucleus does not yet have a cube shape, as assumed for the growth of larger clusters.

\section{The role of cluster-cluster interactions}
\label{app:role_of_cluster_cluster_interactions}

In the analysis in the main text, we have worked under the simplifying assumption of an \textit{ideal} assembly process, meaning that interactions among clusters are negligible. 
This assumption is frequently used in the literature on self-assembly. It is motivated by the observation from experiments and molecular dynamics simulations that intermediate assembly products are typically only present in very low concentrations \cite{prevelige1993nucleation, Rapaport2008, Hagan2014}.
Here, we would like to investigate and quantify to which extent cluster-cluster interactions can influence the time efficiency of self-assembly and whether the morphology of the building blocks is relevant in this regard. 
To this end, we first introduce a simple measure that informs whether cluster interactions can \textit{potentially} influence the assembly process. 
Afterward, we extend our effective model to include cluster interactions and estimate the effect that those might have on the assembly time. 

To estimate the extent to which interactions among oligomers can potentially influence the dynamics, we define the \textit{assembly ideality index} that is calculated under the assumption of an ideal assembly process,
\begin{equation}   
\label{definition_ideality_index}
    I 
    := 
    \frac{S}{C} \, 
    \int\limits_0^{T_{90}} 
     \nu K^2 (t) \,  \text dt
    \, ,
\end{equation} 
where, as above, ${K(t):=\sum\nolimits_{s=\sigma}^{S-1} c_{s}(t)}$ denotes the total concentration of incomplete complexes in the system at time $t$, and we integrate over the time window $[0, T_{90}]$ to estimate the total number of cluster-cluster interactions until time $T_{90}$. 
The index, therefore, estimates the expected number of cluster interactions per completed structure until time $T_{90}$ (note that the concentration of completed structures at 90\% yield is ${0.9 \, C/S\approx C/S}$), assuming the same reaction rate $\nu$ between two clusters as between a cluster and a monomer. 
Small values ${I \ll 1}$ indicate that cluster interactions occur only with a small probability $I$ during the growth of a structure and can thus be neglected. 
Hence, the index provides a simple consistency check for the ideality assumption of a simulation. 
Note that assuming the same reaction rate $\nu$ between clusters as between clusters and monomers most likely overestimates the reaction rate between clusters since larger clusters typically diffuse and therefore react more slowly than monomers, and only a fraction of their interactions would lead to stable configurations. 
Hence, the ideality index must be interpreted as an upper limit for the expected number of cluster interactions. 
Small values ${I \ll 1}$ are considered a sufficient condition for the ideality assumption.
 In contrast, large values ${I > 1}$ indicate that cluster interactions might \textit{potentially} influence the dynamics, depending on the specifics of the system.  
\begin{figure}[!t]
\centering
\includegraphics[width=0.9\linewidth]{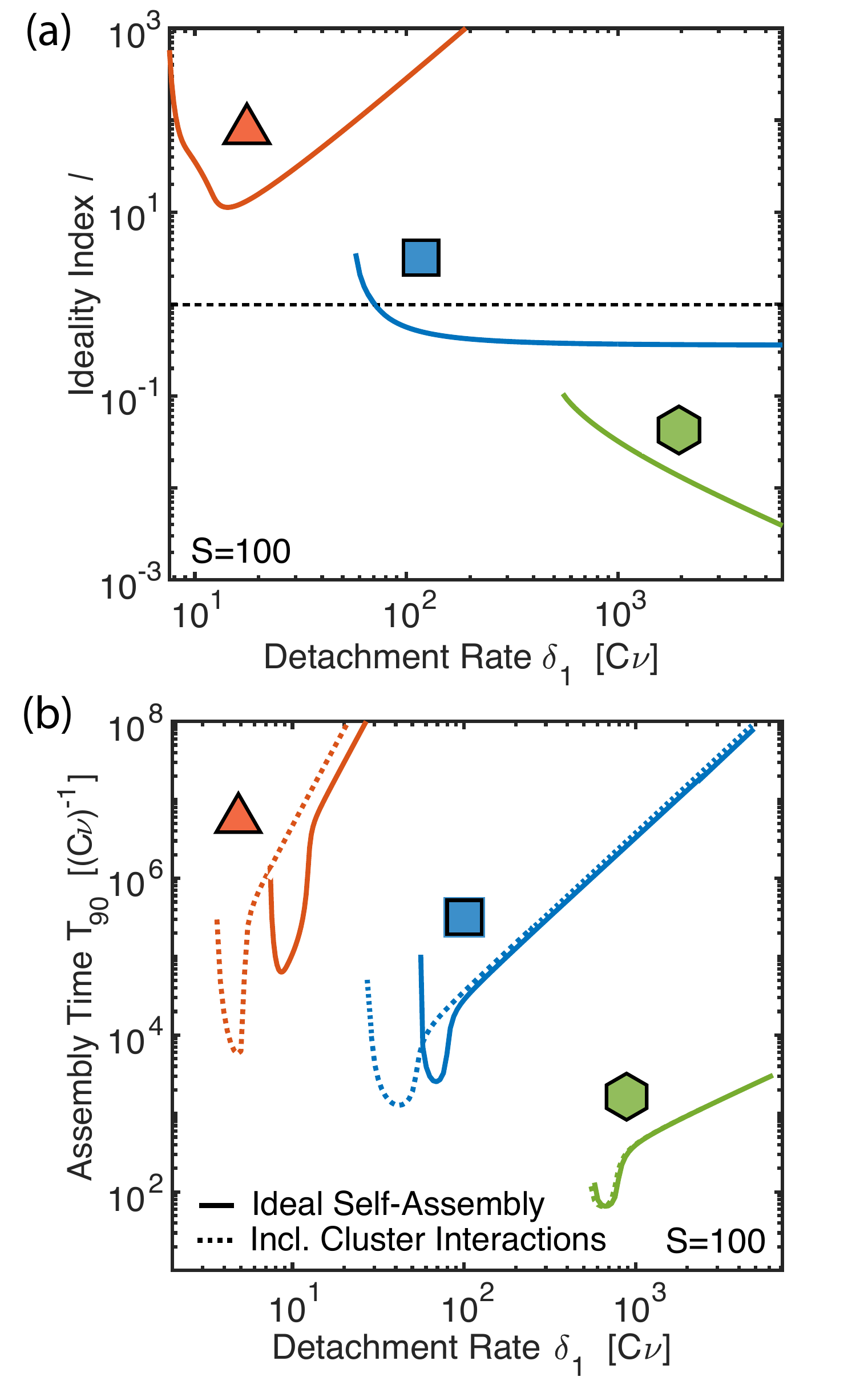}
\caption{\textbf{Beyond ideal self-assembly: the role of cluster-cluster interactions.} 
(a) The ideality index $I$, which estimates the number of interactions of an oligomer with other oligomers until time $T_{90}$ [see definition Eq.~(\ref{definition_ideality_index})], is plotted against the detachment rate $\tilde \delta_1$ for the assembly processes with three different particle morphologies (${S = 100}$). 
The index has been calculated numerically by integrating the effective model, Eq.~\eqref{effective_theory_1st_order}. Values of $ I $ larger than 1 (dashed line) indicate that cluster interactions may be important for the assembly dynamics. 
Thus, a strong dependence of the potential role of cluster interactions on particle morphology is revealed.
(b) Assembly time $T_{90}$ with (drawn line) and without (dotted line) cluster interactions as a function of the detachment rate $\tilde \delta_1$. 
Assembly times were obtained by integrating the effective theory that was extended by additional terms accounting for all possible cluster reactions ${c_s + c_p \overset{\nu }{\to} c_{s+p}}$ subject to the constraint ${s+p \leq S}$ (see App.~\ref{app:effective_model_irr_limit} for details). Since, in an actual system, the rate of reactions between clusters would likely be smaller, this must be considered an upper estimate of the effect expected by cluster interactions. The minimal assembly time is only slightly reduced as a consequence of cluster interactions and -- as in the ideal case -- the assembly time increases like ${T_{90} \sim  \tilde \delta_1^{\sigma-2}}$ for large $\tilde \delta_1$.}
\label{fig:non_ideal_assembly}
\end{figure}

Figure~\ref{fig:non_ideal_assembly}(a) shows the numerically determined ideality index as a function of the detachment rate $\delta_1$ for the assembly processes with the three different particle morphologies.
The index reveals a strong dependence of the potential role of cluster-cluster interactions on the morphology of the constituents: 
While cluster-cluster interactions can be safely neglected for hexagonal-shaped monomers (${I \ll 1}$), the magnitude of the index observed for the triangle-shaped monomers suggests that these interactions play a significant role in the assembly process. 
Furthermore, the index behaves differently as a function of the detachment rate $\tilde \delta_1$: 
While in the system with hexagonal monomers, cluster-cluster interactions become less critical when $\tilde \delta_1$ is increased, their potential influence further increases in the system with triangle-shaped monomers as $\tilde \delta_1$ grows. 
Hence, the ideality index seemingly correlates with the time efficiency of a self-assembly system: Particles with a morphology that assemble time efficiently tend to have a smaller ideality index in general than particles with a morphology that assemble less efficiently. 
  
In order to estimate the extent to which cluster interactions affect the assembly time, we extend the effective model by additional terms accounting for reactions of any two clusters of sizes ${i,j \geq \sigma}$ with $i+j \leq S$ to a cluster of size ${s=i+j}$ at rate $\nu$:
\begin{equation}  \label{eff_theory_cluster_interactions}
    \partial_t c_s = ... \, 
    + 
    \tfrac{1}{2} \, \nu \sum\limits_{\substack{i,j\geq \sigma \\ i+j=s}} c_{i} c_{j} 
    - 
    \nu \, c_{s} \sum\limits_{\substack{i\geq \sigma \\i+s \leq S}} c_{i}   \, ,
\end{equation}
for all ${\sigma\leq s \leq S}$. 
The factor of $\frac{1}{2}$ in front of the first sum avoids double counting and serves as a stoichiometric factor in the case ${i=j}$.
As before, assuming that clusters react at the same rate $\nu$ as monomers will strongly overestimate the effect that cluster interactions have on the assembly time, and thus, the model describes an upper limit for the impact that is to be expected by these additional interactions.

Figure~\ref{fig:non_ideal_assembly}(b) compares the assembly time of the ideal system with that obtained from integrating the extended model accounting for cluster interactions. 
The comparison suggests that cluster-cluster interactions can indeed reduce the assembly time in systems with triangles---and to a much lesser extent also in systems with square-shaped monomers.
However, the reduction of the assembly time is relatively small in general.  
The reason is that, as the size of a cluster increases, the number of its possible binding partners decreases: according to Eq.~\eqref{eff_theory_cluster_interactions}, a cluster of size $i$ can only react with clusters of size ${j \leq S-i}$. 
Therefore, cluster interactions are not effective in avoiding kinetic traps. 

If we also take into account that the model probably overestimates the reaction rate between clusters, the reduction in assembly time due to cluster-cluster interactions will presumably be even smaller in any actual system that uses the respective types of particles. 
On the other hand, cluster-cluster interactions might enhance the occurrence of assembly errors and defects. 
For example, occasionally, clusters might react that do not fit together perfectly, thus leaving defects and holes or distorting the structure. 
Operating at a small ideality index might thus be preferable for general self-assembly systems as it might enhance the robustness of the process against errors. 
In conclusion, the morphology of the constituents could be an essential determinant not only of the time efficiency but also of the reliability of the assembly process.

\section{Invariance of the scaling results to the yield threshold}
\label{app:invariance_to_yield_threshold}

\begin{figure*}[htb]
\centering
\includegraphics[width=0.97\linewidth]{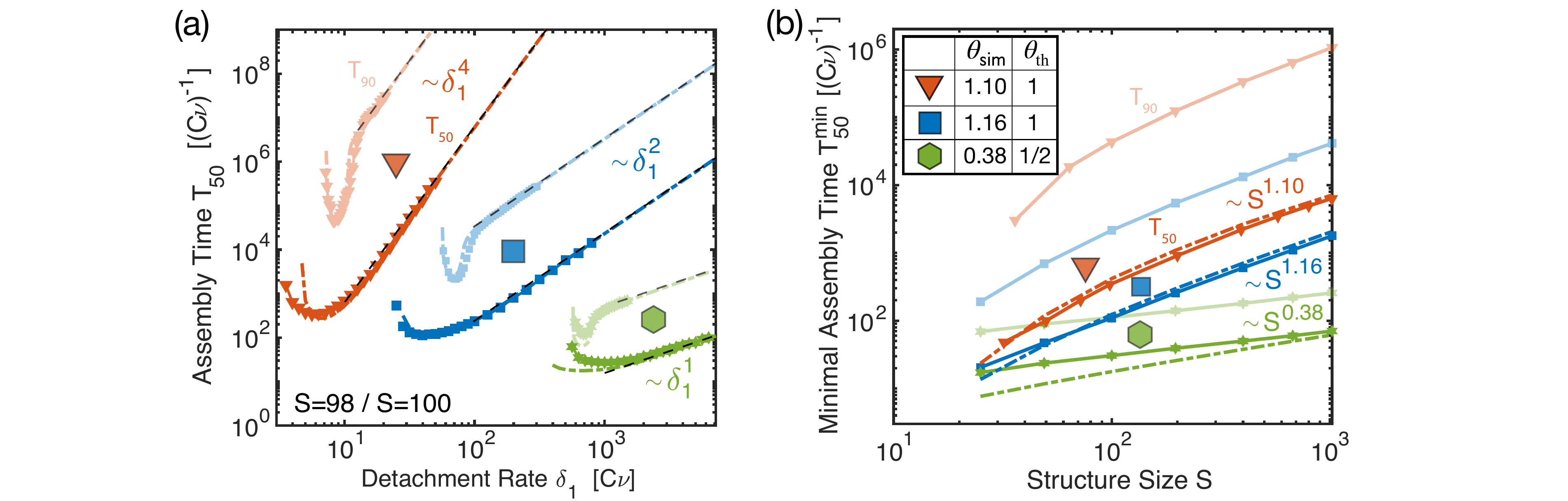}
\caption{
\textbf{Assembly time $T_{50}$ (solid) in comparison with assembly time $T_{90}$ (faded).} 
(a) The assembly times $T_{50}$ (solid) and $T_{90}$(faded) required to achieve 50\% or 90\% yield, respectively, in the strong binding limit are plotted against the detachment rate $\delta_1$ (in units of $C\nu$) for
systems with triangle- (red), square- (blue), and hexagon-shaped (green) building blocks for a fixed target structure size $S = 98$
(triangle-shaped monomers) or $S = 100$ (square and hexagonal monomers). Markers represent averages of 10 independent runs of the stochastic simulation (performed with $N=500\, S$ and $N=300\, S$ (triangle-shaped monomers)). The dash-dotted colored lines represent the prediction of the effective theory obtained by numerically integrating Eq.~\eqref{effective_theory_1st_order}. 
Black dashed lines show the analytic result for the assembly times given by Eqs.~\eqref{Ty_slow_nucleation} and \eqref{qy_slow_nucleation} in the slow nucleation regime (ranges for $\tilde \delta_1$ are chosen manually).
(b) The minimal assembly time $T_{50}^{\text{min}}$ (solid) and $T_{90}^{\text{min}}$ (faded) inferred from the stochastic simulation (markers and solid lines) and effective theory (dashed-dotted lines) are plotted against the size S of the target structure. Although the assembly times $T_{90}$ and $T_{50}$ may differ significantly, they exhibit approximately the same scaling in dependence on the structure size $S$. The scaling exponents $\theta_{\text{sim}}$ inferred from the last three data points of the stochastic simulation for $T_{50}$ (and their theoretical asymptotic values $\theta_{\text{th}}$, cf. Eq.~\eqref{eq:time_complexity_exponent}) are summarized in the table (compare with the analogous exponents $\theta_{\text{sim}}$ for $T_{90}$ exhibited in the table in Fig.~\ref{fig:scaling_laws}(b)).}
\label{fig:T50_scaling}
\end{figure*}

In the main text and the appendix so far, we have studied the behavior of the assembly time $T_{90}$ required to achieve 90\% yield. 
Alternatively, one could also choose a different threshold for the yield, e.g., 50\%, and analyze the behavior of the corresponding assembly time $T_{50}$. 
For our scaling results to be informative and valuable, it is important that these results are invariant to the yield threshold and that the qualitative behavior of the assembly time $T_Y$ is the same, independent of the yield threshold $Y$. 
We will verify this in the following by repeating the same plots shown in the main text for $T_{90}$ now with $T_{50}$. 

Figure~\ref{fig:T50_scaling}(a) shows the assembly time $T_{50}$ as a function of the detachment rate $\delta_1$ in comparison with $T_{90}$ for the different particle morphologies in the strong binding limit. 
We find that $T_{50}$ and $T_{90}$ behave qualitatively similar and, in particular, both exhibit the same scaling $\sim \tilde \delta_1^{\sigma-2}$ for large enough $\delta_1$. 
However, the values of the assembly times $T_{50}$ and $T_{90}$ can differ quite dramatically depending on the particle morphology. 
In particular, differences in the minimal assembly times between the different particle morphologies tend to become less pronounced for a lower yield threshold. 
This is consistent with our analysis of the regime in which nucleation is very slow [cf. section~\ref{sec:slow_nucleation_limit}]. 
In particular, Eq.~\eqref{Ty_slow_nucleation} and \eqref{qy_slow_nucleation} (visualized as black dashed lines in Fig.~\ref{fig:T50_scaling}(a)) predict that the assembly time in this regime depends on the yield threshold $Y$ as $T_{Y} \sim (1-Y)^{1-\sigma}$. 
Thus, the assembly time in this regime depends more strongly on the demanded resource efficiency if the nucleation size $\sigma$ is larger.  
 However, the scaling of the minimal assembly time with the target structure size (i.e., the time complexity exponent $\theta$) remains independent of the yield threshold as we verify in Fig.~\ref{fig:T50_scaling}(b): Comparing the measured time-complexity exponents for $T_{50}$ with those measured for $T_{90}$ (see table in Fig.~\ref{fig:scaling_laws}(b)) shows that they are almost identical. 
Thus, despite the assembly time as such might depend strongly on the yield threshold, all scaling results are invariant to the yield threshold. Differences in the assembly times between the different particle morphologies studied in this work tend to become less pronounced if the requirement for resource efficiency is relaxed.

\section{Seeding}
\label{app:sec:seeding}

\begin{figure*}[!t]
\centering
\includegraphics[width=1.0\linewidth]{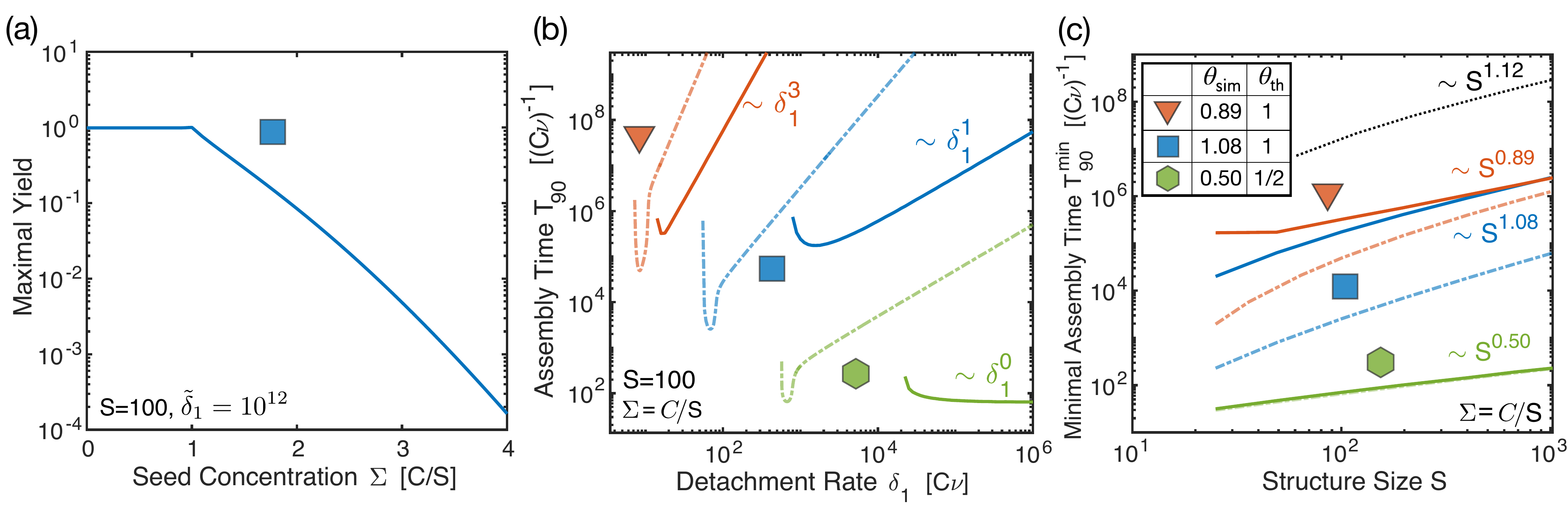}
\caption{\textbf{Yield and assembly time in seeded systems} (according to the effective model). (a) The maximal yield of target structures of size $S=100$ in the effective model in the limit of large $\tilde \delta_1$ (here $\tilde \delta_1=10^{12}$) is plotted against the initial concentration $\Sigma$ of nuclei (seeds). Parameters of the effective model were chosen to simulate the self-assembly of square-shaped monomers ($\sigma=4$, $\gamma=2$, $\omega=1$ and $a=5.3$), and the initial monomer concentration was $m(0)=C-\sigma\Sigma$. (b) Assembly time $T_{90}$ as a function of $\tilde \delta_1$ for seeded systems with initial seed concentration $\Sigma=C/S$ (drawn lines) and corresponding unseeded systems (dash-dotted lines, faded). (c) Minimal assembly time $T_{90}^{\text{min}}$ as a function of $S$ for seeded systems with initial seed concentration $\Sigma=C/S$ (drawn lines) and corresponding unseeded systems (dash-dotted lines, faded). Black dashed line shows the assembly time for a putative system of triangular monomers in which the first assembly step after nucleation has attachment order $\gamma'=3$ instead of $\gamma'=4$. Table shows the time complexity exponents $theta_{\text{sim}}$ measured in the range of target structure sizes $400\leq S\leq 1000$ and their corresponding theoretical values $theta_{\text{th}}$.}
\label{app:fig:seeding}
\end{figure*}

Seeding refers to the experimental method of initially putting a certain number of stable nuclei (seeds) into the system. In systems that enable the formation of different structures, like, e.g., algorithmic self-assembly, seeding is an integral part since the composition of the seed determines the structure or the pattern that grows (i.e., the algorithm that is performed via algorithmic self-assembly) \cite{Evans2017}. From a kinetic point of view, we expect that seeding should reduce the assembly time by eliminating the requirement of the rate-limiting nucleation reactions as the seeds can start growing right from the beginning. First, in order to determine the optimal number (concentration) of seeds, we integrate the effective model with an initial concentration $\Sigma$ of stable nuclei ($c_{\sigma}(0)=\Sigma$) and corresponding monomer concentration $m(0)=C-\sigma \Sigma$ (to guarantee a fixed total number of resources). We find that the maximum yield (obtained in the limit of large $\tilde \delta_1$) is $1$ if the seed concentration is smaller or equal $C/S$ and decreases rapidly if this concentration is exceeded [Fig. \ref{app:fig:seeding}(a)]. This is intuitive since an initial concentration $\Sigma = C/S$ of nuclei exactly equals the concentration of completed structures when the yield is 100\%.

Next, we study the effect of seeding on the assembly time by integrating the effective model with initial concentration $\Sigma = C/S$ of nuclei (corresponding to the number $N/S$ of seeds required for 100\% yield) and corresponding monomer concentration $m(0)=C-\sigma \Sigma$ [cf.~Fig.~\ref{app:fig:seeding}(a),(c)]. 
Surprisingly, contrary to our expectation, we find that the assembly time increases due to seeding if $\gamma>1$ (square- and triangular monomers). This is because if nuclei are already present in the system, the control parameter $\tilde \delta_1$ must be chosen even larger to suppress formation of additional nuclei, which would lead to an excess of clusters. Increasing $\tilde \delta_1$ simultaneously decreases the growth rate if $\gamma>1$, making the overall process less time efficient. In contrast, for hexagon-shaped monomers ($\gamma=1$), the growth rate is not affected by $\tilde \delta_1$ and the minimal assembly time remains the same. With structure growth (rather than nucleation) being rate limiting in seeded systems, increasing $\tilde \delta_1$ beyond the optimal value makes the assembly time increase only proportional to $\tilde \delta_1^{\gamma-1}$ (rather than $\sim \tilde \delta_1^{\sigma-2}$ in the case of unseeded systems [compare Fig.~\ref{fig:scaling_laws}a]). In particular, in the case of hexagonal building blocks ($\gamma=1$), the assembly time is thus constant as a function of $\tilde \delta_1\geq \tilde \delta_1^{\text{opt}}$.  

In the framework of our scaling analysis, the initial condition for seeded systems, $c_{\sigma}(t\!=\!0)\sim S^{-1}$, is consistent with the scale invariance described by the scaling forms for the total concentration of clusters, $K(t,S) \sim S^{-1} \widehat K(S^{-\theta} \tilde t)$ [cf.~App.~Sec.~\ref{app:onset_scaling}], and for the cluster size density, $c(s,t,S)\sim S^{-2}\widehat c (S^{-1}s, S^{-\theta} \tilde t)$ [cf.~Eq.~\ref{eq:scaling_form_c}] (note that by rescaling the cluster size $s\to S^{-1}s$, we get an additional factor of $S^{-1}$ for the cluster size density). Furthermore, the initial condition $m(0)=C-C/S$ is approximately consistent with the scaling form for the monomer concentration, $m(t,S)\sim \widehat m(S^{-\theta} \tilde t)$ [cf.~Eq.~\ref{eq:scaling_form_m}] in the limit of large $S$ where $C/S$ is negligible and $m(0)\approx C$. Therefore, the scaling analysis in App.~Sec.~\ref{app:general_scaling} can be performed identically for seeded systems with the modified initial conditions [cf.~Eqs.~\eqref{app:eq:general_scaling_initial_m} and \eqref{app:eq:general_scaling_initial_c}]. Consequently, the scaling behavior should be the same for seeded systems with initial concentration of seeds $\Sigma \sim S^{-1}$. This is confirmed by Fig. \ref{app:fig:seeding}(c) for square and hexagonal monomers. In the case of triangular monomers, we observe a slightly smaller time complexity exponent compared to the unseeded system (but still close to its theoretical value). This anomalous behavior is likely caused by the higher effective order $\gamma'=4$ of the first assembly step following nucleation [cf. App.~Sec.~\ref{app:effective_model_irr_limit}] (which we did not explicitly account for in the scaling analysis). To see this, we simulated a putative system of triangular monomers in which the attachment order of the first step likewise equals $\gamma=3$ (black dashed line). The assembly time in this case is significantly larger than for the true system and the scaling exponent is very close to that of the unseeded system, confirming that the post-nucleation assembly step is responsible for the drop in the exponent.  \\
Hence, in general, this shows that our scaling results are very robust and apply likewise to seeded systems. In line with our general conclusions, using hexagonal building blocks (or generally monomers with $\gamma=1$) appears to be highly favorable as the minimal assembly time does not increase as a consequence of seeding and the assembly time even becomes a constant function of $\tilde \delta_1$.

\section{Annealing protocols} 
\label{app:sec:annealing}
The assembly dynamics are controlled by the ratio between the frequency of detachment events $\delta_1$ and the frequency of attachment events $\nu m$. Initially, the monomer concentration is $m(0)=C$, but as more and more monomers get consumed during the assembly process, their concentration gradually decreases. Consequently, the ratio between the detachment and attachment rate increases during the assembly process. In order to counteract this effect, a frequently used experimental approach consists in `annealing' the system by decreasing the temperature \cite{Evans2017}. Typically, one starts at a high temperature and gradually cools the system to room temperature. Since the detachment rate decreases with decreasing temperature, $\delta_1 \sim e^{-E_B/(k_B T)}$, if applied optimally, annealing allows to keep the ratio between the detachment rate and attachment rate constant during the assembly process. Here, we show that our scaling analysis also applies to the case of an annealing protocol in which the temperature adapts instantaneously to the momentary concentration of monomers such that the ratio between detachment and attachment rate remains constant. To describe such an annealing protocol with the effective theory presented in Sec.~\ref{sec:general_scaling_theory} and App.~Sec.~\ref{app:general_scaling}, we only need to replace the control parameter $\tilde \delta_1 = \delta_1/(C\nu)$ with a time-dependent control parameter $\tilde \delta_1(t) = \tilde \delta_1(t=0)\frac{m(t)}{C}$ that gradually decreases proportional to the monomer concentration. This replacement only introduces additional factors of $m$ in the dynamic equations, but the monomer concentration does not scale with $S$ since $m(t) = \widehat m(S^{-\theta} \tilde t)$ [cf.~Eq.~\eqref{eq:scaling_form_m}]. Hence, the scaling analysis can be performed in exactly the same way. This implies that all our scaling results apply identically under annealing. Note that we could also use any other annealing protocol of the form $\tilde \delta_1(t) = \tilde \delta_1(0) f(S^{-\theta} \tilde t)$ to maintain the scale invariance. 

Figure~\ref{fig:robustness_of_scaling} shows the average minimal assembly time simulated stochastically with the above annealing protocol plotted against the target structure size (for square-shaped monomers). The minimal assembly time was determined by tuning the parameter $\tilde \delta_1(t=0)$. We find that the self-assembly efficiency is indeed significantly increased as a result of annealing. However, the plot confirms that the scaling of the assembly time remains invariant under annealing.

\section{Systems with unlimited cluster growth}
\label{app:sec:unlimited_growth}

\begin{figure*}[tb]
\centering
\includegraphics[width=1.0\linewidth]{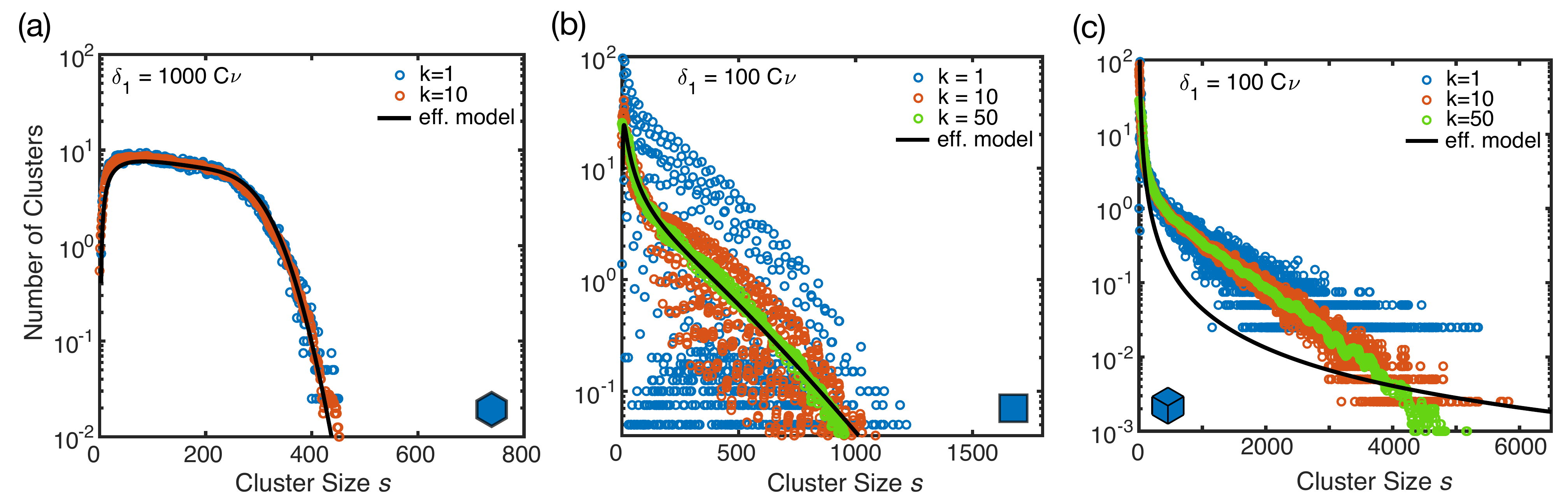}
\caption{\textbf{Final cluster size distribution.} The final cluster size distributions for a) hexagon-shaped ($\omega \approx 1/2>1$), b) square-shaped ($\omega=1$), and c) 3D cube-shaped ($\omega=4/3$) monomers are shown for the respective values of the detachment rate $\delta_1$ indicated in each plot. The target structure size in the simulation was chosen very large ($S=3025$ (a,b) and $S=8000$ (c)) so that the yield was zero in each case, thereby effectively simulating unlimited cluster growth. The initial number of monomers was $N=3 \cdot 10^5$, and the distribution was averaged over 40 independent simulation runs (blue circles). Subsequently, moving averages were taken over windows of $k=10$ (red circles) and $k=50$ (green circles) cluster sizes. The averaged cluster size distribution decreases exponentially for large cluster sizes for square- and cube-shaped monomers (i.e., for $\omega\geq 1$). Black lines show the final cluster size distribution obtained by integrating the effective model. }
\label{fig:cluster_size_distribution_unlimited_growth}
\end{figure*}
So far, we have considered systems with self-limiting cluster growth by assuming periodic boundary conditions of the target structures. However, our scaling analysis can straightforwardly be generalized to systems with unlimited growth of the structures. The only requirement is that the growth rate of large clusters consistently scales with the cluster size as $\sim s^{\omega}$, with some arbitrary growth exponent $\omega$. 

To describe unlimited cluster growth with the effective theory presented in Sec.~\ref{sec:general_scaling_theory} and App.~Sec.~\ref{app:general_scaling}, we only have to drop the absorbing boundary at $s=S$, which removed finished structures from the system, and replace the upper bound in the integral in the equation for the monomer concentration, Eq.~\eqref{eq:advection_eq_monomers}, resp. Eq.~\eqref{app:eq:monomer_conc_dimensionless}, by infinity, to integrate clusters of arbitrary size. Clearly, these changes do not affect the scaling analysis.
The natural generalization of the yield $Y_S$ represents the fraction of resources that are bound in clusters of size larger or equal to $S$. The yield can thereby be written in the same form as Eq.~\eqref{scale_less_yield_condition} by identifying $1-Y_S$ with the number of resources that are either available as monomers or bound in clusters of size smaller than $S$. In the form of Eq.~\eqref{scale_less_yield_condition}, the scale invariance of the yield [cf.~Eq.~\eqref{eq:yield_scaling_form}] and of the corresponding assembly time $T_{90}$ (likewise defined as the time required to achieve 90\% yield) [cf.~Eq.~\eqref{eq:T90_scaling_form}], becomes evident. Hence, with the yield and assembly time defined in this way, all scaling results derived for the case of self-limiting cluster growth can directly be transferred to the case of unlimited cluster growth. 

In principle, we could also define the yield and assembly time in a different way as long as it is consistently scale invariant. For example, the assembly time could alternatively be defined as the time when the average size of a cluster exceeds $S$.
The scale invariance of this condition becomes evident by using the scaling form for the complex concentration, Eq.~\eqref{eq:scaling_form_c} resp. Eq.~\eqref{app:eq:scaling_form_c} 
\begin{equation}
S < \frac{\int_{\sigma}^{\infty} s c(s,t) \, ds}{\int_{\sigma}^{\infty} c(s,t) \, ds} = \frac{\int_{\epsilon}^{\infty} S \, x \, S^{-2} \, \widehat c(x,S^{-\theta}\tilde t) \, S \, dx}{\int_{\epsilon}^{\infty} S^{-2} \, \widehat c(x,S^{-\theta}\tilde t) \, S \, dx} \, ,
\end{equation}
since all factors of $S$ on the right and on the left-hand side cancel. Note that it is not possible to include the monomers as clusters of size 1 in the definition of the mean cluster size since the contribution of the monomers and the complexes in the denominator, $\widehat m + S^{-1} \int_{\epsilon}^{\infty}\widehat c(x)\, dx$, would scale differently with $S$ and would thus violate the scale invariance. However, one can additionally demand that the monomer concentration has dropped below a certain value $X$, which gives the additional scale-free condition $\widehat m(S^{-\theta}\tilde t) \leq X$ for the assembly time. 
This shows that our scaling results are very general and likewise apply to systems with unlimited cluster growth. \\

\textit{Cluster size distribution.} 
It is interesting to characterize the final cluster size distribution in systems with unlimited cluster growth. 
To this end, we choose a target structure size in the simulation that is very large and ensures that the final yield is zero. 
Therefore, the size limit does not affect the system's evolution, effectively mimicking a system with unlimited cluster growth. Figure~\ref{fig:cluster_size_distribution_unlimited_growth} shows the final cluster size distributions for hexagon-, square-, and 3D cube-shaped monomers, which are characterized by different values of the growth exponent (${\omega<1, \omega=1}$ and ${\omega>1}$, respectively). 
Since the growth rates of leading and subleading attachment reactions differ strongly, the distributions are scattered over broad ranges (in number of clusters, respectively cluster sizes) for square- and cube-shaped monomers (also compare with Fig.~\ref{fig:cluster_size_distribution}). 
The distributions can be narrowed down by calculating moving averages in the number of clusters over increasingly larger windows of $k$ subsequent cluster sizes (Fig.~\ref{fig:cluster_size_distribution_unlimited_growth} shows the distributions calculated for $k=1$, $k=10$ and $k=50$). 

We find that these averaged distributions decrease exponentially with large cluster sizes for square- and cube-shaped monomers. 
In contrast, for hexagonal monomers, due to the smaller growth exponent, the distribution is more compact and has a significantly shorter tail. 
In the cases of hexagon- and square-shaped monomers, the averaged final distributions coincide well with the distribution obtained by integrating the effective model, Eq.~\eqref{effective_theory_1st_order} (black curves). However, in the case of cube-shaped monomers, the effective model predicts a distribution that decreases as a power law rather than exponentially, thus implying a significantly longer tail than the stochastically simulated distribution. 
The reason for this discrepancy is that if single clusters quickly grow very large in the three-dimensional system, the time scale of the subleading processes (`domino effect', which scales as ${\sim s^{2/3}}$) can no longer be neglected against that of the leading order processes [cf.~Eq.~\eqref{eq:app:neglect_domino_effect_time_scale}]. 
Therefore, the effective growth rate of large clusters is actually smaller than the growth rate predicted by Eq.~\eqref{eq:app:growth_rate}. To accurately describe the distribution for three-dimensional structures, the time scale of the domino effect would have to be considered explicitly in Eq.~\eqref{eq:app:neglect_domino_effect_time_scale}. On the other hand, self-limitation of cluster growth, as assumed in our original model, prevents single clusters from growing extremely large. Thus, the effective model accurately describes the assembly time for limited cluster growth, as seen in Fig.~\ref{fig:3D_structures}.

\section{Heterogeneous systems}
\label{app:sec:heterogeneous_systems}

\begin{figure}[tb]
\centering
\includegraphics[width=0.38\linewidth]{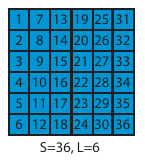}
\caption{
\textbf{Heterogeneous systems.} Simulating the self-assembly of heterogeneous structures (information-rich structures), we assume that each species (labeled by an index $1...S$) is unique and binds only with the respective neighboring species, as shown. Thus, each species appears only once in a finished structure and at a specified position. The underlying kinetics for self-assembly are the same as in the homogeneous case, but the different species can have distinct binding and detachment rates or varying concentrations. In App.~Sec.~\ref{app:sec:heterogeneous_systems}, we investigate how self-assembly behaves for various heterogeneous systems. }
\label{fig:heterogeneous_sketch}
\end{figure}
 Next, we study how our results generalize to heterogeneous systems with different species of monomers, each species binding specifically only with certain other species. Specifically, we consider a fully heterogeneous system of square-shaped monomers such that each species (labeled $1...S$) can bind only with four other species ($i+1$, $i-1$, $i-L$, $i+L$), and each site within a cluster can only be occupied by a specified species; see Fig.~\ref{fig:heterogeneous_sketch}. We investigate five different relevant cases: i) all species have the same reaction rates $\nu$ and $\delta_1$ and are available in the same copy numbers $N$. ii) The reaction rates $\nu_i$ for the different species are chosen randomly. iii) The detachment rates of four specified species that are able to form a nucleus are set to zero. iv) Four specified species that are able to form a nucleus are provided in much larger concentrations. v) Investigating the role of the boundary conditions, we assume that species at the boundary limit the growth of the clusters such that structures with explicit boundaries self-assemble.   

 \subsection{Identical reaction rates}
 \label{app:sec:heterogen_identical_rates}
 
 First, we consider the case in which all species have the same attachment and detachment rates and are available in identical concentrations. In this ideal case, one can show mathematically that the system behaves equivalent to a homogeneous system in the limit of large particle numbers. To establish this equivalence, the key insight is that, assuming periodic boundary conditions of the structures, the concentrations of all species are identical since all species are subject to the same dynamics. We show this rigorously for the evolution of the monomer concentration. To this end, we first have to introduce some notation: We denote the species $i$ of a complex as the species with the lowest label of the constituent monomers of the complex. A complex in the heterogeneous system is then uniquely characterized by its species index and its morphology, which specifies the size of a cluster and its shape. We denote the morphology by another index $\zeta$ that takes values in some countable set of numbers characterizing the morphology via some mapping, which we do not further specify. By $c_{j,\zeta}(t)$, we denote the concentration of complexes of species $j$ and morphology $\zeta$ at time $t$. We use $nn(i)$ as a shorthand notation of the set of indices $\{i+1,i-1,i+L,i-L\}$ with which species $i$ is able to bind. Furthermore, $nn(i,j,\zeta)$ denotes the set of indices of species that are direct neighbors of $i$ and constituents of the complex $(j,\zeta)$. The temporal change of the monomer concentration of species $i$ is then given by
\begin{align}
\begin{split}
 \partial_t m_i = 
 - \mu_{\text{het}} \, m_i \sum_{j\in nn(i)} m_{j} 
 &- 
 \sum_{\substack{(j,\zeta) \\ i\notin (j,\zeta) \\ |nn(i,j,\zeta)|\geq 1}} \nu \, m_i \, c_{j,\zeta}   \\
 &+ 
 \sum_{\substack{(j,\zeta) \\ i\in (j,\zeta) \\ |nn(i,j,\zeta)|= 1}} \delta_1 \, c_{j,\zeta} \, .
 \end{split}
 \end{align}
 The three terms on the right account for the loss and gain of monomers due to nucleation (with rate $\mu_{\text{het}}$), attachment to complexes, and detachment from complexes. For the attachment reactions with monomers of species $i$, we consider all complexes that contain at least one neighbor of species $i$ but not species $i$ itself. Similarly, the complexes from which a monomer of species $i$ can detach are those that contain species $i$ and exactly one neighbor of $i$. Next, we use that the average concentrations of different monomer and complex species must be the same ($m_i=m_j$ and $c_{i,\zeta}=c_{j,\zeta}$) since they are subject to equivalent dynamics. Hence, the dimerization term simply becomes $-4\mu_{\text{het}} m_i^2$. We furthermore note that, for a given cluster morphology $\zeta$, there is exactly one species index for each binding site of the cluster fulfilling the conditions in the sum of the attachment term. Thus, the second term can be written as 
\begin{equation}
-\sum_{\zeta} m_i \, c_{i,\zeta} \, \tilde b(\zeta) \, \nu \, ,
\end{equation}
where $\tilde b(\zeta)$ denotes the number of binding sites of the cluster with morphology $\zeta$, and the sum goes over all possible cluster morphologies. Note that the concentration of a specified cluster species $i$ equals the total concentration of complexes with the same morphology divided by $S$: $c_{i,\zeta}=c_{\zeta}/S$. Finally, for the sum in the last term, there is exactly one species index fulfilling the conditions for each site in the cluster that has exactly one neighbor. Denoting the number of sites with one neighbor for a given cluster morphology by $b_1(\zeta)$, the last term thus becomes
\begin{equation}
\sum_{\zeta} c_{i,\zeta} \, b_1(\zeta) \, \delta_1 \, .
\end{equation}
Hence, by setting $\mu=4 \, \mu_{\text{het}}$, this gives the same dynamic equation for the monomer concentration as one would have for a homogeneous system
\begin{equation}
\partial_t m =
\mu \, m^2 
-\sum_{\zeta} m \, c_{\zeta} \, \tilde b(\zeta) \, \nu 
+ \sum_{\zeta} c_{\zeta} \, b_1(\zeta) \, \delta_1 \, .
\end{equation}
Setting $\mu=4 \, \mu_{\text{het}}$ makes sense intuitively since, in the homogeneous case, two monomers can dimerize in four possible constellations (four possible binding sites). In contrast, in the heterogeneous case, they can only dimerize in one constellation.  \\
Similarly, the evolution of the cluster concentrations can be written independently of the species indices in a form equivalent to that of a homogeneous system. Thereby, we use the notation $\zeta=\zeta'+1$ to indicate that the morphology $\zeta$ equals $\zeta'$ once we add a single particle to the boundary of $\zeta'$. Similarly, $\zeta'+_1 1$ indicates that we add one particle to $\zeta'$ so that it has exactly one neighbor. The evolution of cluster concentrations then reads
\begin{align}
\begin{split}
\partial_t c(\zeta) =
&\sum_{\substack{\zeta' \\ \zeta = \zeta'+1}} \nu \, m c(\zeta')
- \nu \, m \, \tilde b(\zeta) \, c(\zeta)  \\
+
&\sum_{\substack{\zeta' \\ \zeta' = \zeta +_{1} 1}} c(\zeta') \, \delta_1
- \delta_1 \, b_1(\zeta) \, c(\zeta) \, .
\end{split}
\end{align}
Together, this shows that the heterogeneous process decouples into $S$ homogeneous processes, one for each species. In other words, this means that a heterogeneous process with total particle number $N\cdot S$ behaves identically on average as a homogeneous process with particle number $N$. 
Due to this mathematical equivalence, all scaling laws derived for homogeneous systems directly transfer to heterogeneous systems in the ideal case. Next, we check how robust these scaling laws are if we deviate from the ideal case, e.g., by varying the rate constants or the particle numbers of the different species or by assembling structures with explicit boundaries rather than periodic boundary conditions.

\subsection{Random heterogeneous binding rates}
\label{app:sec:heterogen_random_rates}

\begin{figure}[tb]
\centering
\includegraphics[width=0.90\linewidth]{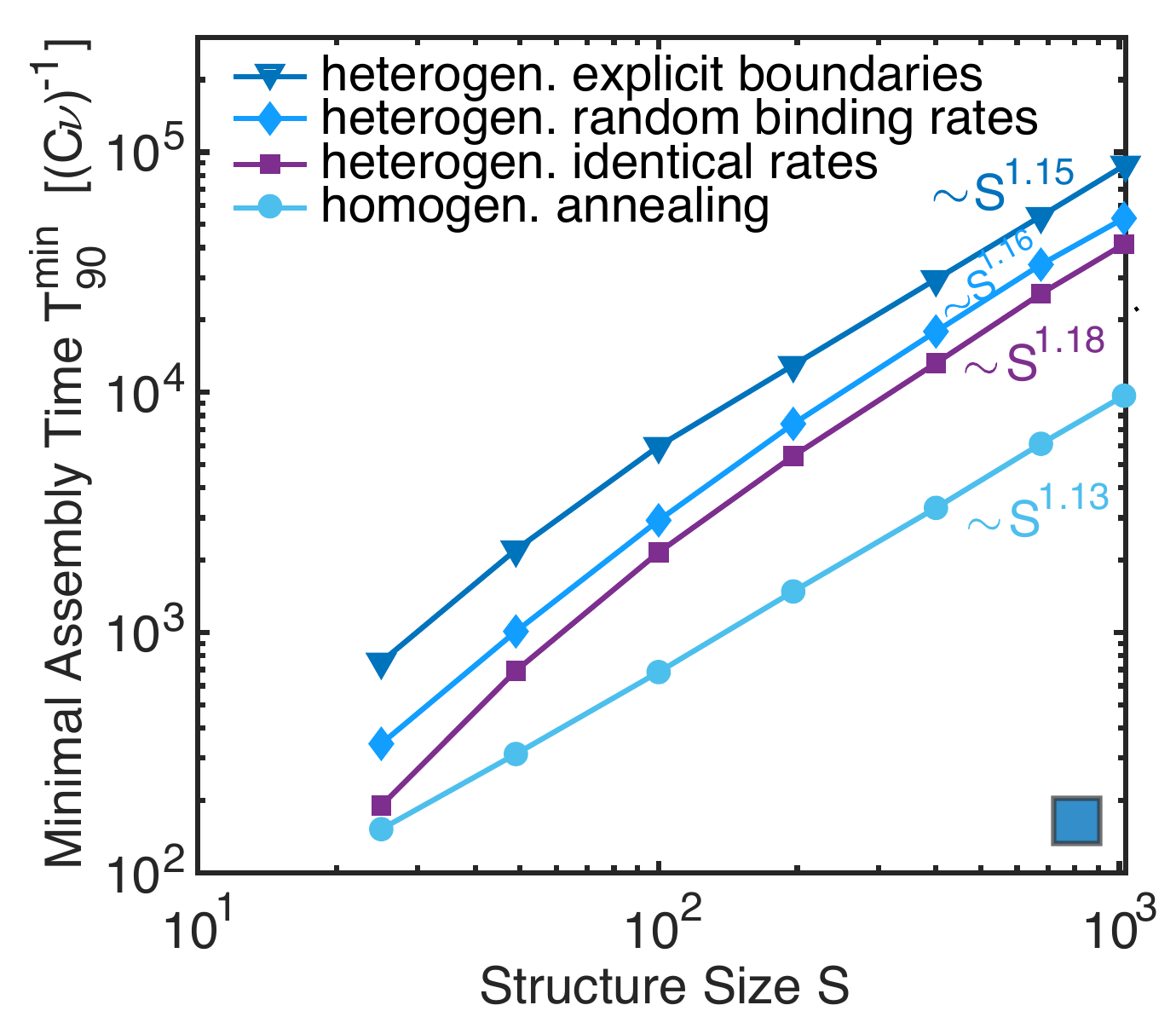}
\caption{\textbf{Robustness of the scaling behavior.} The minimal time required to achieve 90\% yield ($T_{90}^{\text{min}}$) is plotted in dependence of the target structure size $S$ for different variants of the model. Diamond and triangle markers show $T_{90}^{\text{min}}$ for heterogeneous systems with random heterogeneous binding rates drawn from a (truncated) Gaussian with coefficient of variation 50\% [cf.~App.~Sec.~\ref{app:sec:heterogen_random_rates}] or structures with explicit boundaries [cf.~App.~Sec.~\ref{app:sec:heterogen_explicit_boundaries}], respectively. Squares and circles show $T_{90}^{\text{min}}$ for the heterogeneous system with identical rates and periodic boundaries (equivalent to the homogeneous system, cf.~Fig.~\ref{fig:scaling_laws}(b)) as well as for a homogeneous system under annealing (i.e. with a time-dependent detachment rate $\delta_1(t)\sim m(t)$) [cf.~App.~Sec.~\ref{app:sec:annealing}]. Data points were obtained by averaging 10-20 independent runs of the stochastic simulation performed with $N=500$ particles per species (heterogeneous case) or $N=500S$ particles (homogeneous case). While the assembly time varies for the different model variants, the measured time complexity exponents are largely invariant. }
\label{fig:robustness_of_scaling}
\end{figure}
First, we consider a system with random heterogeneous binding rates. Clearly, for distinct species, the assumption of identical binding rates for all species is an idealization. More realistically, the rates will vary to some extent. We, therefore, simulated a system with random heterogeneous nucleation and attachment rates for the different species, drawn independently from a (truncated) normal distribution with a coefficient of variation of $50\%$. The normal distribution was truncated for values below $20\%$ of the mean to ensure that individual rates do not become negative or extremely small. We measure the assembly time accordingly in units of $(C\nu_{\text{avg}})^{-1}$, where $\nu_{\text{avg}}$ is the average attachment rate of all species. For each run, the binding rates were chosen independently, and the assembly times were averaged over 10-20 independent runs. The measured time complexity exponent is identical to the one of the system with homogeneous rates, demonstrating the robustness of the scaling behavior. However, the time efficiency, in general, is reduced as a consequence of heterogeneous binding rates. We attribute this to the fact that the ratio between the average effective nucleation and effective growth rate increases with the coefficient of variation of the distribution of the rate constants (since the average nucleation and growth rate correspond to moments of different order of the distribution). Thus, increasing the coefficient of variation of the rate constants is equivalent to increasing the dimerization barrier $\tilde \mu=\mu/\nu$, which, according to Eq.~\ref{app:eq:complementary_scaling_T_Y}, increases $T_{90}^{\text{min}}$.

\subsection{Structures with explicit boundaries}
\label{app:sec:heterogen_explicit_boundaries}

Heterogeneous systems furthermore offer an opportunity to test the impact of the boundary conditions of the clusters. In the homogeneous case, we considered the self-assembly of closed structures, implemented by periodic boundary conditions of the clusters. The geometry of the clusters thereby limits their further growth. In a heterogeneous system, structure growth is automatically limited if the species required for further growth are missing. Therefore, we think that heterogeneous systems realized, e.g., with DNA bricks as used in references \cite{ke2012three,wei2012complex} provide a valuable testing ground for our scaling results. However, the emerging structures differ in their boundary conditions from the systems we considered in the main text (periodic vs. explicit boundaries)\footnote{Note also that explicit boundaries violate the equivalence between homogeneous and heterogeneous systems [cf.~App.~Sec.~\ref{app:sec:heterogen_identical_rates}] due to the distinct properties of the boundary species.}. It is, therefore, important to understand whether and how the boundary conditions influence the assembly time. Simulating clusters with explicit boundaries, we find that the assembly time slightly increases, but its scaling behavior as a function of the target structure size remains invariant [cf.~Fig.~\ref{fig:robustness_of_scaling}]. Statistically, i.e., when averaging over many simultaneously growing clusters, explicit boundaries merely reduce the effective average growth rate of clusters since clusters stop growing once they reach the boundary. Hence, changing the boundary conditions is effectively equivalent to changing the binding rate $\nu$, which affects the assembly time but not its scaling behavior. This suggests that our scaling results are generally robust with respect to the boundary conditions limiting structure growth.

\subsection{Heterogeneous detachment rates}
\label{app:sec:heterogen_detachment_rates}

The question arises whether, by choosing the reaction rates for the heterogeneous system in a concerted way, the assembly time can be reduced. Since nucleation is the rate-limiting process, a simple idea to achieve this is to increase the effective rate of seed formation by making the bonds between four specified neighboring species very strong. Specifically, to test this scenario, we set the detachment rates for the four bonds between species $1$, $2$, $L+1$, and $L+2$ to zero in the simulation. The intention is that the specified species will quickly form the required seeds that can subsequently grow into complete structures.
However, the simulation reveals that the minimal assembly time increases significantly as a result of this modification of the reaction rates.
In fact, to achieve a yield of 90\%, the uniform detachment rate $\delta_1$ of the remaining bonds had to be chosen significantly larger than in the homogeneous case, making the simulation extremely inefficient. We therefore had to simulate the scenario with a reduced nucleation rate $\mu/\nu=0.1$ (see Sec.~\ref{scaling_theory_dimerization_barrier}) and correspondingly plotted the assembly time in units of $(C\nu)^{-1} \, \mu/\nu$ (assuming a complementary time complexity exponent of 1 as in the homogeneous case [cf.~App.~Sec.~\ref{scaling_theory_dimerization_barrier}]). Figure~\ref{fig:heterogen_detachment_and_concentrations} indeed shows a strong increase of the minimal assembly time as compared to the system with identical rates (purple line in Fig.~\ref{fig:heterogen_detachment_and_concentrations}). The reason for this increase is the same as in the scenario of seeding discussed in App.~Sec.~\ref{app:sec:seeding}: By enhancing the nucleation rate of four selected species, $\delta_1$ has to be chosen much larger for all remaining species to avoid spurious nucleation. This drastically reduces the effective growth rate and hence the overall time efficiency. In fact, if the time scale of nucleation of the four species becomes negligible against the overall assembly time, there is effectively no difference from putting a corresponding number of seeds into the system. Therefore, in the limit of large $S$, the assembly time is approximately the same as for the homogeneous seeded system; compare Fig.~\ref{app:fig:seeding}(c) and dash-dotted line in Fig.~\ref{fig:heterogen_detachment_and_concentrations}. 
Note that the minimal assembly time will not increase if hexagonal instead of square-shaped monomers are used because the growth rate of hexagonal monomers is independent of $\tilde \delta_1$.

\subsection{Heterogeneous concentrations}
\label{app:sec:heterogen_concnetrations}
\begin{figure}[tb]
\centering
\includegraphics[width=0.90\linewidth]{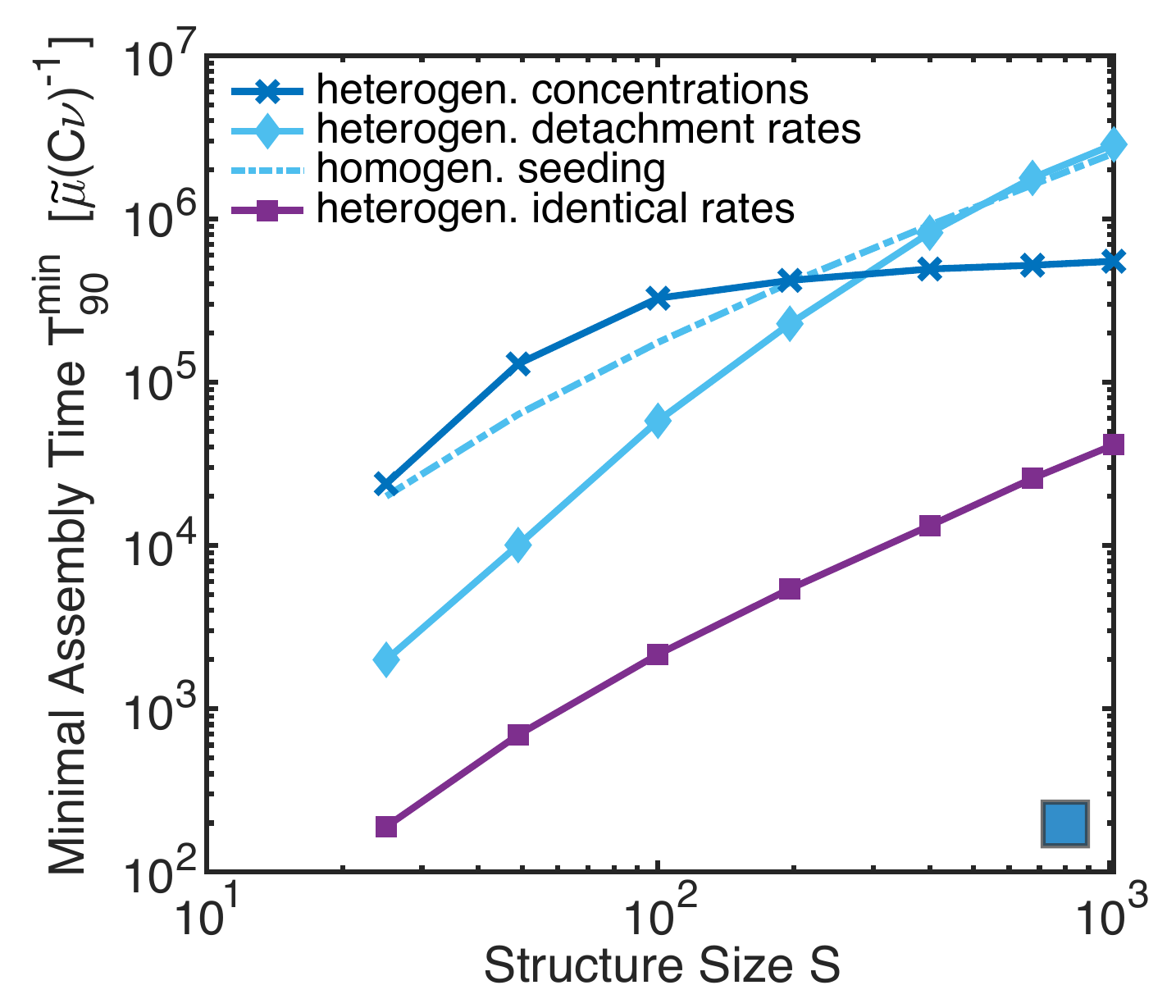}
\caption{\textbf{Heterogeneous concentrations and heterogeneous detachment rates.} Dark blue line with crosses shows the minimal assembly time of a heterogeneous system in which species $1, 2, L+1$ and $L+2$ [cf.~Fig.~\ref{fig:heterogeneous_sketch}] are provided in sixfold higher concentration compared to the other species. Bright blue line with diamond markers shows the minimal assembly time when the detachment rates for bonds between species $1, 2, L+1$ and $L+2$ are set to zero. In both cases, data points were obtained by averaging 10 independent runs of the stochastic simulation performed with $N=500$ particles per species and a dimerization barrier $\tilde \mu:=\mu/\nu=0.1$ to increase the efficiency of the simulations [cf.~App.~Sec.~\ref{scaling_theory_dimerization_barrier}] (note that the y-axis is scaled in units of $\tilde \mu$ accordingly). For comparison, the blue dash-dotted line shows $T_{90}^{\text{min}}$ for the homogeneous system under seeding determined with the effective model (cf.~Fig.~\ref{app:sec:seeding}), and the purple line with square markers shows $T_{90}^{\text{min}}$ for the heterogeneous system with identical rates and concentrations (equivalent to the homogeneous system, cf.~Fig.~\ref{fig:scaling_laws}(b)). }
\label{fig:heterogen_detachment_and_concentrations}
\end{figure}
In ref. \cite{Murugan2015}, the authors describe how the yield in heterogeneous systems can be increased by providing a few neighboring species in significantly larger concentrations compared to the other species. In this way, the authors argued that specific assembly pathways are favored, leading to higher yields as resources are used more effectively. Here, we want to analyze the time efficiency of this strategy. To this end, we set the initial particle number of species $1$, $2$, $L+1$, and $L+2$ to $6N$ while all other species are provided in $N$ copies per species. In order not to restrict the achievable yield, we do not count the additional $20N$ monomers when calculating the yield. Nevertheless, we find that to achieve an elevated yield of 90\%, the detachment rate must be chosen very large, causing the simulation to become extremely inefficient. We therefore again simulate this scenario with a reduced dimerization rate $\mu/\nu=0.1$ (and plot the assembly time in units of $(C\nu)^{-1} \mu/\nu$) as in the last paragraph. As for heterogeneous detachment rates, providing four species in higher concentrations increases the effective nucleation rate. Consequently, the assembly time increases drastically since $\tilde \delta_1$ must be chosen large to suppress spurious nucleation [cf.~Fig.~\ref{fig:heterogen_detachment_and_concentrations}]. Specifically, the rate of nucleation of species $1$, $2$, $L+1$, $L+2$ is $\bar \mu \, (6N)^4$ while the nucleation rate of each of the $S-4$ remaining nucleus species is $\bar \mu N^4$. Hence, providing species in heterogeneous concentrations has a similar effect as scaling the (homogeneous) dimerization rate parameter $\tilde \mu := \mu/\nu$ as $\tilde \mu \to \tilde \mu \, (1+6^4/S)$. With Eq.~\eqref{app:eq:complementary_scaling_T_Y}, this suggests that the assembly time scales roughly as $T_{90}^{\text{min}} \sim 6^4+S$. Hence, as long as $S$ is small compared to $6^4$, we expect the minimal assembly time to increase only very slowly with $S$. 
Indeed, Fig.~\ref{fig:heterogen_detachment_and_concentrations} shows that the assembly time transiently becomes almost constant as a function of $S$. In the limit of large $S$, we expect that the effect of heterogeneous concentrations declines more and more (provided the concentrations remain constant) and the assembly time ultimately behaves similarly as in the homogeneous case (purple line in Fig.~\ref{fig:heterogen_detachment_and_concentrations}). Hence, we find that neither the yield nor the time efficiency can be enhanced through heterogeneous concentrations since heterogeneous concentrations increase the effective nucleation rate, which reduces the efficiency according to Eq.~\eqref{app:eq:complementary_scaling_T_Y}. In their work \cite{Murugan2015}, Murugan et al. considered a system of one-dimensional rings that self-assemble irreversibly. Furthermore, they work in a regime where the yield is very low. We expect that their finding of a yield increase due to heterogeneous concentrations is associated with these system-specificities and has a lot to do with avoiding stochastic effects that arise in irreversible systems \cite{gartner2020stochastic}.

\subsection{Hierarchical heterogeneous self-assembly}
\label{app:sec:heterogen_hierarchical_assembly}

\begin{figure}[tb]
\centering
\includegraphics[width=0.90\linewidth]{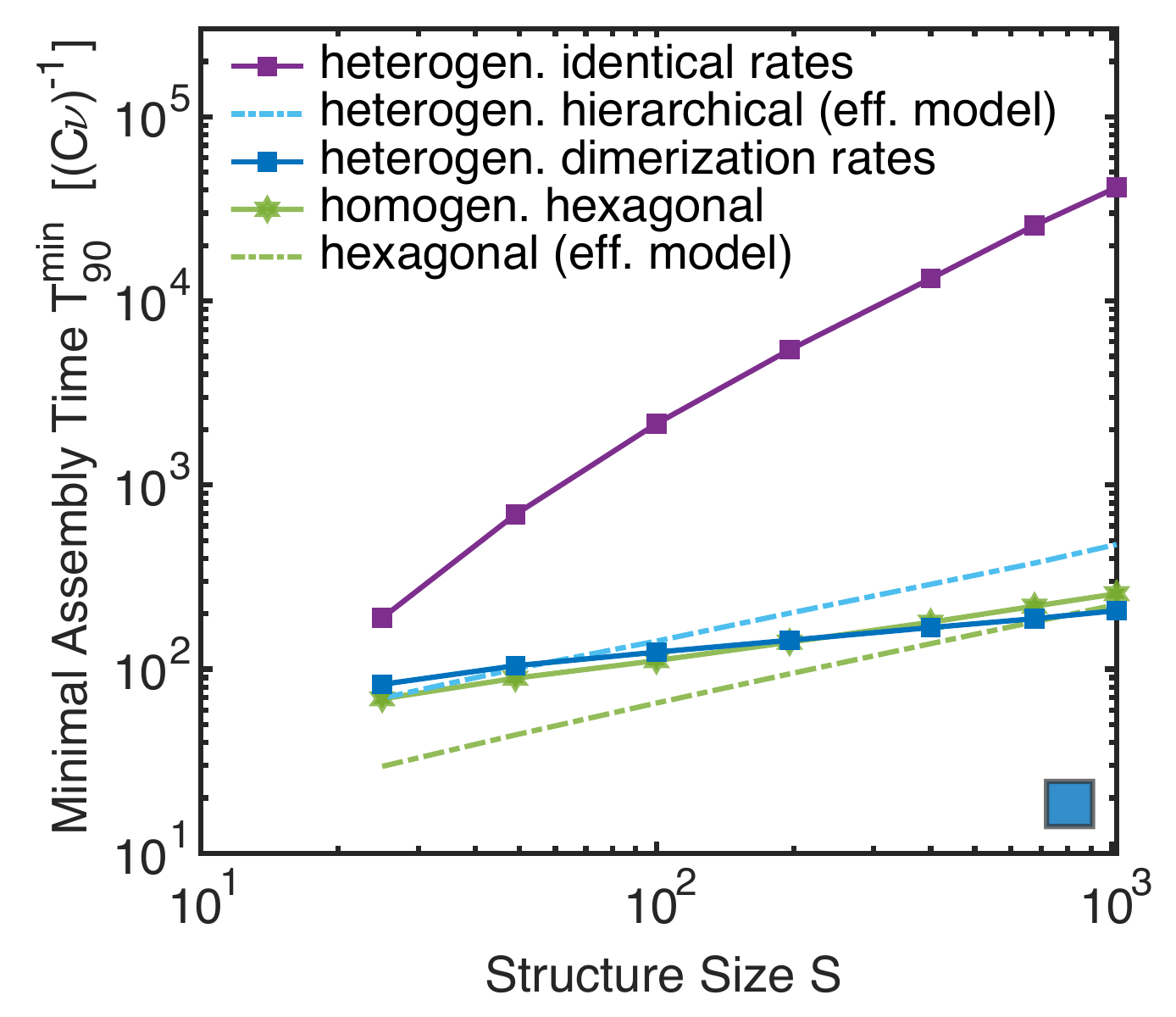}
\caption{\textbf{Improving heterogeneous self-assembly: hierarchical self-assembly and heterogeneous dimerization rates.} Blue dash-dotted line shows the minimal assembly time obtained with the effective model for the hierarchical self-assembly scenario in which square-shaped monomers first form rectangle-shaped dimers ($\delta_1^{\text{strong}}=0$), which then assemble `on gap' as shown in Fig.~\ref{fig:optimal_morphologies_examples}. Notice that this hierarchical scenario can only be realized with heterogeneous building blocks. Dark blue line with square markers shows the minimal assembly time of a heterogeneous system in which only species 1 and 2 can dimerize with rate $\mu_{12}=\nu$ and the dimerization rates of all other pairs of species as well as the detachment rates are zero. Simulations were performed with $N=1000$ particles per species and averaged over 10 independent runs. Purple line with square markers shows the assembly time in a heterogeneous system with identical rates for comparison. Furthermore, green line with star markers and green dash-dotted line show the assembly time of a homogeneous system with hexagonal monomers (stochastic simulation and effective model, respectively) copied from Fig.~\ref{fig:scaling_laws}(b) for comparison. With hierarchical self-assembly or heterogeneous detachment rates the required assembly time can be reduced drastically.}
\label{fig:hierarchical_heterogen_and_heterogen_dim}
\end{figure}

We realize that any measures leading to an increase in the effective nucleation rate are counterproductive in enhancing self-assembly efficiency. The key to increasing time efficiency instead lies in enhancing the speed of cluster growth by reducing the attachment order $\gamma$. Hierarchical self-assembly, as discussed in Sec.~\ref{sec:summary_and_applications}, offers a great opportunity in this regard, particularly for heterogeneous systems. For example, we showed that rectangular building blocks with 6 binding sites that self-assemble `on gap' [cf.~Fig.~\ref{fig:optimal_morphologies_examples} and Sec.~\ref{sec:summary_and_applications}] are topologically equivalent to hexagons and thus self-assemble with the same efficiency as hexagonal monomers. In our heterogeneous system with square-shaped building blocks, this could be realized by making the bonds between every second pair of species (i.e., $1-2$; $3-4$;... and then ($L+2$)-($L+3$); ($L+4$)-($L+5$); etc.) significantly stronger than the other bonds. This will induce the monomers to form rectangle-shaped dimers with 6 binding sites, which then assemble into the final structures with a minimal time complexity exponent $\theta\approx 1/2$. We simulated this scenario with the effective model [cf.~App.~Sec.~\ref{app:effective_model_irr_limit}] analogous to the hierarchical assembly scenarios in Fig.~\ref{fig:hierarchical_assembly} in the main text. Thereby, we simulated the dimerization of monomers with rate $\mu=\nu$ and detachment rate $\delta_1^{\text{strong}}=0$ and the the subsequent self-assembly of the dimers with the same parameters $\sigma=3, \gamma=1$ and $\omega=1/2$ as in the self-assembly of hexagonal monomers. By tuning the detachment rate $\tilde \delta_1^{\text{weak}}$ for the bonds between the dimers, we obtain the minimal assembly time plotted in Fig.~\ref{fig:hierarchical_heterogen_and_heterogen_dim} against the target structure size. Comparing with Fig.~\ref{fig:scaling_laws}(b) (or the green dash-dotted line in Fig. \ref{fig:hierarchical_heterogen_and_heterogen_dim}), we find that the minimal assembly time is roughly a factor 2 larger than in the case of hexagonal monomers (since the dimers first have to form in the first assembly step). 

Hence, the ability to control each bond individually in a heterogeneous system offers plentiful additional possibilities to reduce the attachment order via hierarchical self-assembly and increase the time efficiency relative to a homogeneous system. 

\subsection{Heterogeneous dimerization rates}
Another possibility to increase self-assembly efficiency is by reducing the nucleation rate. In principle, this can be achieved with heterogeneous dimerization rates. Specifically, let us assume that only species 1 and 2 can form a dimer (with rate $\mu_{12}=\nu$), whereas all other pairs of species are unable to form dimers ($\mu_{ij}=0$ for $i,j\neq 1,2$), but the monomers of those species can still attach to existing clusters with rate $\nu$. In nature, such scenarios are quite common. For example, flagellin proteins, the building blocks of flagellae in \textit{E. Coli} do not react freely in the cytosol \cite{kushner1969}. Instead, the molecules only attach irreversibly to the end of an existing flagellum when they are passed through the flagellum \cite{jones1991bacterial}. 

The fact that only species 1 and 2 can dimerize ensures that just the right number of nuclei will form and thus the detachment rates of all species can be chosen arbitrarily small. Thereby self-assembly becomes very efficient. 
Indeed, simulating this scenario with square monomers and detachment rate $\delta_1=0$ [cf.~Fig.~\ref{fig:hierarchical_heterogen_and_heterogen_dim}], we find that roughly the same time efficiency is achieved as with hexagonal monomers where the detachment rate is chosen optimally; compare with Fig.~\ref{fig:scaling_laws}(b). Roughly the same minimal assembly time is also achieved by optimizing the dimerization rate in a homogeneous system [cf.~App.~Sec.~\ref{scaling_theory_dimerization_barrier}].

Controlling the dimerization rate is very effective and, in the heterogeneous case, it does not even require fine-tuning of the rate constants. However, implementing such a scenario with artificial components would be very challenging since it requires a particular molecular design that causes strong allosteric (cooperative) binding effects or the use of specialized enzymes. In contrast, controlling the ratio $\tilde \delta_1$ between detachment and attachment rates is simpler as it can be tuned by several experimentally accessible parameters [cf. discussion in Sec.~\ref{sec:model_description}]. \\

In conclusion, due to the exact equivalence between homogeneous and heterogeneous systems in the ideal case, both types of systems behave very similarly and can be typically related to each other. We found that the scaling properties of the assembly time in heterogeneous systems are robust to noise (e.g., random variations of the rate parameters) and to modifications regarding the boundary conditions. However, the (transient) scaling behavior is affected by concerted modifications, e.g., of the detachment rates or by using heterogeneous concentrations. The resulting effects on the assembly time can qualitatively be well understood on the basis of the results on homogeneous systems.

\clearpage

\bibliography{bibliography.bib}

\end{document}